\documentclass[a4paper,twocolumn,11pt,accepted=2023-09-01]{quantumarticle}
\pdfoutput=1
\usepackage[utf8]{inputenc}
\usepackage[english]{babel}
\usepackage[T1]{fontenc}
\usepackage{amsmath}
\usepackage[colorlinks]{hyperref}
\usepackage[numbers,sort&compress]{natbib}

\UseRawInputEncoding 

\usepackage{tikz}

\usepackage{enumitem}
\usepackage{soul}
\usepackage{mathtools}
\usepackage{amsfonts}
\usepackage{geometry}
\usepackage{wrapfig}
\usepackage{setspace}
\usepackage{titlesec}
\usepackage{graphicx}
\usepackage{amsthm,amssymb,color}
\usepackage{bbm}
\usepackage{bm}
\usepackage{synttree}
\usepackage{caption}
\usepackage{subfig}
\usepackage{multirow}
\usepackage{multicol}
\usepackage{xfrac}
\usepackage{fixltx2e}
\usepackage{comment}

\newcommand{\COMMENT}[1]{}
\newcommand{\blk}{\color{black}}
\newcommand{\david}{\color{red}}
\definecolor{cadmiumgreen}{rgb}{0.0, 0.42, 0.24}
\definecolor{darkmagenta}{rgb}{0.55, 0.0, 0.55}
\newcommand{\rob}{\color{darkmagenta}}
\newcommand{\lorenzo}{\color{blue}}

\newcommand{\ket}[1]{\left| #1 \right>} 

\makeatletter
\def\l@subsubsection#1#2{}
\makeatother

\begin{document}

\title{Why interference phenomena do not capture the essence of quantum theory} 
\author{Lorenzo Catani}\email{lorenzo.catani4@gmail.com}
\affiliation{Electrical Engineering and Computer Science Department, Technische Universit\"{a}t Berlin, 10587 Berlin, Germany}

\author{Matthew Leifer}\email{leifer@chapman.edu}\affiliation{Institute for Quantum Studies and Schmid College of Science and Technology, Chapman University, One University Drive, Orange, CA, 92866, USA}

\author{David Schmid}\email{david.schmid@ug.edu.pl}\affiliation{International Centre for Theory of Quantum Technologies, University of Gdansk, 80-308 Gdansk, Poland}

\author{Robert W. Spekkens}\email{rspekkens@perimeterinstitute.ca}
\affiliation{Perimeter Institute for Theoretical Physics, 31 Caroline Street North, Waterloo, Ontario Canada N2L 2Y5}

\begin{abstract} 
Quantum interference phenomena are widely viewed as posing a challenge to the classical worldview. Feynman even went so far as to proclaim that they are the {\em only mystery} and the {\em basic peculiarity} of quantum mechanics.
Many have also argued that basic interference phenomena force us to accept a number of radical interpretational conclusions, including: that a photon is neither a particle nor a wave but rather a
Jekyll-and-Hyde sort of entity that toggles between the two possibilities, that reality is observer-dependent, and that systems either do not have properties prior to measurements or else have properties that are subject to nonlocal or backwards-in-time causal influences.
In this work, we show that such conclusions are not, in fact, forced on us by basic interference phenomena.  We do so by describing an alternative to quantum theory, a statistical theory of a classical discrete field (the `toy field theory') that reproduces the relevant phenomenology of quantum interference while rejecting these radical interpretational claims.  It also reproduces a number of related interference experiments that are thought to support these interpretational claims, such as the Elitzur-Vaidman bomb tester, Wheeler's delayed-choice experiment, and the quantum eraser experiment. The systems in the toy field theory are field modes, each of which possesses, at all times, {\em both} a particle-like property (a discrete occupation number) and a wave-like property (a discrete phase).  Although these two properties are jointly possessed, the theory stipulates that they cannot be jointly {\em known}.  The phenomenology that is generally cited in favour of nonlocal or backwards-in-time {\em causal influences} ends up being explained in terms of {\em inferences} about distant or past systems, and all that is observer-dependent is the observer's {\em knowledge} of reality, not reality itself. 
\end{abstract}

\maketitle

\tableofcontents


\section{Introduction}

\blk

The third book of Feynman's celebrated Lectures on Physics \cite{Feynman1961} begins with a famous quote concerning the double-slit interference experiment in quantum theory:
\begin{quote}
In this chapter we shall tackle immediately the basic element of the mysterious behavior in its most strange form. 
We choose to examine a phenomenon which is impossible, {\em absolutely} impossible, to explain in any classical way, and which has in it the heart of quantum mechanics. In reality, it contains the {\em only} mystery. We cannot make the mystery go away by ``explaining'' how it works. We will just {\em tell} you how it works. In telling you how it works we will have told you about the basic peculiarities of all quantum mechanics.
\end{quote}
This kind of claim, {\em that the basic phenomenology of quantum interference resists explanation in terms of a classical worldview} and indeed, that it captures the {\em essence} of quantum theory, is widespread.  It is the purpose of this article to dispute it. \blk


What, exactly, is purported to be mysterious about quantum interference?   
Consider the double-slit experiment, where a  quantum system (a photon, for instance) impinges on a plate pierced by two slits, and is subsequently detected on a distant screen.   
The system exhibits wave-like behaviour insofar as it leads to an interference pattern on the screen.  It exhibits particle-like behaviour insofar as it is only ever detected at a single location on the screen and insofar as  the interference pattern on the screen can be made to disappear if a measurement is made of which slit it passed through.  
\footnote{\label{Einsteinfootnote}The puzzle regarding wave-particle duality was central to Einstein's early thinking regarding quanta. 
 Moreover, the standard thought experiment on how to experimentally toggle between particle-like and wave-like behaviour---as with so many thought experiments in the foundations of quantum theory---seems to originate with Einstein.  It is attributed to him by Bohr in the following passage of \cite{Bohr1949}: 
 \begin{quote}
The extent to which renunciation of the visualization of atomic phenomena is imposed upon us by the impossibility of their subdivision is strikingly illustrated by the following example to which Einstein very early called attention and often has reverted. If a semi-reflecting mirror is placed in the way of a photon, having two possibilities for its direction of propagation, the photon may either be recorded on one, and only one, of two photographic plates situated at great distances in the two directions in question, or else we may, by replacing the plates by mirrors, observe effects exhibiting an interference between the two reflected wave-trains. In any attempt of a pictorial representation of the behaviour of the photon we would, thus, meet with the difficulty: to be obliged to say, on the one hand, that the photon always chooses one of the two ways and, on the other hand, that it behaves as if it had passed both ways. 
 \end{quote}
 }  
Thus, a quantum system seems to behave in a manner that is inconsistent with the hypothesis that it is a particle following a classical trajectory, {\em and also} inconsistent with the hypothesis that it is a wave with an amplitude at every point in space.\footnote{Recall that in classical physics, particles and waves are two distinct types of entities.  The essence of a classical particle is to have a definite position in space, while the essence of a classical wave is to have an amplitude at each point in space.} 

 It is often suggested, therefore, that a quantum system is a Jekyll-and-Hyde
  sort of entity, behaving sometimes as a particle and sometimes as a wave, and toggling between these two behaviours depending on the experimental circumstances. 
Furthermore, it is generally thought that there is no possibility of recovering the two sorts of behaviours as two perspectives on a single type of reality for such a system\footnote{That is, it is generally thought that the situation is {\em unlike} what occurs in relativity theory, wherein the differing descriptions of observers moving inertially with respect to one another are recovered as different perspectives on a single reality.}.  
This view can be summarized as endorsing the {\em complementarity} of the wave and particle behaviours, to use the terminology popularized by Bohr. 
 
Moreover, the fact that the experimentalist can freely choose how to measure the system is sometimes taken to imply that they can freely choose whether the system at a given time is a particle or a wave.  This, in turn, is often taken to imply that reality is observer-dependent. 

Finally,  it is often noted that when one has an option of either placing a detector at one of the slits, or not doing so, it is a challenge to provide an explanation of how this choice determines whether there is interference or not.   Specifically, whenever the detector is present at a slit and the quantum system is {\em not} found there, 
one can conclude that it definitely passed through the other slit.  But in that case, 
how could the system know whether or not a detector was in fact present at the first slit, far away?
And yet it seems that it {\em needs} to have this information, as this determines whether or not it is allowed to appear at positions corresponding to the minima of the interference pattern on the distant screen. 
In this sense, the phenomenology of quantum interference seems to resist an explanation in terms of local causal influences. 

To summarize, not only is the basic phenomenology of quantum interference experiments thought to capture the essence of quantum theory, but it is taken by some to support the following three interpretational claims:
\begin{itemize}
\item {\bf Wave-particle complementarity}---i.e., that a photon is neither a particle nor a wave, but rather sometimes one and sometimes the other, with no possibility of a unified description of these two aspects.\footnote{This impossibility of a single unified description is central to Bohr's notion of complementarity, which is why we speak of `wave-particle complementarity'.  The term `wave-particle duality' could also be used here, but its interpretational connotations are less clear.  
 Note that the term `complementarity' is sometimes used in a weak sense to mean a denial of the possibility of jointly {\em measuring} some pair of quantities rather than the stronger sense of a denial of the possibility of their being jointly well-defined.  We have in mind the strong notion here.  Bohr seems to have believed that the weak notion implies the strong notion (textual evidence in favour of this claim is provided in footnote 8 of Ref.~\cite{Bartlett2012}), but the sort of account presented in this article and elsewhere demonstrates that it does not.} 
\item {\bf The observer dependence of reality}---i.e., that an experimenter may determine  whether a photon is a particle or a wave,
 simply by her choice of how to observe the photon.
\item {\bf The failure of explanation in terms of local causes}---i.e., either no causal explanation of quantum interference exists (a kind of anti-realism), or else if such an explanation {\em does} exist, it requires radical types of causal influences, e.g., influences that propagate faster than light  or backwards in time. 
\end{itemize}


 There are several variants of the  standard quantum interference experiment  
   that have been studied and which have been claimed to further elucidate what is mysterious about quantum interference and to further support one or more of the three interpretational claims.

Elitzur and Vaidman have described a dramatic `bomb-tester' thought experiment to argue for the possibility of `interaction-free' measurements and thus for the existence of nonlocal causal influences~\cite{ElitzurVaidman1993}. 

Wheeler proposed what has come to be known as the `delayed-choice experiment' where it seems that one can choose whether a photon behaves like a particle or like a wave based on a decision that is made after the photon is already inside the interferometer~\cite{Wheeler1978}. 
The lesson he took from this was that one must give up on a certain type of realism: namely,
 that the past has no existence until it is recorded in the present. Others have interpreted the experiment as evidence for backwards-in-time causation~\cite{Mohroff1996, Mohroff1999}. 

The quantum eraser~\cite{Scully1982} is an experiment in which information about a photon's path is encoded in an auxiliary quantum system in such a way that one can later choose to measure the auxiliary system and learn which-way information about the photon, in which case there is no interference pattern, {\em or} one can implement a complementary measurement on the auxiliary system which recovers an interference pattern for each of its outcomes.  (In the latter case, the which-way information is thought to be `erased' from the auxiliary system, hence the name.)  The novelty here is that the choice of what to measure on the auxiliary system can be made even {\em after} the interference experiment is over.  This has been interpreted as supporting the conclusion that is often drawn 
from Wheeler's delayed-choice experiment, that whether a quantum system behaves like a particle or a wave inside the interferometer is influenced by a decision that is made in its future. 
In this work, we demonstrate that the basic phenomenology of quantum interference---i.e., the precise phenomenology that has been claimed to capture the essence of quantum theory--- does not, in fact, pose a challenge to the classical worldview, nor force us to accept any of the three interpretational claims. 
We do so by introducing a classical statistical theory that reproduces
  this phenomenology  and that fails to support any of the interpretational claims.
This theory is a version of the `toy theory' introduced by one of the authors in Ref.~\cite{Spekkens2007}.\footnote{A more formal treatment of the toy theory that appeals to symplectic structure and that can be applied to both discrete and continuous-variable systems can be found in \cite{Spekkens2016}.   It is also shown there how for dimensions that are an odd prime, the toy theory reproduces the predictions of stabilizer quantum mechanics \cite{Gottesman1997}, while for continuous variables, it reproduces the predictions of `quadrature quantum mechanics'.   Refs.~\cite{Catani2017,HenautCatani2018} builds on this work by providing a formal characterization of the state update rule for the toy theory and elucidating further the relationship between the toy theory and stabilizer quantum mechanics. Many other works have been motivated in some way by the toy theory~\cite{vanEnk, Pusey2012, Bartlett2012, Larsson2012, Blasiak2013, DeSilvestro, Hausmann2021,Braasch2021}.} Specifically, it is an application of the theory-construction technique of Ref.~\cite{Spekkens2007}, which is a type of `quasi-quantization' procedure, in the language of Ref.~\cite{Spekkens2016}.  This procedure begins with a classical theory of some degree of freedom, considers the statistical version of that theory (wherein an observer may have uncertainty about the physical state) and then imposes an `epistemic restriction', which restricts the allowed states of knowledge of the observer.  We here apply it to the case of a classical discrete field theory.   The fundamental systems of the discrete field theory are {\em  modes}, each of which has {\em both}  a particle-like property (a discrete occupation number) and a wave-like property (a discrete phase).  The epistemic restriction stipulates what states of knowledge about the properties of modes are possible.  For instance, it stipulates that the occupation number and the phase, though simultaneously {\em well-defined}, cannot be simultaneously {\em known}.
We refer to this theory as the `toy field theory'.

We show that if nature was described by the toy field theory rather than quantum theory, then the phenomenology of quantum interference experiments (specifically, those aspects that are typically regarded as problematic) would be reproduced, but none of the interpretational claims would hold.  That is, in the toy field theory,
there is a single reality wherein each system has {\em both} particle-like and wave-like properties at all times, these properties and their dynamics are in no way observer-dependent, and the phenomenology can be explained entirely in terms of local causal influences.\footnote{Along the way, we describe the toy-theoretic counterparts of the second-quantized and first-quantized descriptions of interference experiments and discuss the implications for how best to understand the relationship between these.}

Other realist interpretations of quantum theory, such as Bohmian mechanics~\cite{Bohm1952suggested}, also manage to provide an account of interference phenomenology without positing the observer dependence of reality or wave-particle complementarity and they manage to provide 
an account of {\em some} of the interference phenomenology 
  without violating locality.  A recent article by Blasiak~\cite{Blasiak2015} provides another example of such an approach.
In Appendix~\ref{OnticQInterference}, we discuss the ways in which the toy field theory differs from these models.  We note, in particular, that it preserves a stronger notion of classicality.   Furthermore, we show that it provides a {\em local} account of aspects of interference phenomenology that these models require nonlocality to reproduce, such as the quantum eraser experiment. 
 The toy field theory is thereby distinguished from these prior models insofar as it establishes that this richer interference phenomenology does not imply the failure of explanation in terms of local causes.

It is important to note that there {\em do} exist 
 aspects of the phenomenology of quantum interference that resist classical explanation relative to most 
  notions of classicality, including our own preferred notion (articulated in Sec.~\ref{beyondTRAP}).  For instance, the particular operational predictions appearing in 
  Hardy's proof of Bell's theorem using a pair of overlapping interferometers~\cite{Hardy1992}
  are inconsistent with a locally causal ontological model.  Another such aspect is the precise functional form of the wave-particle duality relation~\cite{catani2022aspects} which resists explanation in terms of a noncontextual ontological model (see Sec.~\ref{beyondTRAP}).  
  What we are arguing herein is that there is nonetheless a large subset of interference phenomena that {\em do not} in fact pose any challenge to a classical worldview, and that, ironically, this subset includes the aspects of interference phenomenology that have been so often touted as ``containing the only mystery''. 
We hope that this work may serve as a corrective to the conventional view on this matter.

The broader lesson, \blk also argued for in Refs.~\cite{Spekkens2007,Spekkens2016,SpekkensPirsaTalk}, is that {\em interpretational claims should be supported by mathematically rigorous no-go theorems}. 
One should not be credulous of statements that a given operational phenomenology implies some interpretational claim {\em unless} the statement is backed up by a rigorous no-go theorem {\em proving} the implication (typically against the backdrop of additional assumptions).


Any such theorem must begin, therefore, with a careful consideration of how to mathematically formulate any given interpretational thesis and its negation, and then proceed to demonstrate that the two alternatives have distinct consequences for the operational phenomena that can be observed.  This is a high bar, but it is the one to which quantum foundations research should be held.

 Note, however, that proving a no-go theorem is not a panacea.   It does not necessarily resolve the question of what are the interpretational implications of a given phenomenology.  Such a theorem is significant only to the extent that its assumptions are reasonable and of broad applicability.  Indeed, the {\em most} significant no-go results are those that elicit an extended investigation into the status of their assumptions.  Such investigations lead to improved versions of the no-go result, or new research directions.  This is where the real value of such results lie.  \blk

John Bell was a proponent of this methodology.  Indeed, Bell's theorem~\cite{Bell1964} is one of the best and earliest no-go theorems of this sort.  Nonetheless, Bell also maintained a healthy scepticism about the best way to get around a given no-go result, famously identifying with ``those who suspect that what is proved by impossibility proofs is lack of imagination''~\cite{BellPilotWave}. 
\blk
 Moreover, Bell also argued for the value of explicit models like the one we will present here. In reference to Bohmian mechanics~\cite{Bohm1952suggested} (a hidden variable model which can reproduce all of the operational predictions of quantum theory),
 Bell stated \cite{BellPilotWave}:
\begin{quote}
Why is the pilot wave picture ignored in text books? Should it not be taught, not as the only way, but as an antidote to the prevailing complacency? To show that vagueness, subjectivity, and indeterminism, are not forced on us by experimental facts, but by deliberate theoretical choice?
\end{quote}
 Bell did not claim that Bohmian mechanics necessarily provides the fundamental description of reality. \blk
Nonetheless, he drew deep foundational insights from it, as is evident from the quote above and the fact that it  set him down the path that led to the discovery of his famous theorem~\cite{BellUnspeakable}.

We take the same attitude here.  The toy field theory is not presented as a candidate for the true theory of nature nor as a candidate for an interpretation of the quantum formalism (indeed, it cannot reproduce the full scope of quantum phenomenology).   That is not its purpose.  Its purpose is merely to show that standard interpretational claims about quantum interference, such as wave-particle complementarity, the observer-dependence of reality, the failure of explanations in terms of local causes,
and the notion that the only possible accounts of interference require a rejection of the classical worldview are ``not forced on us by experimental facts, but by deliberate theoretical choice.'' 




\section{Quantum interference in the Mach-Zehnder interferometer}
\label{Section_3}


The standard arguments that are presented in favour of the three interpretational claims 
  can be made  equally well if one appeals instead to aspects of the phenomenology of a Mach-Zehnder interferometer~\cite{Zehnder1891,Mach1892}, depicted in Fig.~\ref{MachZehnder}, rather than a double-slit experiment.\footnote{The Mach-Zehnder interferometer  is, in fact the original set-up wherein discussions of the puzzling aspects of quantum interference took place, as noted in footnote~\ref{Einsteinfootnote}.
 }
  Indeed, the mathematical similarities are such that the idea that one can use {\em either} to argue in favour of the three interpretational claims is uncontroversial.\footnote{Furthermore, as emphasized in Ref.~\cite{Englert1999}, any two-path interferometer---regardless of whether the paths correspond to positions in space, spin states, energy levels of an atom, or any other pair of distinguishable states---has the same mathematical structure and so it makes no difference which of these one uses to make the argument.  Note, however, that if the paths do not differ in space, then the purported challenge to local causal explanation arises only when one considers the quantum eraser experiment.}
  The Mach-Zehnder interferometer involves only a {\em pair} of spatial paths rather than the continuum of spatial paths involved in the double-slit experiment, while exhibiting an analogous phenomenology of interference. 
It consequently has the advantage of being much simpler to analyze formally, and it is the one that has been the focus of most modern foundational discussions. 
 For these reasons, we will make our argument in the context of the Mach-Zehnder interferometer.
\blk

\subsection{The phenomenology that is traditionally regarded as problematic (TRAP)}

In this section, 
 we will introduce the particular aspects of the phenomenology of quantum interference that are {\em traditionally regarded as problematic}, which we abbreviate as the {\em TRAP phenomenology}.  More precisely, we use this term to refer to the precise aspects of quantum interference phenomenology that appear in arguments in favour of the three interpretational claims listed in the introduction.  The TRAP phenomenology therefore fails to include many aspects of the full phenomenology of quantum interference.  

An example clarifies the point.  Mach-Zehnder interferometers with beamsplitters that yield {\em equal} amplitudes for reflection and transmission (so-called 50-50 or `balanced' beamsplitters) are sufficient for the arguments that are made in favour of the three interpretational claims (as will become apparent below).   As such, even though the full phenomenology of quantum interferometers includes the quantum predictions for experiments with {\em unbalanced} beamsplitters, the latter predictions {\em play absolutely no part} in the argument for the interpretational claims, and therefore they are not 
part of the TRAP phenomenology. 
\blk



\begin{figure}[h!]
\centering
{\includegraphics[width=.5\textwidth,height=.2\textheight]{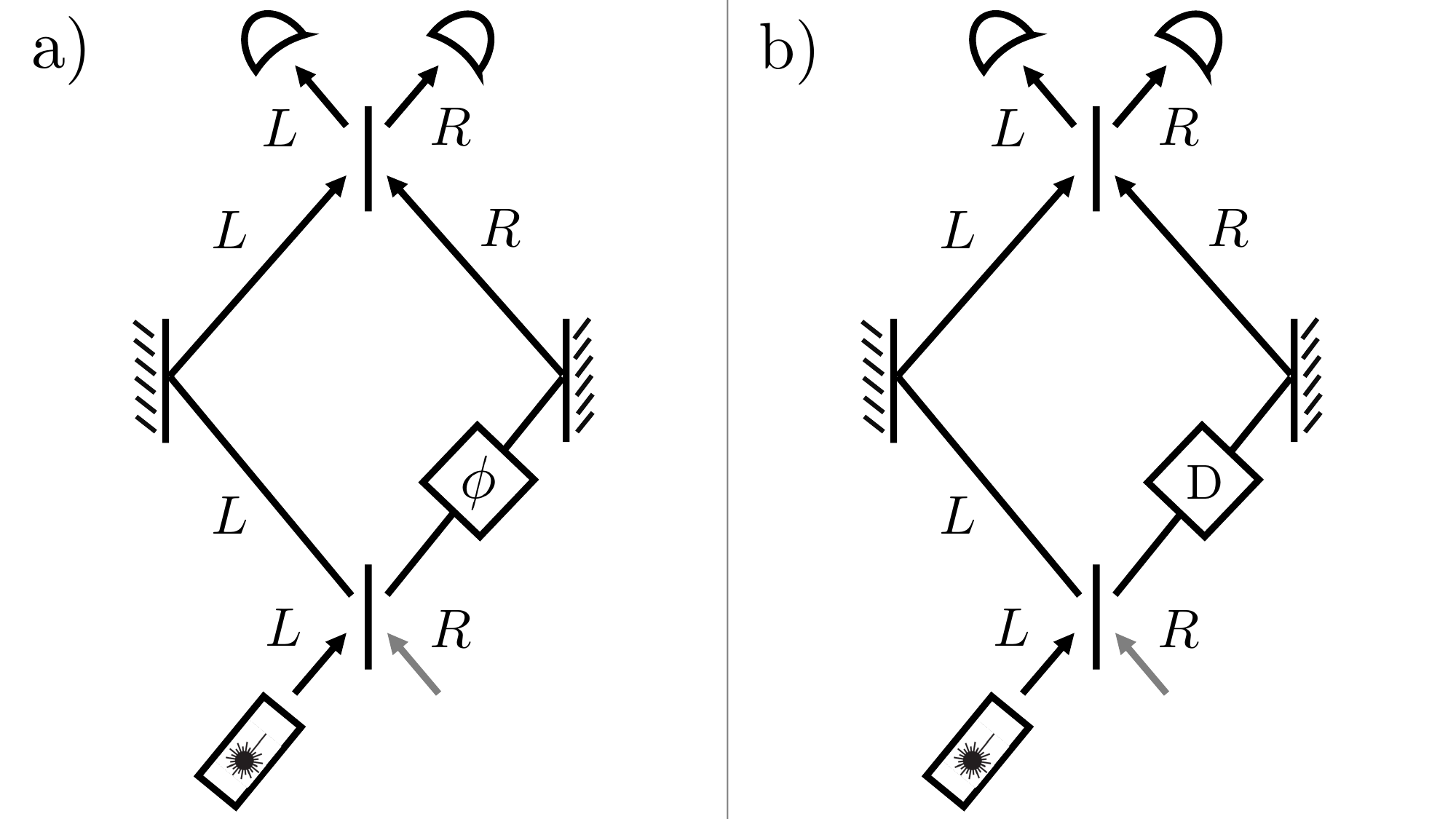}}
\caption{
The set-up of the Mach-Zehnder interferometer, depicting the two possible choices for the device in the $R$ arm: (a) a phase shifter, and (b) a which-way detector. In case (a), the probability of detection in a given output port of the interferometer is a function of the phase shift $\phi$ applied in path $R$.  (In particular, $\phi=0$ leads to the photon always being detected in the $L$ output port, while $\phi=\pi$ leads to the photon always being detected in the $R$ output port). When such constructive and destructive interference occurs, the photon is said to show wave-like behaviour.  In case (b), the probability of detection in either output  port is $\frac12$, such that there is no interference and the photon is said to show particle-like behaviour.
}
\label{MachZehnder}
\end{figure}

We will now consider the two operating modes of a Mach-Zehnder interferometer, illustrated in Fig.~\ref{MachZehnder}, highlighting the TRAP phenomenology and demonstrating explicitly how it is predicted by the formalism of quantum theory.
 \blk


To do so, it is sufficient to consider a which-way degree of freedom associated to the photon, and to model this as a 2-level quantum system, spanned by vectors $\{\ket{L},\ket{R}\}$ associated to paths $L$ (for left) and $R$ (for right).


In the experiments we consider, it is presumed that there is only a single photon in the interferometer at a given time.   For each beamsplitter, the two input ports are labeled $L$ and $R$, and similarly for the two output ports.  
In every run of the experiment, the photon is presumed to be fed into the $L$ input port of the first beamsplitter, and 
to ultimately be detected in one of the output ports of the second beamsplitter.


We presume that there are no phase shifts induced by the photon traveling along a free-space path. Consequently, the only nontrivial evolutions are through each of the two beamsplitters and through the phase shifter or the which-way detector. 
 It is therefore sufficient to consider the quantum state at these four points.   In this sense, we are conceptualizing  the Mach-Zehnder interferometer as a quantum circuit.

We take the 50-50 beamsplitter to be described by the unitary transformation\footnote{Note that the exact unitary describing an actual beamsplitter depends on the physical implementation thereof. For example, the relative optical length seen by the transmitted and reflected beams, and thus the relative phase between them, depends on the physical implementation. 
\blk
}
\begin{align}\label{BSunitary}
&\ket{L} \mapsto \frac{1}{\sqrt{2}}(\ket{R}-\ket{L}),\nonumber\\
&\ket{R} \mapsto \frac{1}{\sqrt{2}}(\ket{R}+\ket{L}).
\end{align}

Note that the action of this unitary on the basis $\{ \frac{1}{\sqrt{2}}(\ket{R}{-}\ket{L}), \frac{1}{\sqrt{2}}(\ket{R}{+}\ket{L})\}$
 is
\begin{align}\label{BSunitaryadditional}
&\tfrac{1}{\sqrt{2}}(\ket{R}-\ket{L}) \mapsto \ket{L},\nonumber\\
&\tfrac{1}{\sqrt{2}}(\ket{R}+\ket{L}) \mapsto \ket{R},
\end{align}
so that it can be conceived of as implementing a swap between the $\{\ket{L},\ket{R}\}$ and 
$\{ \frac{1}{\sqrt{2}}(\ket{R}{-}\ket{L}), \frac{1}{\sqrt{2}}(\ket{R}{+}\ket{L})\}$ bases (and consequently is its own inverse).\footnote{In quantum information theory, such a unitary is called a `Hadamard gate'.}  




In the experiments we consider, it is presumed that the photon is fed into the $L$ input port of the first beamsplitter, so that the initial quantum state is $\ket{L}$. 
The quantum state at the output ports of the first beamsplitter is the image of $\ket{L}$ under the beamsplitter transformation of Eq.~\eqref{BSunitary}, that is, $\frac{1}{\sqrt{2}}(\ket{R}-\ket{L})$.

In the case where the Mach-Zehnder interferometer has a phase-shifter in the $R$ arm, depicted in Fig.~\ref{MachZehnder}(a), the quantum state is then subjected to the
unitary transformation describing the phase shift,
\begin{align}\label{phaseshift}
&\ket{L} \mapsto \ket{L},
&\ket{R} \mapsto e^{i\phi}\ket{R}.
\end{align}
 For the purposes of discussing the arguments presented in favour of the interpretational claims, 
  it is sufficient to consider the case where the 
  phase shift, $\phi$, is either 0 (i.e., no phase shift) or $\pi$, so that the phase factor $e^{i\phi}$ is either 1 or -1.  
  
    In the case of no phase shift, the unitary is just the identity map, so that the quantum state after the phase shifter 
    remains $\frac{1}{\sqrt{2}}(\ket{R}-\ket{L})$.  The second beamsplitter then simply undoes the unitary applied by the first beamsplitter, and the quantum state is mapped back to 
       $\ket{L}$---the photon always emerges  in the $L$ output port of the second beamsplitter.
  
  In the case of a $\pi$ phase shift, the unitary is
  \begin{align}
&\ket{L} \mapsto \ket{L},
&\ket{R} \mapsto -\ket{R},
\end{align}
so that the phase shifter maps 
$\tfrac{1}{\sqrt{2}}(\ket{R}-\ket{L})$ to $\tfrac{1}{\sqrt{2}}(\ket{R}+\ket{L})$ (up to a global phase). 
 Recalling that the action of the second beamsplitter is given by Eq.~\eqref{BSunitaryadditional}, the quantum state at the output of the interferometer 
 is $\ket{R}$.  Thus, under a $\pi$ phase shift, the port at which the photon emerges after the second beamsplitter is the {\em opposite} to the one into which it was fed at the first beamsplitter.

The detectors at the output ports of the second beamsplitter of course implement a measurement of the which-way degree of freedom, that is, a measurement of the basis $$\{ |L\rangle, |R\rangle \},$$  so that we get the $L$ outcome with probability 1 if there is no phase shift and the $R$ outcome with probability 1 if there is a $\pi$ phase shift.

The fact that one output port shows a probability minimum (probability 0 of the photon exiting at this port) and the other a probability maximum (probability 1 of the photon exiting at this port) and that one can vary between these by changing the relative phase in the two arms of the interferometer is the manifestation of destructive and constructive interference  in this experiment.\footnote{The dependence of the output port on the relative phase is analogous to how, in the double-slit experiment, whether a given point on the screen is an intensity maximum or an intensity minimum depends on the path-length difference between the two slits, and thus depends on the phase difference between these two paths. } 


Thus, a photon passing through the Mach-Zehnder interferometer with phase-shifter
 exhibits wave-like behaviour.


We now turn to the Mach-Zehnder interferometer of Fig.~\ref{MachZehnder}(b), wherein there is a detector rather than a phase shifter in the $R$ arm.  The detector is presumed to implement a nondestructive measurement of the photon's which-way degree of freedom,\footnote{Using detectors that implement destructive (rather than nondestructive) measurements would not make any difference in terms of the arguments we run in this article, as we show in appendix~\ref{DestructiveMeasurements}.} that is, a measurement of the $\{|L\rangle, |R\rangle \}$ basis, where the state-update rule is presumed to be the usual projection postulate, that is, 
\begin{align}\label{PPL}
|\psi\rangle \mapsto \frac{ \left( |L\rangle \langle L| \right) |\psi\rangle }{ | \langle L  |\psi\rangle|} = |L\rangle
\end{align}
if the outcome is $L$ and
\begin{align}\label{PPR}
|\psi\rangle \mapsto \frac{ \left( |R\rangle \langle R| \right) | \psi\rangle }{ |\langle R|\psi\rangle|} = |R\rangle
\end{align}
if the outcome is $R$, where the final equalities in Eqs.~\eqref{PPL} and \eqref{PPR} hold up to a global phase factor. 
\blk

%
%
%
%
%
%

  \blk



Recall that after the first beamsplitter, 
 the quantum state of the photon is $\frac{1}{\sqrt{2}}(\ket{R}-\ket{L})$.  Therefore, the probability of each outcome of the which-way measurement is $|\frac{1}{\sqrt{2}}|^2=\frac12$. 

The state of the photon {\em after} it passes through the which-way detector 
  depends on whether or not the detector fired.
If the detector {\em does not} fire, corresponding to the $|L\rangle$  outcome of the $\{ |L\rangle, |R\rangle\}$ basis measurement, then 
the quantum state of the photon becomes
 $|L\rangle$, and the second beamsplitter then maps the state, via Eq.~\eqref{BSunitary}, to $\frac{1}{\sqrt{2}}(\ket{R}-\ket{L})$.
If the detector {\em does} fire, then  the state of the photon becomes $|R\rangle$ and 
the second beamsplitter then maps the state, via Eq.~\eqref{BSunitary}, to 
$\frac{1}{\sqrt{2}}(\ket{R}+\ket{L})$. 
  In both cases, the final measurement is equally likely to find the photon at either output port.



Having a uniform distribution over the output ports corresponds to having neither destructive nor constructive interference.  Thus, a photon passing through a Mach-Zehnder interferometer with the which-way detector in place exhibits no interference, that is to say, particle-like behaviour. \blk


\subsection{The purported implications of the TRAP phenomenology}



We now provide a more detailed 
 account of the arguments conventionally given in defense of the claim that the TRAP phenomenology
 supports the three interpretational claims discussed in the introduction.  This time around, however, we consider these arguments in the context of the Mach-Zehnder interferometer rather than the double-slit experiment.



{\bf Wave-particle complementarity.} 
We have noted that in the version of the Mach-Zehnder interferometer where no which-way information is acquired, the photon exhibits interference, while the interference disappears in the case of the which-way measurement being done.

 As a result of these considerations, it is sometimes argued   that one cannot account for the behaviour of the photon in the two experiments either under the assumption that it is a particle or under the assumption that it is a wave.  The notion that these exhaust the possibilities has led to the view that the photon must be understood to sometimes be a particle and to sometimes be a wave,  depending on how it is probed.



{\bf The observer-dependence of reality.}
Whether or not the photon is subjected to a which-way measurement inside the interferometer is a choice which is at the discretion of the experimenter.   Consequently, if it is thought that the nature of the photon (particle or wave) depends on how it is measured, then the experimenter has the power to determine its nature.  



{\bf The failure of explanation in terms of local causes.} 
Consider the Mach-Zehnder interferometer with the which-way detector on the $R$ arm and imagine  that in a particular run of the experiment, the detector does not fire.
In this case, the photon is inferred to be in the $L$ arm inside the interferometer.  Furthermore, under the assumption that all causes are local, the photon must have {\em already been} in the $L$ arm prior to the which-way detector's outcome having been registered. 
If the particle passed along the $L$ arm, however, then it cannot know whether or not the detector is or is not present {\em in the $R$ arm}, some distance away.  But it requires this information, since whether there is a detector in the $R$ arm or not determines whether it is allowed to sometimes exit the interferometer at output port $R$ (this is allowed when the detector is present) or whether it must always exit at output port $L$ (this is required when the detector is absent).  
 In light of these considerations, it is sometimes argued 
  that  something beyond local causation 
   is required to explain what is observed.  For instance, one might imagine that the photon in the $L$ arm of the interferometer learns about whether the detector is present or not in the $R$ arm through an instantaneous and {\em nonlocal} influence.\footnote{ 
   The idea that quantum interference phenomena imply a kind of nonlocality has been endorsed by many researchers. For instance, Elitzur and Vaidman~\cite{ElitzurVaidman1993} explicitly describe their bomb-tester experiment (which is simply a dramatization of the TRAP phenomenology of interference, as we show in Sec.~\ref{Section_5}) as  ``a novel manifestation of  nonlocality of quantum mechanics''.  The phenomenology of the quantum eraser is also described in this manner by Chiao {\em et al.}~\cite{Chiao1995}: ``Although the nonlocality of quantum mechanics is most apparent in tests of Bell's inequalities, it also plays a central role in experiments exploring complementarity. One such, the quantum eraser, was discussed by Scully, Englert and Walther [...]''.  Finally, Aharonov {\em et al.}~\cite{Aharonov2017} state that the double-slit experiment exhibits a notion of nonlocality termed ``dynamical nonlocality'' and which is described as follows: ``the particle has both a definite location and a nonlocal modular momentum that can `sense' the presence of the other slit and therefore, create interference.''}
In summary, the arguments proceeding from the phenomenology of the double-slit experiment to the three interpretational claims can equally well be made starting from the phenomenology of the Mach-Zehnder interferometer.  Only cosmetic details have changed.

\subsection{Transitioning from a first-quantized to a second-quantized description}\label{First2Second}



Before presenting the toy field theory, we pause to 
present an alternative mathematical formalism for deducing the quantum predictions for the Mach-Zehnder interferometer.
Whereas up until now we have given an account wherein the system of interest is presumed to be something which has spatial location as a degree of freedom (what we have called `the photon' in our discussion thus far),
we turn now to providing an account wherein the system of interest is presumed to be something that can be {\em  parameterized by the points in space} and which has a {\em field strength} as a degree of freedom (i.e., a `mode' of the photon field). 
 In the conventional jargon, we reformulate the quantum predictions within a {\em second-quantized} rather than a {\em first-quantized} description.  This reformulation will be useful to make clear the analogy to the toy field theory that we present in the next section.

We begin by saying a few words about the first-quantized description.  

Consider the double-slit experiment within this description.  
The system of interest is said to have a {\em motional} degree of freedom, describing the position of the system, as well as its momentum (which is canonically conjugate to position).
In quantum theory, such a motional degree of freedom can be represented by the Hilbert space of square-integrable functions, with one basis corresponding to the different possibilities for the position and a complementary (or `mutually unbiased')
 basis corresponding to the different possibilities for the momentum.  

When we shift attention from a double-slit experiment to a Mach-Zehnder interferometer, the first-quantized description still posits a motional degree of freedom, but this can now 
be taken to be significantly simpler, as it only needs to describe a {\em binary} position (a `which-way' or `which-path' variable)  and a canonically conjugate {\em binary} `momentum' variable (stipulating whether the relative phase between the two paths is 0 or $\pi$).  The Hilbert space of square-integrable functions is replaced by a two-dimensional Hilbert space, which is nothing more than a two-dimensional complex vector space. This is the description that was used in the previous section.

We now contrast this with the second-quantized description.

For the double-slit experiment, the systems of interest in this case are a set of field modes, parameterized by locations in space, and each of these has what one might call an `excitational' 
 degree of freedom (to contrast with `motional'). 
For each mode, the infinite-dimensional Hilbert space associated to this degree of freedom is termed its `Fock space', 
 with one basis corresponding to the different possibilities for the occupation number and a complementary 
basis corresponding to the different possibilities for the phase.   
 
 If one focuses on the Mach-Zehnder interferometer rather than the double-slit experiment, it becomes 
 sufficient to consider just a {\em pair} of field modes, corresponding to the left and right paths through the interferometer.   Furthermore, it is sufficient for our purposes to associate to each mode a two-dimensional Hilbert space, 
 wherein one basis corresponds to whether the occupation number of the mode is 0 or 1 (occupied or unoccupied),
  and a complementary 
  basis corresponds to whether the phase of the mode is 0 or $\pi$. We refer to this 
   as a {\em qubit Fock space}, and we refer to the subtheory of the full field theory that is sufficient to describe the Mach-Zehnder interferometer experiments we consider here as the {\em qubit field theory}.
   \footnote{A more accurate terminology would be the {\em stabilizer qubit field theory} or the {\em Clifford qubit field theory}, since the theory allows only the subsets of the states, transformations, and measurements that are 
described in the Stabilizer formalism for quantum computation~\cite{Gottesman1997}, where the symmetry group is the Clifford group.}

Transitioning from a first-quantized to a second-quantized description
 is generally only considered {\em necessary} when one is modelling processes where excitations of the field (i.e., `particles') can be created or destroyed.  Although we here consider only processes that  
 conserve the number of excitations, we will nonetheless find that the second-quantized formalism is central to understanding how the phenomenology of interference can be accounted for with only local causal influences.

We now demonstrate how the quantum predictions about the Mach-Zehnder interferometer, obtained in the previous section within the first-quantized description, can also be obtained within the second-quantized description.




In the qubit Fock space for mode $L$, we denote the quantum states with occupation numbers 0 and 1 by $|0\rangle_L$ and $|1\rangle_L$ respectively, and similarly for mode $R$.   
It is conventional to refer to the quantum state with occupation number 0 as the {\em vacuum quantum state}, and we follow this convention here.

If, in the first-quantized description, one assigns the $|L\rangle$ quantum state to
  the photon, then   in the second-quantized description, one assigns
  the quantum state $|1\rangle_L \otimes |0\rangle_R$ to the 
  pair of modes. Henceforth, we will suppress the tensor product symbol when describing qubit Fock space vectors, so that this state will be denoted simply by $|1\rangle_L |0\rangle_R$.
  
Note that if one is considering only a single photon in the first-quantized description, as we do throughout this work, then in the second-quantized description, the quantum state of the modes will be confined to the subspace having total occupation number 1.  With this in mind, the translation between the first-quantized and second-quantized descriptions is represented by the following map:
\begin{align}\label{translation}
&|L\rangle \mapsto  |1\rangle_L  |0\rangle_R,\nonumber\\
&|R\rangle \mapsto  |0\rangle_L |1\rangle_R.
\end{align}
Other translations follow in a straightforward way from this map, such as the counterpart of the $\{ \tfrac{1}{\sqrt{2}} (\ket{R}-\ket{L}), \tfrac{1}{\sqrt{2}} (\ket{R}+\ket{L})\}$ basis:
\begin{align}\label{translation2}
&\tfrac{1}{\sqrt{2}} (\ket{R} {-} \ket{L}) \mapsto  \tfrac{1}{\sqrt{2}} ((|1\rangle_L  |0\rangle_R {-} |0\rangle_L |1\rangle_R),\nonumber\\
&\tfrac{1}{\sqrt{2}} (\ket{R} {+} \ket{L}) \mapsto  \tfrac{1}{\sqrt{2}} (|1\rangle_L  |0\rangle_R {+} |0\rangle_L |1\rangle_R).
\end{align}

From the fact that the unitary representing the 50-50 beamsplitter in the first-quantized framework is the one given in Eq.~\eqref{BSunitary}, we infer from Eq.~\eqref{translation} that the unitary representing the 50-50 beamsplitter in the second-quantized framework is the map
\begin{align}\label{BSunitaryFock}
&\ket{1}_L \ket{0}_R \mapsto \frac{1}{\sqrt{2}}( \ket{0}_L \ket{1}_R - \ket{1}_L \ket{0}_R),\nonumber\\
&\ket{0}_L \ket{1}_R \mapsto \frac{1}{\sqrt{2}}(\ket{0}_L \ket{1}_R +\ket{1}_L \ket{0}_R).
\end{align}
(As all quantum states we consider will be confined to the total occupation number 1 subspace,
 we do not need to specify how the beamsplitter acts on $\ket{0}_L \ket{0}_R$ and $\ket{1}_L \ket{1}_R$.)

Notice that in the first-quantized description,  the beamsplitter creates a {\em superposition} of quantum states of the photon, whereas in the second-quantized description, it creates {\em entanglement} between the two modes.  





In the second-quantized description, a $\phi$ phase shift in arm $R$ of the Mach-Zehnder interferometer is described by the following unitary transformation on the qubit Fock space of mode $R$:
\begin{align}\label{Qpiphasshift}
&\ket{0}_R  \mapsto \ket{0}_R,
&\ket{1}_R \mapsto 
e^{i\phi} \ket{1}_R.
\end{align}



We now consider the which-way measurement. 
This is associated to the basis $\{ |L\rangle, |R\rangle\}$ in the first-quantized
description, while in the second-quantized
 description it is associated  to the following basis  of the total occupation number 1 subspace: 
  $$\{ \ket{1}_L \ket{0}_R, \ket{0}_L \ket{1}_R\}.$$  
  The latter measurement can be implemented by either implementing a measurement of the occupation number of the left mode, associated to the basis $\{ \ket{0}_L,  \ket{1}_L \}$ of the qubit Fock space of mode $L$, or a measurement of the occupation of the right mode, associated to the basis $\{ \ket{0}_R, \ket{1}_R\}$ of the qubit Fock space of mode $R$. 
 

The state update rule for a repeatable measurement of occupation number on a mode is as follows.  
If the occupation number is found to be 0, the state of the mode being measured updates as
\begin{align}\label{stateupdateFock}
|\psi\rangle \mapsto \frac{\left( |0\rangle \langle 0|\right)|\psi\rangle }{ | \langle 0  |\psi\rangle |} = |0\rangle,
\end{align}
up to a global phase.
If the joint quantum state of the pair of modes is entangled prior to the measurement, 
then the state update rule
 is written as
\begin{align}\label{stateupdateFock0}
|\psi\rangle_{LR} \mapsto \frac{ ( I_L \otimes |0\rangle \langle 0|_R)\; |\psi\rangle_{LR} }{ | \langle 0  |\psi\rangle |}.
\end{align}

The case where the occupation number is found to be 1 is represented analogously.  

The toy field theory is best understood in relation to this second-quantized description of the quantum predictions.
\section{The toy field theory}
\label{Section_4}



\subsection{Description of the toy field theory and its account of the TRAP phenomenology}
\label{ToyTrap}

We now present the details of the toy field theory together with how it reproduces the TRAP phenomenology.
As the definition of the toy field theory follows closely that of the toy theory presented in Ref.~\cite{Spekkens2007} (see also Ref.~\cite{Spekkens2016}), we do not to present it in full generality, but rather focus on just those aspects which are relevant to understanding the TRAP phenomenology.  Consequently, we will interleave together both the account of certain general features of the toy field theory and the description it provides of the Mach-Zehnder interferometer. 

To conceptualize the toy field theory properly, it is useful to consider it in relation to a theory-construction scheme proposed in Ref.~\cite{Spekkens2016}.
The scheme begins with a classical physical theory. Such a theory stipulates the systems that are posited to exist and the attributes that they are posited to possess, hence the system's space of physical states\footnote{What we refer to here as {\em physical states} are often called {\em ontic states} in the quantum foundations literature~\cite{Spekkens2007}, 
from the greek {\em ontos}, meaning {\em reality}.}.  This is its kinematics.   Such a theory also stipulates the possible deterministic evolutions of the physical state space.  This is its dynamics.   The next step in the scheme is to consider the {\em statistical theory} associated with this classical theory of physics.  This is the theory that describes the statistical distributions over the space of physical states and how these change under deterministic evolution or upon acquiring new information about the system (such as by learning the outcome of a measurement).  
The third, and critical step in the theory-construction scheme is to impose a constraint on the statistical theory.  This constraint is usefully understood as a restriction on the statistical distributions that can be prepared,\footnote{While we adopt a view on probabilities as states of knowledge of an agent in a single run of the experiment, our arguments do not rely on this view. We discuss in appendix \ref{EnsembleEpistemicTalk} how to interpret the toy field theory if adopting a view on probabilities as describing the relative frequencies in an infinite ensemble of runs of the experiment.} or more generally as a restriction on the knowledge that an agent may have about the system.\footnote{The states of knowledge that an agent may have about the physical state are typically called {\em epistemic states} in the quantum foundations literature~\cite{Spekkens2007}, from the greek {\em epist\={e}m\={e}}, meaning {\em knowledge}.}
It is termed an {\em epistemic} restriction. The result of this theory-construction scheme, therefore, is an epistemically-restricted statistical theory of classical systems.

The toy field theory is a theory of this sort built on a classical physical theory of discrete fields.  We begin by describing its kinematics in detail.

 In the toy field theory, like the qubit field theory, the systems are modes.\footnote{In appendix \ref{ModesNotParticles} we further discuss the difference between modes and particles, and the importance of adopting the former as the systems of the toy field theory.}  Each mode is assumed to have an occupation number and a phase, and the pair of these properties together constitute the analogue of the `excitational'  degree of freedom of a mode in the qubit field theory.  The toy field theory, however, is classical, meaning that  the occupation number and the phase of a mode are understood as two attributes of the mode that have {\em simultaneously well-defined values.}  A complete specification of the values of these two attributes will be termed the {\em physical state } of the mode. 
 
\blk


We assume that the phase of a mode is discrete, only taking value $0$ or $\pi$.  This implies that the associated phase factor $e^{-i\theta}$ can only take the value $1$ or $-1$.
Noting that the phase factor associated to a phase of $2\pi$ is equivalent to the one associated to a phase of $0$, it is clear that we can represent a discrete phase by a binary variable taking values in the integers modulo 2.
   That is, the phase is assumed to take a value in the set $\{0,1\}$ and addition of phases is computed via modulo-2 arithmetic~\footnote{Modulo-2 arithmetic works like ordinary arithmetic, 
   but loops back to zero when one reaches 2, so that $1 \oplus 1 = 0$.\blk}.  We denote such a binary phase variable by $\Phi$, and addition modulo 2 by $\oplus$ throughout, so that the sum of binary phases $\Phi$ and $\Phi'$ is denoted $\Phi \oplus \Phi'$. Note that in modulo-2 arithmetic, sums and differences are equivalent, so $\Phi \oplus \Phi'$ can also be understood as the phase {\em difference}.  We denote the binary phase variable associated with mode $L$ by $\Phi_L$ and the one associated with mode $R$ by $\Phi_R$.

The occupation number of a mode is also assumed to be a discrete variable, taking the value  $0$ or $1$, which we again represent as taking values in the integers modulo 2 (rather than the natural numbers).
Throughout, we presume that the sum of the discrete occupation number of mode $L$ and mode $R$ is fixed to be 1 and that all transformations preserve this condition.  
It follows that if one mode has occupation number 1, then the other mode has occupation number 0, and vice-versa. 
We denote the binary occupation number variable associated with mode $L$ by $N_L$ and the one associated with mode $R$ by $N_R$. The constraint that the total occupation number is 1 can therefore be represented as the constraint $N_L \oplus N_R = 1$.\footnote{A note is in order regarding the use of the integers modulo 2 rather than the natural numbers.  First, the operation $N \mapsto N \oplus 1$ is the analogue of the unitary associated to the $X$ Pauli operator within the $\{ |0\rangle, |1\rangle\}$ subspace, and {\em not} the analogue of the operation that increases the occupation number by one (in the quantum formalism, the latter would be represented by the creation operator, which is not unitary).  Similarly, the variable  $N_L\oplus N_R$ is not the sum but the {\em parity} of the discrete occupation numbers of modes $L$ and $R$, that is, it is the toy field theory analogue of the quantum observable $Z_L \otimes Z_R$ where $Z_L$ denotes the $Z$ Pauli operator within the  qubit Fock space of the $L$ mode, and similarly for $Z_R$.\blk}


 Concerning the dynamics of the toy field, it is sufficient for our purposes to stipulate that it is local and deterministic. Some concrete examples of valid dynamics will be described further down.

We turn now to describing the epistemic restriction that is imposed on the statistical theory of these discrete fields. 


For a single mode, the restriction on knowledge is straightforward to express: no agent can have knowledge about more than one property.\footnote{When their information about the mode is based purely on data in its causal past, or purely on data in its causal future.}  The agent can know the value of $N$, or the value of $\Phi$, or the value of $N \oplus \Phi$ (the {\em parity} of number and phase), but not the value of more than one of these.  The agent is required to be maximally ignorant otherwise; e.g., if $N$ is known, then the agent assigns a uniform distribution over $\Phi$ and consequently also assigns a uniform distribution over $N \oplus \Phi$.

Note that the epistemic restriction implies that not only can one not {\em prepare} a mode with simultaneously known values of occupation number and phase, one cannot {\em measure} occupation number and phase simultaneously either.

There are also restrictions on what any agent can know about the properties of a {\em pair} of modes.  Specifically, they can only have knowledge of the values of  certain {\em pairs} of variables (as opposed to all four of the variables that define the physical state of the pair of modes).  For instance, they might know one variable about the left mode and one about the right (e.g., the occupation number of both).   Alternatively, they might know a pair of variables that are nontrivial functions of variables describing properties of the left and right modes (e.g., the relative occupation number $N_L\oplus N_R$ and the relative phase $\Phi_L\oplus \Phi_R$).

\subsubsection{Accounts of the initial preparation and of the beamsplitter}

We are now in a position to describe the analogue in the toy field theory of the quantum state $|1\rangle_L |0\rangle_R$ at the input ports of the Mach-Zehnder interferometer.  It is a state of knowledge wherein the occupation number of both modes is known, while their phases are unknown.  More specifically, it is the probability distribution wherein with unit probability, $N_L=1$ and $N_R=0$, while $\Phi_R$ and $\Phi_L$ are uniformly distributed.

The evolution of the occupation number and phase of each mode as the photon traverses the interferometer are governed by simple dynamical laws that are local and deterministic. In particular, the occupation number and phase of one mode only affect those of the other mode when the two are spatially contiguous
 and interact, e.g., at a beamsplitter.  

The dynamics on the physical state of the pair of modes that is induced by the beamsplitter 
is given by the following function:
\begin{align}\label{BSfunction}
&N^{\rm out}_{L} = \Phi_L^{\rm in} \oplus \Phi_R^{\rm in} \nonumber  \\
&N^{\rm out}_{R} =  N_L^{\rm in} \oplus N_R^{\rm in} \oplus \Phi_L^{\rm in} \oplus \Phi_R^{\rm in}\nonumber  \\
&\Phi^{\rm out}_{L} =  N_L^{\rm in} \oplus \Phi_R^{\rm in}\nonumber \\
&\Phi^{\rm out}_{R} =  \Phi_R^{\rm in}.
\end{align}
 This can be equivalently expressed as the following simple rule:
\begin{quote}
{\em The Swap Rule:} The 50-50 beamsplitter swaps the values of $N_L$ and $\Phi_L \oplus \Phi_R$,
 while keeping the values of $\Phi_R$ and $N_L \oplus N_R$ constant.  
\end{quote}
 In the case where $N_L\oplus N_R=1$, so that only one of the modes is occupied, this rule implies that 
 the output mode of the beamsplitter which comes to be occupied is determined by the relative phase of its input modes, and the relative phase of the output modes is determined by which input mode was occupied.

It is straightforward to see how the beamsplitter dynamics transforms an agent's state of knowledge about the physical states of the pair of modes.  At the input of the first beamsplitter, the agent assigns equal likelihood to the phase of the $L$ mode being 0 or 1, and similarly for the phase of the $R$ mode.  Consequently, the agent assigns equal likelihood to the relative phase between the modes being 0 or 1. 
 By the swap rule, there is equal likelihood of each output mode of the first beamsplitter becoming occupied.   
 

Furthermore, because 
it is certain that $N_L=1$ and $N_R=0$ at the input of the first beamsplitter, the Swap Rule dictates that at the output of the first beamsplitter, it is certain that the relative phase of the two modes is 1, $\Phi_L \oplus \Phi_R =1$. 
Meanwhile, the local phase $\Phi_R$ is uniformly distributed at the input to the first beamsplitter and is left unchanged by the swap rule, so that it remains uniformly distributed at the first beamsplitter's output. 


\subsubsection{ Mach Zehnder interferometer with phase shifter}
  We now consider the effect of a phase shifter on the $R$ arm of the interferometer, depicted in Fig.~\ref{MachZehnder}a).

In the case where $\phi=0$, the discrete phase of the $R$ mode is unchanged, so that the two modes meet at the second beamsplitter with their phases undisturbed. Recalling that after the first beamsplitter the relative phase was 1 with unit probability, it follows that it remains 1 with unit probability.  To determine what occurs after the second beamsplitter, it suffices to apply the Swap Rule again.  Swapping the values of $N_L$ and $\Phi_L \oplus \Phi_R$,
 we deduce that the distribution will assign unit probability to $N_L=1$ (and hence $N_R=0$).
   Thus, the detector at the $L$ output port of the interferometer fires with unit probability.
  
On the other hand, in the case where $\phi=\pi$, the discrete phase of the $R$ mode is flipped, and so the relative phase between the modes is flipped as well.  Applying the Swap Rule, one deduces that it is the detector at the $R$ output port of the interferometer that now fires with unit probability.


In short, the toy field theory is seen to reproduce the operational signature of complete constructive and destructive interference.  By varying whether $\phi$ is $0$ or $\pi$, one can vary the output port at which the detection occurs, which is the operational counterpart of varying the location of constructive interference by varying the applied phase shift.

\subsubsection{ Mach Zehnder interferometer with which-way detector}\label{MZIwWhichWay}
  We now consider the case where there is a which-way detector on the $R$ arm, depicted in Fig.~\ref{MachZehnder}b).

Recall that we established earlier that, for the assumed preparation, there is equal likelihood of each output mode of the first beamsplitter becoming occupied.   Consequently, a measurement of the occupation number of mode $R$ is equally likely to have outcome 0 or 1.   The toy field theory therefore reproduces the prediction that the which-way detector in the $R$ arm is equally likely to fire as not. 
We must now consider the impact of a detector in the $R$ arm for what is observed {\em downstream} at the output ports of the interferometer. That is, we consider the measurement update rule for an ideal measurement of occupation number in the toy field theory. \blk
\blk
\begin{quote}
{\em Measurement update rule:} After a nondestructive repeatable measurement of the occupation number of a mode, one assigns zero probability to physical states that are inconsistent with the outcome of the measurement.  Furthermore, the discrete phase of the mode is randomized, i.e., with probability $\frac12$, it is left unchanged and with probability $\frac12$, it is flipped. 
\end{quote}



In other words, one's probability distribution over physical states is updated in two steps: the first step is purely an update of knowledge based on learning the occupation number of the mode, whereas the second step is a disturbance to the phase of the mode. 

This update rule follows from the epistemic restriction and the stipulation of repeatability (i.e.,  if repeated, it yields the same outcome), as is explained in Refs.~\cite{Spekkens2007,Spekkens2016}.  In short, repeatability implies that upon measuring one variable, its value becomes fixed with unit probability to the value revealed by the measurement; consequently, the canonically conjugate variable must be randomized so as to ensure that the two variables are not both known simultaneously.\footnote{It is worth noting that the measurement update rule does {\em not} imply that the toy field theory posits objective stochasticity. This is because the systems that make up the measurement device {\em also} are subject to the epistemic restriction, and it can be shown that whether the discrete phase of the mode is flipped or not is a deterministic function of a property of those systems, a property whose value must be unknown if the measurement device is to function properly. See, e.g., the discussion in Sec.~III.B.8 of Ref.~\cite{Bartlett2012} for more details on how this works in a continuous-variable context; the discrete case is analogous.}

We return now to describing the evolution through the Mach-Zehnder interferometer when the detector in the $R$ arm is in place. 
 
In the case where the detector on arm $R$ {\em does not} fire, i.e., where the measurement reveals that $N_R=0$ (and hence $N_L=1$), one updates the probability distribution by setting the probability of every physical state for which $N_R=1$ (and thus for which $N_L=0$) to zero.    Recalling that it is unknown whether it is the $L$ or the $R$ output port of the first beamsplitter that becomes occupied, 
the first step of the update rule is to resolve this uncertainty.
  In the case where the detector  on arm $R$ {\em does} fire, the only thing that changes is that the distribution evolves by the  knowledge update appropriate for learning that $N_R=1$ (rather than $N_R=0$).  In short, the first step of the update rule {\em resolves one's uncertainty} about $N_R$.\blk 

The second step of the update rule is to randomize $\Phi_R$, the phase of the $R$ mode.  (Note that this randomization occurs regardless of which outcome is obtained in the measurement.) Meanwhile, the phase of the $L$ mode is left unchanged by the measurement on the $R$ mode. 
It follows that the {\em relative phase}, $\Phi_R \oplus \Phi_L$, is also randomized.  


Thus, whereas at the output ports of the first beamsplitter the relative phase is 1 with unit probability, 
 the presence of the detector on the $R$ arm (whether it fires or not) causes the relative phase to become equally likely to be 0 or 1. 

To determine what occurs after the second beamsplitter, it suffices to apply the Swap Rule again.  Swapping the values of $N_L$ and $\Phi_L \oplus \Phi_R$, we deduce that at the output port of the second beamsplitter, it is equally likely to be the case that $N_L=1$ (and hence $N_R=0$) as it is to be the case that $N_L=0$ (and hence $N_R=1$).

It follows that it is equally likely for the detector at the $R$ output port of the interferometer to fire as it is for the detector at the $L$ output port to fire. 




This is the operational counterpart of interference being destroyed by a which-way measurement, and the toy field theory is seen to reproduce it.

Note that both the phase shifter on arm $R$ and the which-way detector on arm $R$ are presumed to only have a causal influence on the physical state of mode $R$.  In this sense, they involve only {\em local} causal influences.

Although the marginal {\em probability distribution} over the physical states of mode $L$ changes as a result of learning the outcome of the measurement on mode $R$, this is not a causal influence, but rather merely an updating of one's {\em knowledge} about the distant system.  In particular, given that the occupation numbers of the pair of modes are known to be anticorrelated, resolving one's uncertainty about one leads one to immediately resolve one's uncertainty about the other.  

\subsubsection{Summary}



At the end of the day, the manner in which the toy field theory reproduces the TRAP phenomenology of the Mach-Zehnder interferometer is very simple, and can be summarized quite briefly.  
The output port of the interferometer that becomes occupied is determined purely by the relative phase between the two modes inside the interferometer.  Because this relative phase is determined by which input port of the interferometer is initially occupied, in every run of the experiment it takes the same value.  It follows that the output port of the interferometer that becomes occupied is also the same in every run of the experiment.  
A $\pi$ phase shift applied in an arm of the interferometer flips the relative phase in comparison to the case where no phase shift is applied, and consequently one can toggle the output port that becomes occupied in every run by toggling the applied phase shift.
 This corresponds to wave-like behavior.  However, when a which-way detector is placed in one arm of the interferometer, the phase of that mode is randomized, 
 thereby randomizing
  the relative phase, 
  thereby randomizing which output port becomes occupied. This corresponds to particle-like behavior.
\blk

The analysis presented in this section is sufficient to understand all that follows in this article.  However, there may be some readers who wish to see a {\em more formal} development of the toy field theory and how it reproduces the TRAP phenomenology. 
  We therefore provide this in Appendix~\ref{toyfieldtheoryformalaccount}.  Specifically, we describe the evolution of an agent's state of knowledge through the Mach-Zehnder interferometer in terms of formal expressions for the probability distribution over physical states.  We also provide a diagrammatic representation of these distributions and their evolution.   Readers who wish to see this development should turn to
  Appendix~\ref{toyfieldtheoryformalaccount} before continuing on.

\subsection{Undermining the standard interpretational claims}

In the previous section, we demonstrated that the TRAP phenomenology of the Mach-Zehnder interferometer can be reproduced exactly in the toy field theory, which is a classical statistical theory with local and deterministic evolution. 
We now explain in more detail how this fact 
undermines the  standard interpretational claims about quantum interference.

\subsubsection{Wave-particle complementarity}
In the standard view, interference phenomena are explained in terms of a Jekyll-and-Hyde
system, one that is sometimes a particle, with all and only the properties of a particle, and sometimes a wave, with all and only the properties of a wave.
In the toy field theory, on the other hand, the systems are modes of a field, and it is presumed that every mode has the same possible properties at all times, namely, an occupation number and a phase.  
While the standard picture requires variability in the nature of the system under investigation,
  the toy field theory posits that the nature of the systems do not change.  At all times, a mode has {\em both} a particle-like property, occupation number, {\em and} a wave-like property, phase. The only variability is in the {\em values} of these two properties and in {\em what is known} about them.

Note that a different alternative to the standard approach is to maintain that there is {\em both} a particle {\em and} a wave.\footnote{\label{Bohmianfootnote} The picture that Bohmian mechanics provides of a single photon in an interferometer is of this sort, with the wave influencing the motion of the particle~\cite{Philippidis1979}.  
As such, the fact that Bohmian mechanics reproduces the predictions of quantum theory is also sufficient to undermine the claim that the TRAP phenomenology forces one to accept wave-particle complementarity.  For further discussion of this point, see Appendix~\ref{OnticQInterference}.}
 But this is {\em not} the approach taken by the toy field theory, because in the latter there is not a pair of things (a particle and a wave), but a single thing (a mode) simultaneously having particle-like and wave-like properties.\footnote{\label{Bohmianfootnote2}More formally, in the toy field theory, for a given mode, the occupation number (the particle-like property) and the phase (the wave-like property) are canonically conjugate to one another, whereas if one posits a particle and a wave, then one is positing different canonically conjugate pairs of variables: the position and momentum of the particle, as well as the amplitude and rate of change of amplitude of the wave at every point in space.}


 Notice that there is nothing particularly novel about assuming that the entities at play are modes of a field, rather than particles.  Indeed, the notion that the only systems appearing in our physical theories should be modes of a field, not particles, is arguably implied by the usual manner of modelling both bosons and fermions in quantum field theory---by associating different types of particles to different types of classical fields and then quantizing these.
The toy field theory conforms to this paradigm, admitting only field modes among its entities.\footnote{What {\em is} unconventional about the toy field theory is that the field is a vector field where the components of the vector, the occupation number and phase, take only {\em discrete} rather than continuous values. }

\subsubsection{Observer-dependence of reality} 
Consider now the claim that the physical reality inside the interferometer depends on what the experimenter decides to observe.  In the toy field theory, while it is true that the values of certain variables are disturbed in different ways depending on whether there is a phase shifter or a which-way detector in place (specifically, whether the phase is changed deterministically or randomized), the variables describing reality do not change. 
Each mode has both a  particle-like property (occupation number) and a wave-like property (phase) at all times.  The changes in the phenomenology associated to changing between a which-way detector and a phase shifter do not force one to accept that the property that is well-defined must change between a particle-like property and a wave-like property, but only that which of these two properties is {\em known} and which is unknown changes.
In the case of the phase shifter, the relative phase is known, while the which-way information (i.e., which mode is occupied) is unknown.  In the case of the which-way detector, where the outcome has been registered, the situation reverses: the which-way information is known, while the relative phase becomes unknown. 

%


\subsubsection{Failure of explanation in terms of local causes} 
Accounting for the TRAP phenomenology has been taken by some to require
an anomalous sort of causal influence (such as a nonlocal one)
or else a 
 rejection of the aspect of realism that 
  stipulates~\cite{Schmid2021unscrambling} that correlations need to be explained causally.  This implication is shown to be invalid, however, because the toy field theory provides a realist causal 
explanation of the TRAP phenomenology that is explicitly local.

The presence of a detector in mode $R$ implies that  a phase flip is implemented with probability 1/2.
The information about whether a phase flip occurred or not
is encoded in the {\em phase} of mode $R$, regardless of the value of the {\em occupation number} of mode $R$.  In particular, this phase information is encoded even if mode $R$ is unoccupied.
Because the physical state of mode $R$ inside the interferometer impacts the physical state at the outputs of the second beamsplitter through a sequence of local interactions, it follows that whether a phase flip was applied or not can come to determine which output port becomes occupied without any recourse to nonlocality. 





\subsection{The significance of the move from the first-quantized to second-quantized descriptions}\label{significanceofthemove}

We presented the toy field theory as an analogue of the second-quantized description of the Mach-Zehnder interferometer.  As it turns out, however, one can also develop a toy theory that is associated to the {\em first-quantized} description of the Mach-Zehnder interferometer.  Because we will want to contrast these two toy theories in this section, and we wish to emphasize that they are analogues of the first-quantized and second-quantized descriptions respectively, we will call them the `first-quantized toy theory' and the `second-quantized toy theory'.  Note that `second-quantized toy theory' is just another name for what we have previously called the `toy field theory'.

 A consideration of the first-quantized toy theory, and how it relates to the second-quantized toy theory,
 holds some important lessons for 
how one ought to interpret the relationship between the first- and second-quantized descriptions, both in the toy theories and in quantum theory.   As we noted in subsection \ref{First2Second}, the standard conception of the first-quantized description of an experiment in quantum theory is that the system is a photon with a motional degree of freedom, rather than a set of modes with excitational degrees of freedom, as the second-quantized description would have it.  However, an examination of the toy-theoretic analogue of the distinction between first-quantized and second-quantized descriptions will reveal that it is perhaps better to think of the first-quantized description as also being about modes, but merely providing a {\em coarse-grained} description of these.

A consideration of the relationship between the first-quantized and second-quantized toy theories 
{\em also} highlights why 
a consideration of the second-quantized description is critical for considerations about locality, in particular, for our argument against the third interpretational claim.  We consider each point in turn.

We will begin by presenting the first-quantized toy theory
in a manner that conforms to the standard conception of the first-quantized description, namely, as one wherein the entities are photons which are the bearers of motional degrees of freedom. 

Thus, we imagine that the system of interest in a Mach-Zehnder interferometer is a single photon and we assume that it has two properties: a discrete which-way property describing whether the photon is on the left or on the right, denoted by the binary variable $W\in \{L,R\}$, and a discrete momentum that is canonically conjugate to this which-way property, which we will call the photon's phase and denote by $\Theta \in \{ 0,1\}$.  One can then define a toy theory by assuming an epistemic restriction, stipulating, for instance, that $W$ and $\Theta$ cannot be known simultaneously.  This theory is of precisely the same form as the one developed in Ref.~\cite{Spekkens2007}.  Using Ref.~\cite{Spekkens2007}, it is straightforward to find the first-quantized-toy-theoretic counterparts of the quantum states, unitaries, and quantum measurements appearing in the first-quantized description of the Mach-Zehnder interferometer.  



Just as the TRAP phenomenology of the Mach-Zehnder interferometer can be accounted for quantumly in either the first-quantized or second-quantized descriptions, so too can it be accounted for using either the second-quantized toy field theory (presented earlier) or the first-quantized toy theory.



The shortest route to proving that this is the case is to simply note that the first-quantized toy theory can be {\em derived} from the second-quantized toy theory by adding the restriction that the total occupation number is 1, in a manner precisely analogous to how, in quantum theory, the first-quantized description of the phenomenology of the Mach-Zehnder interferometer can be derived from the second-quantized description.

The translation is achieved via the following map: 
\begin{align}\label{translationtoy}
 N_L=1,N_R=0 &\mapsto W = L\nonumber\\
  N_L=0,N_R=1 &\mapsto W = R\nonumber\\
  \Phi_R \oplus \Phi_L=0  &\mapsto \Theta=0\nonumber\\ 
    \Phi_R \oplus \Phi_L=1  &\mapsto \Theta=1.
\end{align} 


The variable $N_L \oplus N_R$ is not considered a dynamical variable in the first-quantized toy theory because it is presumed to be constant.  Similarly, although $\Phi_L \oplus \Phi_R$ {\em is} treated as a dynamical variable in the first-quantized toy theory, neither of the local phase variables (i.e., $\Phi_L$ or $\Phi_R$) are.  
For instance,  the joint phase flip on the $L$ and $R$ modes, $(\Phi_L,\Phi_R) \mapsto (\Phi_L\oplus 1,\Phi_R\oplus 1)$, is equivalent to the identity map within the first-quantized toy theory, even though it is clearly inequivalent to the identity map within the second-quantized toy theory. 
  This is the sense in which the first-quantized toy theory is a coarse-graining of the second-quantized toy theory.

As an example of how a state of knowledge gets translated,  consider the state of knowledge in the second-quantized toy theory corresponding to the occupation number of the $L$ mode being known to be 1, the occupation number of the $R$ mode being known to be 0, and the phases of both modes being unknown.  This maps to the state of knowledge in the first-quantized toy theory corresponding to the $W$ variable being known to take the value $L$, and the $\Theta$ variable being unknown. 

As another example, consider the translation of the Swap Rule for the beamsplitter dynamics.
 To describe the translation
  succinctly, it is useful to define a binary variable $\bar{W}\in \{0,1\}$ such that $\bar{W}=0 \leftrightarrow W=R$ and $\bar{W}=1 \leftrightarrow  W=L$.  Then we have
\begin{quote}
{\em Swap Rule in the first-quantized toy theory:} Under the 50-50 beamsplitter, the values of the binary which-way variable $\bar{W}$ and the binary phase variable $\Theta$ are swapped. 
\end{quote}

Similarly, we have
\begin{quote}
{\em Measurement update rule in the first-quantized toy theory:} After a nondestructive repeatable measurement of the which-way variable $W$, one assigns zero probability to the physical states that are inconsistent with the outcome of the measurement.  Furthermore, the phase variable $\Theta$ is randomized, i.e.,  with probability $\frac12$, it is left unchanged and with probability $\frac12$, it is flipped. 
\end{quote}

Accounting for the TRAP phenomenology of the Mach-Zehnder interferometer is very straightforward in the first-quantized toy theory. 

Consider first the case where there is a phase shifter in place. The photon is prepared so that it is certain that the which-way variable has value $L$, i.e., $W=L$.
The first beamsplitter causes the phase variable $\Theta$ at its output
 to track the which-way variable $W$ at its input, and so it is certain that this phase variable has value 1, i.e., $\Theta=1$, after the first beamsplitter. 
The phase shifter then either leaves the value of $\Theta$ unchanged, or flips it, and the action of the second beamsplitter ensures that the which-way variable $W$ at its output tracks the phase variable at its input.   Consequently, the value of the which-way variable at the output  ports of the whole interferometer is unchanged or flipped (relative to its value at the input ports of the whole interferometer) according to whether the phase shifter implemented a 0 or $\pi$ phase shift. This ability to shift the location of the dark output port by changing the phase shift inside the interferometer is the operational signature of interference. 

 The case where there is a which-way detector in place is also simple to analyze.  The detector simply reveals the value of the which-way variable $W$ inside the interferometer and is equally likely to fire or not to fire given that the first beamsplitter causes $W$ to track the value of $\Theta$ at the input ports of the interferometer and given that the latter is uniformly distributed.  Meanwhile, the measurement causes the value of $\Theta$ to be randomized.  This in turn implies that the value of $W$ at the output of the interferometer is random, given that under the action of the second beamsplitter, it tracks the value of $\Theta$.  We have the operational signature of the loss of interference.


A diagrammatic version of this account is provided in Appendix~\ref{lessontoyfieldtheoryformalaccount}.

So the first-quantized toy theory can reproduce the TRAP phenomenology just as easily as the second-quantized toy theory can.

Nonetheless, as we now explain, the second-quantized description 
 is critical for undermining the third interpretational claim.
Specifically, if one focusses {\em only} on the coarse-grained variables of the first-quantized toy theory and one {\em forgets} their definitions in terms of the variables of the second-quantized toy theory (given in Eq.~\eqref{translationtoy}), then it becomes impossible to justify the claim that this account of the TRAP phenomenology is one that appeals only to {\em local causal influences}. 


When a detector is placed on the $R$ arm and does not fire, one can seek to explain the resulting loss of interference in the first-quantized toy theory by imagining that the value of $\Theta$ is randomized, as stipulated by the measurement update rule of the first-quantized toy theory,
 but what is unclear is whether one is warranted in interpreting this randomization as a {\em local} influence of the detector on the photon.  
Indeed, the standard story about the first-quantized description makes it tempting to interpret $\Theta$ as an {\em internal degree of freedom} of the photon, and that consequently it is a property that can only be accessed {\em at the location of the photon}.  If one does so, then when the detector on the $R$ arm {\em does not} fire, meaning that the photon took the $L$ arm inside the interferometer (i.e., $W=L$), one is led to conclude that one {\em cannot} imagine the randomization of the value of $\Theta$ to be the result of a local causal influence from the detector on the $R$ arm.

By contrast, if one interprets $W$ and $\Theta$ simply as coarse-grainings of the variables $N_R, N_L, \Phi_R$ and $\Phi_L$ of the second-quantized toy theory, then $\Theta$ is simply the relative phase of the two modes, i.e., it is defined to be $\Phi_R \oplus \Phi_L$, and consequently $\Theta$  can be randomized by randomizing either $\Phi_R$ or $\Phi_L$, and each of the latter operations {\em can} clearly be achieved by acting on a single arm through a local causal influence. In this approach, $\Theta$ is clearly a global property of the pair of modes. Note that it is not a {\em holistic} property, since its definition is given entirely in terms of the properties of mode $R$ and of mode $L$. Such global but nonholistic properties are common in physics.  The centre of mass of a collection of particles is a good example---it can be modified only if the location of one or more of the particles is modified.  Just as the concept of centre of mass poses no challenge to the notion of reductionism, neither does the property $\Theta$ when it is understood as the relative phase of the pair of modes. 



To sum up, the possibility of a local explanation of the phenomenology is {\em only manifest} in a second-quantized or field-theoretic description.  The fact that most previous discussions of interference phenomenology (and accounts thereof) have made use of a first-quantized description, and that it was {\em not} standard to understand this as a coarse-graining of the second-quantized description, may partially explain why researchers previously did not recognize the possibility of a local causal explanation of the TRAP phenomenology.  These considerations highlight the significance of the field-theoretic perspective in discussions of locality in quantum theory.


\section{Implications for some related experimental scenarios}

\subsection{The Elitzur-Vaidman bomb tester}
\label{Section_5}



The Elitzur-Vaidman bomb-tester, introduced in Ref.~\cite{ElitzurVaidman1993}, is a well-known way of making the puzzling features of quantum interference phenomena more dramatic.

Imagine that a way of constructing bombs has been devised such that the detonator is a transparent trigger that is activated whenever light passes through it and which is so sensitive that even a single photon is sufficient to cause the bomb to explode.  Obviously, such bombs must be kept in complete darkness until they are ready to be used.
Now suppose that the manufacturing technique is imperfect, so that some of the bombs are faulty. One can imagine, for instance, that faulty bombs fail to have a trigger. Unlike the functional bombs, the faulty bombs do {\em not} explode when a photon passes through the region where the trigger should be.
The task of interest is to find a means of verifying that a bomb is functional {\em without causing it to explode}. 

\begin{figure}
\centering
{\includegraphics[width=.5\textwidth,height=.21\textheight]{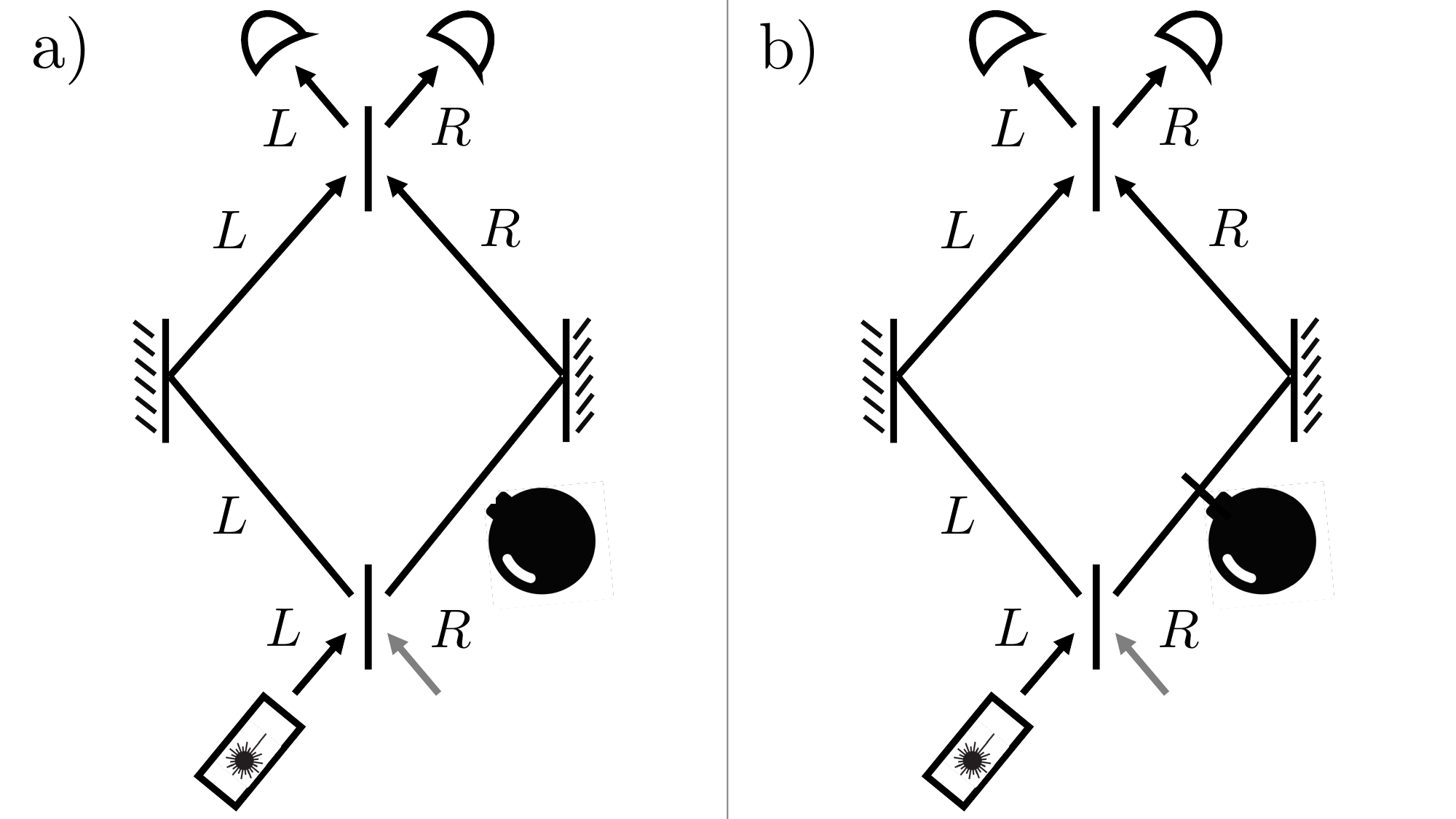}}
\caption{ The Elitzur-Vaidman bomb tester dramatizes the puzzling features of interference. 
 A Mach-Zehnder interferometer can be used to determine whether a bomb is faulty or functional, purportedly without any local interaction between the photon and the bomb.  The case of a faulty bomb (having no trigger) is depicted in (a), while that of a functioning bomb (having a transparent trigger sensitive to a single photon) is depicted in (b).  These correspond to a Mach-Zehnder interferometer with or without a detector on the $R$ arm. 
}
\label{MachZehnderBomb}
\end{figure}

What Elitzur and Vaidman showed is that the phenomenology of the Mach-Zehnder interferometer
 can be leveraged to solve this problem. 

To begin, one places the bomb in such a way that if it has a trigger, then this intercepts the $R$ arm of the interferometer, as shown in Fig.~\ref{MachZehnderBomb} .  
  Because a functional bomb explodes if and only if a photon impinges on it, a functional bomb with its trigger placed in the $R$ arm behaves precisely like a which-way detector placed in the $R$ arm, and consequently one has precisely the same operational behaviour as in Fig.~\ref{MachZehnder}b): there is no interference, so that the photon is equally likely to be detected in either output port of the second beamsplitter.
If, by contrast, a faulty bomb is placed in the $R$ arm, the situation is that of Fig.~\ref{MachZehnder}a) with a standard Mach-Zender interferometer implementing no phase shift. It follows that for a faulty bomb, one has interference, with the photon always emerging in the  
$L$ output port of the second beamsplitter.

In summary, for a functional bomb that does not explode, 
there is probability $\frac12$ of the photon being found in the $R$ output port of the second beamsplitter,
whereas for a faulty bomb, there is probability $0$ of this occurring. Consequently, if there is no explosion and a detection is made at the $R$ output port, one can conclude with certainty that the bomb is functional.
 Given that there is a probability $\frac12$ that a functional bomb will not explode (corresponding to the which-way detector on the $R$ arm not firing), and there is a probability $\frac12$ that subsequently the photon is found in the  $R$ output port of the second beamsplitter, 
it follows that, using this scheme, there is a probability $\tfrac{1}{4}$ that a functional bomb can be identified as such without exploding it. 


The Elitzur-Vaidman bomb tester provides a particularly clear demonstration of why the TRAP phenomenology of the Mach-Zehnder interferometer is often  thought to resist explanation in terms of local causal influences.
\blk
In a case where the bomb is functional but does not explode, one can conclude with certainty that inside the interferometer, the photon took the path along the $L$ arm rather than the $R$ arm, since if it {\em had} taken the path along the $R$ arm, the bomb, being functional, {\em would have} exploded.  But if the photon took the $L$ arm, then it seems that it could not have acquired any information about whether the bomb was functional or faulty.  However, the photon {\em needs} to have acquired this information in order to know whether or not it is allowed to exit the second beamsplitter via the outport port $L$. 

Based on this account, Elitzur and Vaidman described the bomb-tester as implying `interaction-free measurement', on the grounds that the photon has gained information about the bomb (namely, whether it is functional or faulty) without interacting with it. 
They also argue that because the photon does not interact locally with the bomb, no local causal explanation of the phenomenon is possible.  
Indeed, they describe the bomb-tester 
 in the abstract of Ref.~\cite{ElitzurVaidman1993} as ``a novel manifestation of the nonlocality of quantum mechanics.''

Others have taken 
the phenomenology of the Elitzur-Vaidman bomb-tester
 to imply something different from the existence of nonlocal causal influences.  Penrose, for instance, has suggested that it
 implies what we will term the {\em causal efficacy of possibility}~\cite{Penrose1994}.  Suppose that in actuality, the bomb did not explode and the photon took the $L$ path.  Nevertheless, there is a counterfactual world where the bomb {\em did} explode and the photon took the $R$ path.  
If the photon can be influenced not only by what is occurring in reality (at its location), but by what occurs in counterfactual worlds (at its location), then one can explain the different phenomenology resulting from functional and faulty bombs without necessitating nonlocal influences.  Penrose indeed suggests that the make-up of {\em possible} but {\em counterfactual} worlds has a causal influence on what occurs in the actual world.\footnote{
He puts it as follows (note that in his account of the bomb-tester, a faulty bomb is one whose detonator is jammed):
\begin{quote}
Classically, as the problem is phrased, there is no way of deciding whether the bomb detonator has jammed other than by {\em actually} wiggling it---in which case, if the detonator is not jammed, the bomb goes off and is lost. 
Quantum theory allows for something different: a physical effect that results from the possibility that the detonator {\em might} have been wiggled, even if it was {\em not} actually wiggled!  What is particularly curious about quantum theory is that there can be actual physical effects arising from what philosophers refer to as {\em counterfactuals}---that is, things that might have happened, although they did not in fact happen. (Ref.~\cite{Penrose1994}, p. 240)
\end{quote}
}
This view can clearly be subsumed under
  the overarching claim that whatever the causal explanation might be, it is not one in terms of simple local causes among actual entities, but something more radical.

The account provided by the toy field theory, however, shows that one is not forced to such radical interpretational conclusions. 


As noted above, the bomb-tester experiment is simply the usual Mach-Zehnder interferometer where the distinction between a functional and a faulty bomb is the distinction between a which-way detector being in place in the $R$ arm of the interferometer and no such detector being in place there.
 Consequently, we have already seen how the phenomenology of the bomb-tester can be reproduced in the toy field theory.  Specifically, if the bomb is functional and does not explode, one applies the measurement update rule for the outcome $N_R=0$.  This stipulates that the phase of mode $R$ is randomized, which implies that the relative phase between the modes prior to the second beamsplitter is randomized, which in turn implies (by the Swap Rule) that the occupied output port after the second beamsplitter is randomized. 
Note that the phase is randomized even though the $R$ mode is unoccupied (for further discussion of this point, see Sec.~\ref{whynotearlier}).
 Consequently, for a functional bomb,  there is a  $\frac12$ probability that the $R$ output port of the second beamsplitter becomes occupied and signals that the unexploded bomb is not faulty.
 
Therefore, the phenomenology of the bomb-tester does not force one to accept that a measurement has occurred without interaction.  In the toy field theory, when the bomb is functional, there {\em is} a nontrivial interaction, indeed a completely {\em local} interaction between mode $R$ and the bomb.  Furthermore, half of the time, this local interaction results in a change to the physical state of the $R$ mode, namely, a flip of its discrete phase.  

So we have seen that the precise aspects of the phenomenology of the Elitzur-Vaidman bomb-tester that are purported to be evidence for interaction-free measurement and nonlocality are in fact reproduced in the toy field theory using only local influences and genuine interactions. 
This demonstrates that radical notions such as interaction-free measurement, nonlocal influences, or the causal efficacy of possibility {\em are not required} to explain the phenomenology in question.



\subsection{Wheeler's delayed-choice experiment}
\label{Section_6}

In 1978, Wheeler introduced a version of the double-slit interference experiment now known as the ``delayed-choice'' experiment~\cite{Wheeler1978}. In his own words, the motivation was the following question: 
 ``Can one choose whether the photon {\em shall have} come through both of the slits, or only one of them, after it has {\em already} transversed the screen?''  His thought experiment is meant to provide evidence that the answer to this question is `yes'.  
Wheeler offered several versions of his experiment, all equivalent for the purposes of his argument. In each of these, the key feature is that the experimenter ultimately makes the choice of whether or not to measure the which-way information 
{\em after} the photon has already passed through the slits.
We here present a version of Wheeler's argument that retains this key feature, but is framed in terms of the particular Mach-Zehnder interferometer we have focussed on here.

Wheeler's starting point is Bohr's notion that the behaviour of a photon depends on the whole experimental arrangement.  He grants that Bohr is correct to say that the photon behaves like a wave in the context of the experimental arrangement of Fig.~\ref{MachZehnder}a) (Mach-Zehnder with phase shifter) while it behaves like a particle in the context of  the experimental arrangement of Fig.~\ref{MachZehnder}b) (Mach-Zender with which-way detector).  The question he wishes to pose, however, is this: what if the choice between these two experimental arrangements is made {\em after} the photon has already passed through the first beamsplitter? 

 The delayed-choice experiment that we consider, therefore, is the standard Mach-Zehnder interferometer but where the experimenter makes a last-minute decision of whether to insert a which-way detector or a phase shifter on the $R$ arm. 
The two possibilities in question are exactly those illustrated in Fig.~\ref{MachZehnder}(a) and Fig.~\ref{MachZehnder}(b).

Suppose one is of the opinion that (i) the photon must have passed through both arms of the interferometer
  for there to be interference, and (ii) the photon must have passed through  just one arm of the interferometer 
   in order for there to be {\em no} interference.  In this case, it would seem that a last-minute decision of whether or not to insert a which-way detector on the $R$ arm must somehow determine the behaviour  of the photon inside the interferometer {\em at an earlier time}.
Wheeler calls this ``an apparent inversion of the normal order of time''.

For some, this inversion is not merely `apparent', but actual.  Indeed, Wheeler's experiment has been interpreted as evidence of retrocausality, that is, of backwards-in-time causal influences~\cite{Mohroff1996, Mohroff1999}. Wheeler himself did not endorse this conclusion, but the moral he did draw from it was equally radical.  He concluded that the past of the particle {\em does not exist} until the measurement is performed. In his own words \cite{Wheeler1978}:
\begin{quote}
Does this result mean that present choice influences past dynamics, in contravention of every formulation of causality? Or does it mean, calculate pedantically and don't ask questions? Neither; the lesson presents itself rather as this, that the past has no existence except as it is recorded in the present. It has no sense to speak of what the quantum of electromagnetic energy was doing except as it is observed or calculable from what is observed. More generally, we would seem forced to say that no phenomenon is a phenomenon until---by observation, or some proper combination of theory and observation---it is an observed phenomenon. The universe does not ``exist, out there,'' independent of all acts of observation. Instead, it is in some strange sense a participatory universe. 
\end{quote}
The lesson that Wheeler drew from the experiment, in other words, is that one is forced to reject realism, 
at least at the level of quantum systems.


We now consider what the toy field theory implies about Wheeler's delayed-choice experiment.

In the toy field theory, there is a matter of fact about {\em both} the which-way property (the occupation numbers of each mode) and the relative phase property (the parity of the phases of the two modes) 
after the first beamsplitter but before the which-way detector.  The choice of whether to put the which-way detector in arm $R$ or not only determines which of these properties {\em can be inferred}.

Specifically, in the case where the which-way detector on arm $R$ is in place, whether the detector fires or not allows one to infer which arm of the interferometer is occupied both after and before the measurement, 
but it causes one to become ignorant about the relative phase after the measurement.
By contrast, in the case where the which-way detector on arm $R$ is left out, one knows the relative phase to be zero 
inside the interferometer
 but one remains ignorant of the which-way property. 

The choice to put the which-way detector into arm $R$ does not {\em cause} the which-way property at an earlier time to become well-defined or to take a definite value, it merely provides the experimenter the opportunity to {\em learn} the value that it possessed all along.  Similarly, the choice to leave the which-way detector out does not {\em cause} the which-way property at an earlier time to {\em fail} to be well-defined or to take a definite value,
 it merely means that one {\em does not  learn} this value.

Consequently, the phenomenology of the delayed choice experiment does not force us to accept retrocausality\footnote{An argument in favour of this conclusion is also provided in Ref.~\cite{Chaves2018}. 
} or to give up on realism.

\subsection{The quantum eraser}
\label{Section_7}

%
%
%
%
%
%

\blk



The final example we consider is the quantum eraser experiment, proposed by Scully, Englert and Walther~\cite{Scully1991}.  For a presentation that focusses on its abstract structure, see Refs.~\cite{Englert1999,hillmer2007yourself}.





\begin{figure}[h!]
\centering
{\includegraphics[width=.5\textwidth,height=.2\textheight]{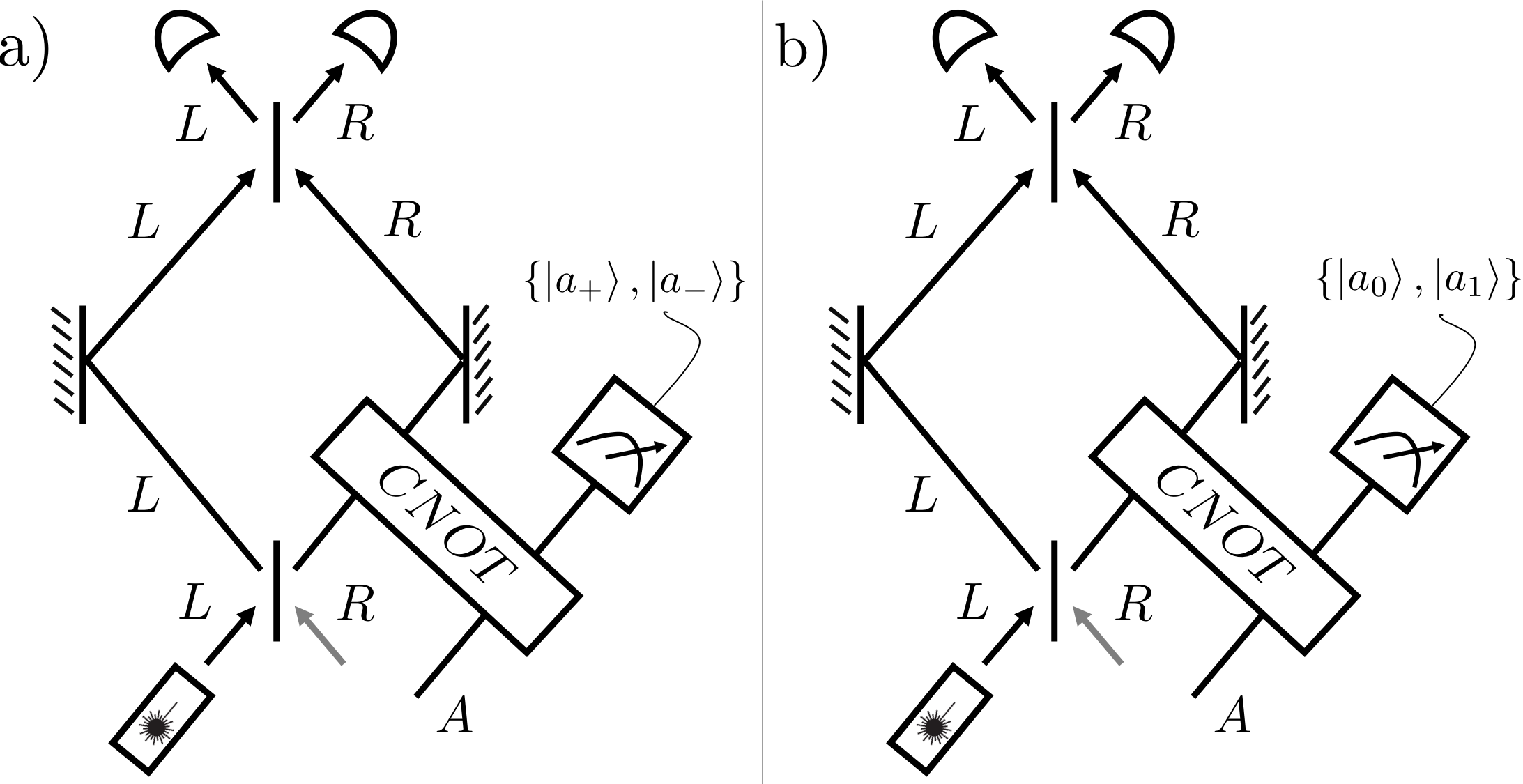}}
\caption{Quantum eraser. Depending on which measurement is performed on the ancilla $A$ coupled to mode $R,$ the which-way information about the photon can be obtained -- thus inducing a particle behavior, or the interference pattern can be preserved/restored -- witnessing a wave behavior. 
}
\label{MachZehnderAncilla}
\end{figure}

We introduce it here as a modification of the basic Mach-Zehnder set-up.  Note, first, that in the latter, the measurement implemented by a which-way detector on the $R$ arm can be modelled as a two-stage procedure consisting of (i) a particular interaction between the photon and an auxiliary system, and (ii) a subsequent measurement of a particular sort on the auxiliary system.  But once this model is contemplated, it becomes apparent that one could also choose to implement a different measurement on the auxiliary system, one that does not reveal the which-way information.  The quantum eraser experiment includes both of these possibilities.  The set-up is depicted in Fig.~\ref{MachZehnderAncilla}.




We begin with an analysis in the first-quantized description.
The auxiliary system, denoted by $A$, is imagined to be a qubit.  We denote the state in which it is initialized by $\ket{a_0}_A$.

Next, we imagine a controlled-NOT gate (CNOT), with the which-way degree of freedom of the photon as the control qubit and the auxiliary system $A$ as the target qubit.  Specifically, the states in the orthogonal basis $\{ \ket{a_0},\ket{a_1}\}$ of $A$ are interchanged (i.e., a NOT operation is applied in this basis) if and only if the photon is in the $\ket{R}$ state.  
Explicitly, the CNOT gate is defined as the following linear map 
\begin{align}\label{CNOT}
&\ket{L}\ket{a_0}_A\mapsto \ket{L} \ket{a_0}_A\nonumber\\
&\ket{L} \ket{a_1}_A\mapsto \ket{L} \ket{a_1}_A\nonumber\\
&\ket{R} \ket{a_0}_A\mapsto \ket{R} \ket{a_1}_A\nonumber\\
&\ket{R} \ket{a_1}_A\mapsto \ket{R} \ket{a_0}_A.
\end{align}
Note that such a map can be achieved even if $A$ is localized to the $R$ arm of the interferometer.
If $A$ starts in the $\ket{a_0}$ state, then after the CNOT gate, the $\{ \ket{a_0},\ket{a_1}\}$  basis of $A$ becomes perfectly correlated with the which-way observable of the photon.


In the scenario of Fig.~\ref{MachZehnderAncilla}, as with the other scenarios considered in this article,
the initial state of the photon 
 is $\ket{L}$, while after 
 the first beamsplitter \blk it is $\frac{1}{\sqrt{2}}(\ket{R}{-}\ket{L})$. Because the auxiliary system $A$ is initialized in state $\ket{a_0}$, the joint state of the photon and $A$ after the first beamsplitter
  is $\frac{1}{\sqrt{2}}(\ket{R}{-}\ket{L})\ket{a_0}$.  
  After the CNOT gate, their joint state is 
 \begin{equation}\label{ent1}
 \tfrac{1}{\sqrt{2}}(\ket{R}\ket{a_1}_A{-}\ket{L} \ket{a_0}_A).
 \end{equation}
Now suppose that the auxiliary system $A$ is measured in the basis $\{ \ket{a_0},\ket{a_1}\}$. If the outcome $a_0$ is obtained, then the quantum state of the photon collapses to $\ket{L}$.  In this case, there is no interference, i.e., equal likelihood of finding the photon in the $L$ and $R$ output ports of the second beamsplitter.  Similarly, if the outcome $a_1$ is obtained, then the quantum state of the photon collapses to $\ket{R}$, and again one sees no interference.

In short, the measurement of the $\{ \ket{a_0},\ket{a_1}\}$ basis on $A$ constitutes an {\em indirect} measurement of the which-way observable of the photon inside the interferometer, and hence destroys the interference.

Now suppose that instead of measuring $A$ in the $\{ \ket{a_0},\ket{a_1}\}$ basis, one measures it in the basis $\{ \ket{a_+},\ket{a_-}\}$, where $\ket{a_{\pm}} \equiv \frac{1}{\sqrt{2}}(\ket{a_0}\pm \ket{a_1})$.  

Note, first of all, that Eq.~\eqref{ent1} can be rewritten as 
 \begin{equation}\label{ent1b}
  \tfrac{1}{\sqrt{2}}\left[ \tfrac{1}{\sqrt{2}}( \ket{R}{-}\ket{L})\ket{a_+}{-} \tfrac{1}{\sqrt{2}}( \ket{R}{+} \ket{L})\ket{a_-}_A\right].
 \end{equation}
It is clear, therefore, that there is equal probability for the $a_+$ and $a_-$ outcomes to occur in a measurement of $\{ \ket{a_+},\ket{a_-}\}$.  Second, note that when the $a_+$ outcome is obtained, the state of the photon collapses to $\tfrac{1}{2}(\ket{R}{-}\ket{L})$, while when the $a_-$ outcome is obtained, it collapses to $\tfrac{1}{2}(\ket{R}{+}\ket{L})$.  So we see that the case of obtaining the $a_+$ outcome leaves the photon in the same state as the one in which it is left after the phase shifter in 
 the Mach-Zehnder interferometer of Fig.~\ref{MachZehnder}(a) when there is no phase shift. 
 Meanwhile, the case of obtaining the $a_-$ outcome leaves the photon in the same state as the one in which it is left in  the Mach-Zehnder interferometer of Fig.~\ref{MachZehnder}(a) after a
 $\pi$ phase shift. As described in Sec.~\ref{Section_3}, the unitary associated to the second beam splitter is such that in the case of no phase shift, the photon is always detected at the $L$ output port of the second beamsplitter, while in the case of a $\pi$ phase shift, the photon is always detected at the $R$ output port. Consequently, for the runs where the $a_+$ outcome is obtained, the photon is always detected at the $L$ output port of the second beamsplitter, while for the runs where the $a_-$ outcome is obtained, the photon is always detected at the $R$ output port of the second beamsplitter.  In both cases, we have complete interference.

Note that if one ignores the outcome of the $\{ \ket{a_+},\ket{a_-}\}$ measurement, then one has an equal mixture of the two types of interference pattern just described, which corresponds to equal probability of the photon emerging at either output port of the second beamsplitter---hence no interference.  This is consistent with what is found if one ignores the outcome of a measurement of the $\{ \ket{a_0},\ket{a_1}\}$ basis, or if one does no measurement on $A$ at all.\footnote{The above observations constitute a version of the Einstein-Podolsky-Rosen thought experiment~\cite{EPR1935} in the context of an interferometer.  It is well known that with a maximally entangled state such as Eq.~\eqref{ent1}, the basis of states to which one system collapses can be determined by the basis that is measured on the other system.  The {\em average} state on the one system, however, is independent of the basis measured on the other.}

The reason why the $\{ \ket{a_+},\ket{a_-}\}$ measurement on $A$ is commonly said to ``erase'' the which-way information encoded in $A$ is that it leads to a randomization of the outcome of a measurement in the $\{\ket{a_0},\ket{a_1}\}$ basis on $A$.
More precisely,  if one implements a measurement of $\{\ket{a_0},\ket{a_1}\}$ after a measurement of $\{ \ket{a_+},\ket{a_-}\}$, then whether one obtains the $a_0$ or $a_1$ is completely {\em independent} of the state of  the photon at the time prior to the measurement of $\{ \ket{a_+},\ket{a_-}\}$.  





We now comment on how these predictions arise in the second-quantized description, since this is the one that will map most closely to the account given by the toy field theory.  As before, the difference to the first-quantized description is that the system of interest is no longer taken to be a photon, but a pair of modes.  This impacts how one models the CNOT operation.  
Specifically, one can imagine a CNOT gate with the mode $R$ as the control qubit and $A$ as the target qubit, where the NOT operation is applied to the basis $\{ \ket{a_0},\ket{a_1}\}$ of $A$ if and only if mode $R$ is in state $\ket{1}$ of the occupation number basis.
Explicitly, 
\begin{align}
&\ket{0}_R \ket{a_0}_A\mapsto \ket{0}_R \ket{a_0}_A\nonumber\\
&\ket{0}_R \ket{a_1}_A\mapsto \ket{0}_R \ket{a_1}_A\nonumber\\
&\ket{1}_R \ket{a_0}_A\mapsto \ket{1}_R \ket{a_1}_A\nonumber\\
&\ket{1}_R \ket{a_1}_A\mapsto \ket{1}_R \ket{a_0}_A.\label{CNOTFock}
\end{align}

 The first pair of mappings in Eq.~\eqref{CNOTFock} trivially imply that
\begin{align}
&(\ket{1}_L\ket{0}_R) \ket{a_0}_A\mapsto (\ket{1}_L\ket{0}_R) \ket{a_0}_A\nonumber\\
&(\ket{1}_L\ket{0}_R) \ket{a_1}_A\mapsto (\ket{1}_L\ket{0}_R) \ket{a_1}_A,
\end{align}
while the last pair trivially imply that
\begin{align}
&(\ket{0}_L\ket{1}_R) \ket{a_0}_A\mapsto (\ket{0}_L\ket{1}_R) \ket{a_1}_A\nonumber\\
&(\ket{0}_L\ket{1}_R) \ket{a_1}_A\mapsto (\ket{0}_L\ket{1}_R) \ket{a_0}_A,
\end{align}
so that 
under the mapping from the second-quantized to the first-quantized description (defined in Eq.~\eqref{translation}), we recover the CNOT gate of Eq.~\eqref{CNOT}.

In short, the analysis of the experiment in the second-quantized description goes through exactly as it did in the first-quantized description with the substitutions implied by Eq.~\eqref{translation}.

Nonetheless, the second-quantized description provides additional clarity on the issue of locality.  In particular, it demonstrates that it is sufficient for the auxiliary system $A$ to interact with the mode $R$ alone. 



Let $A$ be initialized in the state $ \ket{a_0}_A$, and let mode $R$ be initialized in an arbitrary state $\alpha \ket{0}_R + \beta \ket{1}_R$.  The CNOT gate of Eq.~\eqref{CNOTFock} then maps the joint state of $RA$ to the  entangled state $\alpha \ket{0}_R\ket{a_0}_A + \beta \ket{1}_R\ket{a_1}_A$.  A subsequent measurement of the $\{ \ket{a_0},\ket{a_1}\}$ basis, therefore, achieves an indirect measurement of the occupation number in mode $R$.  It follows that one can understand the loss of interference as being due to this indirect measurement of the occupation number of mode $R$.  


By contrast, for a subsequent measurement of the $\{ \ket{a_+},\ket{a_-}\}$ basis on $A$, obtaining the $a_+$ outcome collapses the quantum state of $R$ back to $\alpha \ket{0}_R{+}\beta \ket{1}_R$, i.e., no disturbance to the quantum state of $R$, while obtaining the $a_-$ outcome collapses the quantum state of $R$ to $\alpha \ket{0}_R{-}\beta \ket{1}_R$, 
corresponding to an effective $\pi$ phase shift being implemented on the state (Eq.~\eqref{Qpiphasshift} where $\phi=\pi$).
Recall that in the second-quantized description, the joint state of the pair of modes and the auxiliary system at timestep 3 is obtained by translating 
Eq.~\eqref{translation} via the map of Eq.~\eqref{ent1b}, that is, 
\begin{align}
 & \tfrac{1}{\sqrt{2}} [ \tfrac{1}{\sqrt{2}}( \ket{0}_L\ket{1}_R{-}\ket{1}_L \ket{0}_R)\ket{a_+}\nonumber\\
&{-} \tfrac{1}{\sqrt{2}}( \ket{0}_L\ket{1}_R{+} \ket{1}_L \ket{0}_R)\ket{a_-}_A ].
\end{align}
It follows that
the pair of modes are collapsed to the state $\tfrac{1}{\sqrt{2}}( \ket{0}_L\ket{1}_R{-}\ket{1}_L \ket{0}_R)$ or $\tfrac{1}{\sqrt{2}}( \ket{0}_L\ket{1}_R{+}\ket{1}_L \ket{0}_R)$ depending on whether the outcome is $a_+$ or $a_-$. 
The latter can now be understood as follows: whether the outcome is $a_+$ or $a_-$ simply reveals whether 
 the $R$ mode was, by its interaction with $A$ and the subsequent postselection on the given outcome, effectively subjected to the unitary describing a $\pi$ phase shift or to the identity map. 
 

Importantly, the fact that one sees interference or not depending on whether one measures $\{ \ket{a_+},\ket{a_-}\}$ or $\{ \ket{a_0},\ket{a_1}\}$  on $A$ is true  {\em even if} the decision is delayed until long after one registers which output port of the interferometer the photon was detected in.

Much like in Wheeler's delayed-choice experiment, the claim that is conventionally thought to be supported by the quantum eraser experiment is that the experimenter can choose whether or not the photon behaves like a wave (traveling through both arms) or behaves like a particle (traveling through one arm) after the particle has already entered the interferometer. The purported novelty of the quantum eraser (relative to the experiments we have discussed so far) is that this choice can even be made {\em after the photon has already been detected}~\cite{herzog1995complementarity}.

Under the conventional perspective, the quantum eraser also provides a novel challenge to the possibility of local causal explanations of the phenomenology.  To understand the novelty, it is useful to compare the quantum eraser with the Elitzur-Vaidman bomb tester.  

We begin with a minor difference between the two.
We have already noted that in the conventional interpretation of the Elitzur-Vaidman experiment, it is claimed that if the change of the interference pattern is 
due to a causal influence from the decision of whether or not to learn the which-way observable, then this influence must be nonlocal.  Note, however, that the claimed nonlocal influence was one that acted from the $R$ arm of the interferometer to the $L$ arm.  
Because this decision is made even further afield in the case of the quantum eraser, 
 namely, in the choice of measurement on the auxiliary system $A$, 
it must be that the nonlocal causal influence is from $A$ to {\em both} the $R$ mode and the $L$ mode.\footnote{Relatedly, in the bomb-tester experiment, where the which-way observable is measured {\em directly} by a measurement on mode $R$, one can imagine that at least the change to the physical state of the $R$ mode is achieved by a local disturbance, even if the change to the physical state of the $L$ mode cannot be understood in this fashion.  In the case of the quantum eraser, on the other hand, if changes in the interference pattern are to be explained by an influence {\em from} the decision of what to measure on $A$
 {\em to} the physical state of the photon (or modes), then all such disturbances are nonlocal. }


The more significant difference, however, is a consequence of the fact that one can postpone the decision of whether and how to measure $A$ until {\em after} the photon has emerged from the interferometer and been detected at one of the output ports. This fact implies that if the difference in interference patterns is to be explained in terms of a causal influence---with the decision of what to measure on $A$ being the cause and the photon being the effect---then this causal influence must not only be nonlocal but must, in fact, propagate {\em backwards in time}.  

Of course, in the face of the radical nature of any such causal account, 
one can also consider a more Wheelerian response, namely, that the quantum eraser provides yet more grist for the mill of anti-realism. 
Either way, 
the quantum eraser is thought to support the interpretational claim that it is impossible to provide a local causal explanation of the phenomenology of quantum interference.

{\bf Toy field theory account.} To describe the toy field theory account of the quantum eraser experiment, we need to supplement the  account of the Mach-Zehnder interferometer (presented in Sec.~\ref{Section_4}) with an account of the auxiliary system $A$ and the CNOT interaction between mode $R$ and system $A$.

The system $A$ is presumed to be the toy analogue of an arbitrary two-level quantum system (a qubit).  This is taken to be a system with two properties, one property being the counterpart of the observable associated to the $\{ \ket{a_0},\ket{a_1} \}$ basis, and the other being the counterpart of the observable associated to the $\{ \ket{a_+},\ket{a_-} \}$ basis.  These properties are represented by binary variables, denoted $Q_A \in \{ a_0,a_1\}$ and $P_A \in \{a_+,a_-\}$ respectively. 
$Q_A$ can be interpreted as a discrete coordinate and $P_A$ as its canonically conjugate momentum.  (See Refs.~\cite{Spekkens2007,Spekkens2016} for more details.)
\footnote{
 If one likes, one can conceptualize the auxiliary system $A$  as a single mode, so that $Q_A=N_A$, a discrete occupation number, and $P_A = \Phi_A$, a discrete phase.
In this case, the CNOT operation corresponds to a dynamics that fails to conserve the total occupation number (unlike the beamsplitter and phase-shift operations), and to realize it experimentally would require nonlinear optics. 
But taking $A$ to be a mode is just {\em one} possibility for a concrete physical realization of $A$. 
A different possibility is that $A$ is analogous to the qubit defined by a two-level atom.  If these two levels are treated as two modes of the electron's field (relative to the nucleus), say $G$ (for `ground') and $E$ (for `excited'), then one can take $Q_A = N_G$ and $P_A = \Phi_G \oplus \Phi_E$.  In this case, $A$ is associated to coarse-grained degrees of freedom defined for the pair of modes, and the CNOT operation corresponds to an interaction that preserves particle number.  
Our argument---that the toy field theory account of the quantum eraser refutes the conventional interpretational claims---{\em does not depend} on the particular concrete realization of $A$.  The physical nature of $A$ is {\em irrelevant} for our purposes. As such, we prefer to treat it abstractly in what follows rather than commit to any specific concrete realization.
}


The toy field theory analogue of the initial quantum state $\ket{a_0}$ is the state of knowledge wherein $Q_A$ is known to take value $a_0$, while $P_A$ is unknown.


Now consider the toy field theory analogue of the CNOT of Eq.~\eqref{CNOTFock}.  To describe it succinctly, it is useful to define binary variables $\bar{Q}_A\in \{0,1\}$ and $\bar{P}_A\in\{0,1\}$ such that $Q_A=a_0 \leftrightarrow \bar{Q}_A=0$, $Q_A=a_1 \leftrightarrow \bar{Q}_A=1$, $P_A=a_+ \leftrightarrow \bar{P}_A=0$, and $P_A=a_- \leftrightarrow \bar{P}_A=1$.  Then we have
\begin{align}
&N^{\rm out}_R=N^{\rm in}_R \nonumber \\
&\Phi^{\rm out}_R=\Phi^{\rm in}_R \oplus \bar{P}_A^{\rm in}\nonumber\\
&\bar{Q}^{\rm out}_A=\bar{Q}^{\rm in}_A \oplus N_R^{\rm in} \nonumber \\
&\bar{P}^{\rm out}_A=\bar{P}^{\rm in}_A. \label{cnot}
\end{align}

If one measures $\bar{Q}_A$ after the CNOT dynamics, then if the initial value of $\bar{Q}_A$ is known,
the fact that the value of $\bar{Q}_A$ is shifted by $N_R^{\rm in}$, implies that one can infer  $N_R^{\rm in}$ from the measurement outcome (since $N_R$ is constant under the CNOT dynamics, one thereby learns its final value as well). This is the sense in which one can acquire information about the occupation number of $R$ by learning whether the outcome of the measurement of $Q_A$ is $a_0$ or $a_1$. 

Meanwhile, because $\bar{P}_A$ (and hence $P_A$)  is constant under the CNOT dynamics, a measurement of the final $P_A$ reveals the initial $P_A$, and given that $P_A$ prior to the CNOT dynamics 
is uniformly distributed,
  this measurement is equally likely to yield the $a_+$ or $a_-$ outcome.  
One can infer nothing about the initial physical state of mode $R$ from such a measurement, because the final $P_A$ does not depend on any property of mode $R$.   Nonetheless, the phase property of mode $R$, $\Phi_R$, is {\em disturbed} by the CNOT dynamics. Specifically, it is shifted by an amount corresponding to the initial value of $\bar{P}_A$ (hence correlated with that of $P_A$). Because the final $P_A$ is equal to the initial $P_A$ under the CNOT dynamics, a measurement of the final $P_A$ allows one to infer
the value by which mode $R$'s phase, $\Phi_R$, is shifted: if it is the $a_+$ outcome that is obtained, one infers that $\Phi_R$ 
 has not changed, while if it is the $a_-$ outcome that is obtained, one infers that 
 $\Phi_R$
  has been flipped.
  
 In essence, the CNOT dynamics is such that the final value of $Q_A$ (the coordinate variable of $A$) comes to encode information about the initial occupation number of $R$ and this involves a back-action on $R$, wherein the final phase of $R$ comes to encode information about the initial value of $P_A$ (the conjugate momentum variable of $A$). If one measures $Q_A$ after the interaction, one learns about the occupation number of $R$ (and foregoes the possibility of learning about what the back-action was), whereas if one measures $P_A$ after the interaction, one learns what the back-action was (but foregoes the possibility of learning anything about the initial physical state of $R$).
 
As we noted in our toy field theory analogue of the basic Mach-Zehnder interferometer with a phase-shifter, whether or not $\Phi_R$ has been flipped determines whether or not $\Phi_L \oplus \Phi_R$ has been flipped and consequently
 whether it is always the $R$ output port of the second beamsplitter that is occupied, or always the $L$ output port.  If one sorts the data into those runs where the $P_A$ measurement gives outcome $a_+$ and those where it gives outcome $a_-$, then in the first data set it is the $R$ output port that is always found to be occupied, while in the second it is the $L$ output port.  In other words, interference is observed in each data set. 
 
 A formal account of the toy field theory analogue of the quantum eraser is presented in Appendix~\ref{formalquantumeraser}.



 Having demonstrated that the toy field theory can reproduce the TRAP phenomenology of the quantum eraser, we turn to the question of what light this sheds on the standard interpretational claims.
 

In the toy field theory version of the quantum eraser, the choice of whether to measure $Q_A$ or $P_A$ on the auxiliary system does not have a causal influence on any of the properties of mode $L$ or mode $R$.   Rather, all that depends on this choice is {\em what one can infer} about the properties of the modes, specifically, whether one can infer $N_R$ (and hence also $N_L$) or $\Phi_L \oplus \Phi_R$.  But this is all that is required to explain the phenomenology.

We see, therefore, that a nonlocal influence between the choice of what to measure and the properties of the photon (or the distant modes) is not necessary to explain the phenomenology of the quantum eraser, and it is unnecessary for the same reason that backwards-in-time causal influences are unnecessary to explain the phenomenology of Wheeler's delayed choice experiment: in both cases, what is presumed to require a causal {\em influence} can be explained in terms of an {\em inference}, that is, an updating of one's knowledge.\blk

\section{Discussion}\label{Section_8}

\subsection{Take-away}


\subsubsection{Summary}
The three interpretational claims that are conventionally taken to follow from the TRAP phenomenology are radical when viewed relative to classical physics, and this radicalness is taken by many to imply that quantum phenomena are  somehow irreducibly mysterious.  The arguments we have presented, however, demonstrate that it is possible to account for such phenomena quite naturally, in particular, {\em without} being forced to accept wave-particle complementarity, the observer-dependence of reality, or the failure of local causal explainability.  \blk

Feynman's claim that the phenomenology of quantum interference is `impossible, {\em absolutely} impossible, to explain in any classical way' is proven to be simply false.   His claim that `We cannot make the mystery go away by ``explaining'' how it works' is also refuted---the toy field theory provides a straightforward and unambiguous account of the TRAP phenomenology in terms of the statistical mechanics of a classical (discrete) field theory.
His claim that the phenomenology of quantum interference ``contains the {\em only} mystery'' must also be rejected. 

\blk




\subsubsection{A proposal about methodology}
Ultimately, this work aims to shed light on the general question:
\blk
\begin{quote}
Which aspects of the phenomenology of quantum theory {\em force us} to reject the classical worldview?
\end{quote}

It has been common for researchers to attack this question as Feynman did: by identifying phenomena for which they (and others) cannot see how it is possible to provide an explanation in terms of a classical worldview, and then taking the {\em failure to identify such an explanation} as grounds for concluding that {\em there is no such explanation}.  


We propose a more rigorous methodology.  First, one must articulate what, precisely, one considers to be `the classical worldview', by articulating the principles that one believes {\em define} classicality.  Next, one must formalize these principles mathematically.  Finally, one must prove a rigorous no-go theorem that demonstrates a contradiction between the formalized principles and the aspects of the quantum phenomenology in question.

 
 In the next section, we describe what we take to be the best  notion of classicality, and describe what it implies under  our proposed methodology. \blk

\subsubsection{Beyond the TRAP phenomenology: what is really nonclassical?}\label{beyondTRAP}

Whatever principle of classicality one adopts, it is clear that the toy field theory should come out as a classical theory by its lights.  It then follows that the TRAP phenomenology {\em can} be reproduced in a classical theory and therefore cannot capture the {\em basic peculiarity of quantum theory}, contrary to Feynman's claim.
But what then {\em is} the essence of quantum theory?  What operational phenomena {\em do} imply a departure from the classical worldview?

Our own view on how best to characterize classicality is this: operational predictions are classically explainable if they can be realized in a theory where (i) the kinematical state space of a system (i.e., its space of physical states) is represented by a set while dynamics are represented by functions from this set to itself, (ii) inferences are done using Bayesian probability theory and Boolean propositional logic, and (iii)  a methodological principle for theory construction termed {\em Leibnizianity}~\cite{Schmid2021unscrambling} is satisfied.  

A few details about the notion of Leibnizianity.  Historically, Leibniz proposed a methodological principle that can be summarized as follows:
 situations that are empirically {\em indiscernible} in principle according to one's theory should be modelled as physically (i.e., ontologically) {\em identical} within the theory.  
Ref.~\cite{Leibniz} makes the case for the credentials of this principle in physics, given the use to which it was put by Einstein.  The principle of Leibnizianity proposed in Ref.~\cite{Schmid2021unscrambling} is a generalization of this methodological principle to the context of a theory wherein one is making probabilistic rather than deterministic inferences.\footnote{Ref.~\cite{CataniLeifer2020} proposes an alternative framework to that of Ref.~\cite{Schmid2021unscrambling}, but articulates a principle of classicality that is in the same spirit as Leibnizianity, namely, that the map from the ontological level to the empirical level should not be many-to-one, a principle termed `operational no fine-tuning' therein.}  

It is worth noting that our preferred notion of classicality is closely connected to the notion of classicality at play in Bell's theorem~\cite{Bell1964} and the one at play in the Kochen-Specker theorem~\cite{KochenSpecker1967}.  It can be shown that if operational predictions are realizable in a theory satisfying criteria (i), (ii) and (iii), then they are realizable 
    by a generalized-noncontextual ontological model~\cite{Spekkens2005}.   The latter, in turn, is a generalization of the notion of a noncontextual hidden variable model appearing in the Kochen-Specker theorem~\cite{KochenSpecker1967}.  
Furthermore, insofar as Bell's notion of a {\em locally causal} model is a special case of the notion of a generalized-noncontextual ontological model, it too is subsumed under our preferred notion of classicality.
Thus we see that the two most stringent notions of nonclassicality in quantum foundations, the one associated with the nonexistence of a locally causal model (Bell's theorem and its variants) and the one associated with the nonexistence of a noncontextual ontological model (the Kochen-Specker theorem and its variants) come out as instances of nonclassicality relative to our preferred notion.  Connected to this is the fact that our preferred notion of classicality is intimately related to the positivity of quasi-probability representations~\cite{Spekkens2008,schmid2020structure}, is natural from the perspective of a general framework for describing possible theories~\cite{SchmidGPT,ShahandehGPT,schmid2020structure}, and emerges in the presence of sufficient noise, whether this arises from limited experimental precision or from decoherence (see, e.g., the discussion at the end of Appendix C.1 in Ref.~\cite{Ravi1} as well as the results of Refs.~\cite{marvian2020inaccessible} and ~\cite{baldijao2021emergence}). 

Our preferred notion of classicality also satisfies two other reasonable desiderata for any notion of classicality: 
it  is empirically testable~\cite{Mazurek2016,mazurek2017experimentally,Ravi1,Ravi2,Mazurek2016,robust,Schmid2018}, 
and possession of a resource of nonclassicality (by its lights) is associated to 
having a quantum-over-classical advantage for  information processing~\cite{POM,RAC,RAC2,Saha_2019,comp1,comp2,schmid2021only,SchmidSpekkens2018,lostaglio2020contextual,contextmetrology,Pusey2014,KunjwalLostaglioPusey,YK20}.

 
It is straightforward to verify that the toy field theory satisfies
 our preferred notion of classicality.  It satisfies criterion (i) because 
 its kinematical state space is simply a set representing the possible physical states of one or more modes, and its dynamics is given by 
  functions on this set.\footnote{In fact, its kinematical state space is a discrete phase space and its dynamics is given by {\em symplectic} functions, that is, functions which are discrete analogues of Hamiltonian evolutions (these are all bijective, hence also {\em reversible}). In this sense, it satisfies an even stronger notion of classicality than the one we have articulated above.\blk}  It satisfies criterion (ii) because Bayesian probability theory and Boolean propositional logic are sufficient to describe inferences within this theory.  (The epistemic restriction only constrains what particular probability distributions can represent an agent's knowledge of the physical state of the system and what particular logical propositions about this physical state can correspond to questions that can be asked.)  
The toy field theory can also be shown to satisfy criterion (iii), that is, the principe of Leibnizianity (and hence generalized noncontextuality as well) for the same reason that other epistemically restricted classical statistical theories do~\cite{Spekkens2007,Spekkens2016}. Together with our demonstration that the toy field theory can reproduce the TRAP phenomenology, this implies that the TRAP phenomenology is classically explainable according to our preferred notion of classicality.\footnote{We pause here to comment on how our claim can be made to square with Bohr's pronouncement (echoed by many others) that we are forced in quantum theory to acknowledge complementarity, that is, we are forced to acknowledge that noncommuting observables cannot be represented by classical random variables that are jointly well-defined. 
What we wish to emphasize is that Bohr did not {\em demonstrate} that complementarity is forced upon us by the operational phenomena of quantum theory.  He merely {\em asserted} it.
 Bohr's pronouncements, in other words, did not meet the standards of the rigorous methodology we are advocating here.  It is {\em only} with the advent of certain rigorous no-go theorems, such as the Kochen-Specker theorem~\cite{KochenSpecker1967,Bell}
and the no-go theorem for the generalized notion of noncontextuality proposed in Ref.~\cite{Spekkens2005},
that one is warranted in concluding that  it is impossible to represent certain sets of noncommuting quantum observables by classical variables, at least not without violating 
the assumption that the values must be assigned context-independently.
  The fact that the TRAP  phenomenology 
     is not rich enough to support such a no-go theorem
   is {\em why} a noncontextual model such as the toy field theory has the capacity to reproduce this phenomenology.}

 Nonetheless, there will be other, more nuanced, aspects of interference phenomena that are not classically explainable according to our preferred notion of classicality. 
These aspects will arise as operational predictions that are not realizable by a generalized-noncontextual ontological model.  (The notion of a locally causal model is likely not sufficient for this purpose, because most interference experiments do not have the appropriate causal structure for deriving a Bell-type theorem.)
 To determine to what extent interference phenomena present a challenge to our preferred notion of classicality,
  therefore, one should seek to identify those aspects of interference phenomena that imply the failure of a generalized-noncontextual ontological model.


Some progress on this front has been made since the appearance of a draft version of this article.   Specifically, it was shown in Ref.~\cite{catani2022aspects} that the precise functional form of a particular wave-particle duality relation resists explanation  in terms of a generalized-noncontextual ontological model.
Another example is described in Ref.~\cite{wagner2022coherence}.\blk\footnote{Similar projects have been undertaken for other types of quantum phenomena.  For instance, the fact that nonorthogonal quantum states cannot be discriminated with certainty is sometimes thought to be an intrinsically quantum feature of state discrimination.  However,  if quantum states are represented by probability distributions that overlap on the physical state space, lack of perfect discriminability is to be expected.  Nonetheless, it was shown in Ref.~\cite{SchmidSpekkens2018} that the particular {\em tradeoff} that quantum theory predicts between the degree of nonorthogonality and the probability of successful discrimination is inconsistent with the principle of generalized noncontextuality. A similar analysis has also been done for state-dependent cloning~\cite{lostaglio2020contextual} and uncertainty relations \cite{Catani2022UR}.
As another example, although weak values in pre- and post-selected experiments can be given an account within a generalized-noncontextual ontological model (see Ref.~\cite{Karanjai2015} for instance), it was shown in Ref.~\cite{Pusey2014,KunjwalLostaglioPusey} that {\em anomalous weak values}~\cite{Aharonov1988} cannot.
}




\subsubsection{Interpretational lessons}

A motivation for studying 
classical statistical theories of the sort described in this paper (and characterizing the principles of classicality that they embody)
 is that the insights obtained by doing so are useful in guiding our search for a more compelling interpretation of quantum theory.  
 
 In particular, Ref.~\cite{Schmid2021unscrambling} articulates a research program that seeks to secure an interpretation of quantum theory that salvages the spirit of locality and generalized noncontextuality by taking the quantum state to describe knowledge, 
while going beyond the usual framework of ontological models
 by allowing for intrinsically quantum notions of causation and inference.   It proposes a formalism that seeks to unscramble the omelette of ontology and epistemology in quantum theory, to use the evocative imagery of E.T. Jaynes~\cite{Jaynes}.
For various interference phenomena, we have here shown that what was previously attributed to exotic types of {\em causal influence} could be explained within the toy field theory by 
mere {\em  inference}, i.e., merely resolving some uncertainty based on the acquisition of new information.  In this sense, the toy field theory can also be understood as a small contribution to the program~\cite{Schmid2021unscrambling,Schmid2019,schmid2021guiding,Bartlett2012,fuchs2002quantum} of unscrambling the quantum omelette.



  The toy field theory also suggests a novel take on 
   the relationship between the first-quantized and second-quantized descriptions in quantum theory.  As we noted in Sec.~\ref{First2Second}, the first- and second-quantized descriptions are typically understood as making different assumptions about the nature of the systems and their properties: particles with motional degrees of freedom on the one hand, or modes with excitational degrees of freedom on the other. Our analysis, however, suggests that it is better to understand {\em both} descriptions as being about modes with excitational degrees of freedom, with the first-quantized description being merely a more {\em coarse-grained} description than the second-quantized one, describing only a subset of the degrees of freedom that are described in the latter.\footnote{We provide some further details about how to conceptualize a theory with an ontology of modes in Appendix~\ref{ModesNotParticles}.  We also demonstrate that one can recast the toy field theory in terms of arrays of spatially localized modes and a rule for free propagation of excitations in Appendix~\ref{SpatiallyLocalizedModes}.} \blk




\subsection{Further remarks}

\subsubsection{Why was the toy field theory not recognized earlier?} \label{whynotearlier}
A natural question regarding the toy field theory is why it was not recognized earlier that a local classical statistical theory of this type could reproduce the basic phenomenology of quantum interference.  One reason is that interference phenomena have been studied predominantly in the first-quantized rather than the second-quantized description, and this has exacerbated the problem of seeing how they admit 
 of a local explanation.  The more significant reason, however, is  that a certain (largely unrecognized) interpretational assumption---the reality of the quantum state---prevented researchers from contemplating theories of this sort. \blk


As has been emphasized in Refs.~\cite{Spekkens2007,Harrigan2010} most attempts to provide realist accounts of operational phenomena presume from the outset that quantum states represent states of reality rather than representing states of incomplete knowledge about reality.  This assumption implies, in particular, that the {\em vacuum quantum state} represents a state of reality rather than representing a state of incomplete knowledge about reality.  It then follows that if a mode of an interferometer is described by the vacuum quantum state, then its physical state is fixed.  But if there is no possibility of {\em varying} the physical state of that arm, then there is no possibility of encoding information in its physical state.   According to this view, then,  an arm of the interferometer that is in the vacuum quantum state {\em cannot propagate forward any information about devices on that arm} because it does not have the capacity to encode any such information.  

One of the keys to the success of the toy field theory, therefore, is that the analogue of a pure quantum state (and in particular, the analogue of the vacuum quantum state) is a state of incomplete knowledge about reality, i.e., a probability distribution that has support on more than one physical state.\footnote{This is called a $\psi$-epistemic model in the foundations literature~\cite{Harrigan2010}. For further reading on the view that quantum states are states of incomplete knowledge, 
see Refs.~\cite{Spekkens2007,Spekkens2016}.  Note that the interpretational program of QBism~\cite{QBism1,QBism2} also takes as its starting point the idea that the quantum state does not represent reality.  Indeed, early work on Quantum Bayesianism laid some of the groundwork for Refs.~\cite{Spekkens2007,Spekkens2016}.  Nonetheless, there are ways of pursuing an epistemic view of quantum states that are quite distinct from the QBist research program. See, e.g., Ref.~\cite{Schmid2021unscrambling}.}
 Using concepts from statistical mechanics, one might say that the vacuum quantum state is analogous to a macrostate and hence that it is consistent with many microstates. 
 Specifically, even if the occupation number of a mode is 0, there are two possible values that its discrete phase might take, and hence such a mode can still encode one bit of information.
This is what opens up the possibility  that  information about a device (e.g., whether it implements a which-way measurement or not) can be propagated to other devices (such as the final detectors) through a mode which, in the quantum account, is in the vacuum quantum state.   In particular, in the case of the Elitzur-Vaidman bomb-tester, it is what opens up the possibility that information about whether the bomb is functional or faulty can be propagated to the final detectors through the physical state of the $R$ mode even though in the quantum account the $R$ mode is in the vacuum quantum state. \footnote{\label{onticindifference}This possibility is closely related to another feature of classical statistical theories with an epistemic restriction, namely,  that a transformation may leave the probability distribution over physical states invariant even though it modifies the physical state.   This occurs whenever the transformation implements a permutation within a set of physical states that are assigned equal probability.  In the toy field theory, for example, when the occupation number is point-distributed, so that (by the epistemic restriction) the phase is uniformly distributed, a phase flip transformation, $N\mapsto N, \Phi \mapsto \Phi\oplus 1$,  leaves the distribution invariant even though it modifies the physical state.
In summary, classical statistical theories with an epistemic restriction render implausible 
what has been termed the
 `ontic indifference principle' in Ref.~\cite{Hardy2013}, namely, that if a pure quantum state is left invariant, then so too is {\em any} physical state that is assigned nonzero probability by that quantum state.  
}

The toy field theory consequently provides yet another example (in addition to those presented in earlier work~\cite{Spekkens2007,Spekkens2016,Bartlett2012}) of how treating pure quantum states as states of incomplete knowledge opens up new interpretational possibilities.



\subsubsection{Inference, not influence}
Another way in which our analysis has made use of the epistemic view of quantum states 
 is how the remote state update rule in quantum theory has come out
 as analogous to  the updating of a  state of knowledge upon acquisition of novel information. Because such updating is governed by Bayes' rule, it is typically termed {\em Bayesian conditioning}.   This analogy has also been noted in many previous works~\cite{Spekkens2007,Spekkens2016,LeiferSpekkens,fuchs2002quantum}. 


Recall that the remote state update rule in quantum theory stipulates how the quantum state of a system is updated when one learns the outcome of a measurement (either destructive or nondestructive) on a different system that is correlated with the first (by virtue of a common cause acting on both).  
This is distinct from the usual state update rule in quantum theory, which stipulates how the quantum state of a system is updated when one learns the outcome of a direct (nondestructive) measurement on the system.

We noted that 
 the counterpart in the toy field theory of the usual quantum state update rule  
  is
  a Bayesian conditioning followed by a disturbance.  For a nondestructive measurement of the occupation number of mode $R$, for instance, there is a Bayesian conditioning based on the outcome of the measurement, followed by a randomization of the phase of mode $R$.

On the other hand, the counterpart of the {\em remote} update rule involves {\em only} a Bayesian conditioning, i.e., it is {\em purely} a resolution of some of the agent's uncertainty about the physical state (i.e., with no disturbance to the latter).
  Specifically, given that the probability distribution prior to the measurement is such that the occupation numbers of the pair of modes are anticorrelated,
  it follows that learning the occupation number of mode $R$ causes one to update one's knowledge of the occupation number of mode $L$.  Furthermore, it is {\em only} the agent's knowledge about mode $L$ that changes as a result of a measurement on the $R$ mode; the physical state of mode $L$ is not changed. This clarifies how, in the toy field theory account of the TRAP phenomenology, there is no need for a causal influence from the $R$ arm to the $L$ arm of the interferometer.\footnote{In Appendix~\ref{EnsembleEpistemicTalk}, we elaborate on the importance of an interpretation of probabilities in terms of credences in our approach.  In Appendix~\ref{ensemblereframing}, we point out that one can nonetheless reframe the predictions of the toy field theory in terms of an infinite ensemble of repetitions of the experimental set-up.  We show that in this reframing, Bayesian updating about a single system is replaced by an updating of the ensemble describing the set of copies of the system. In particular, we explain the counterpart of the remote quantum state update rule in this ensemble language, which provides another way of seeing why it is local.}

 In a similar manner,  in the quantum eraser experiment, the remote state update rule for $L$ and $R$ based on $A$ (i.e, the update rule for the composite system comprising the pair of modes given a measurement on the auxiliary system) corresponds, in the toy field theory, to simply updating one's knowledge of $L$ and $R$ based on what is learned about $A$ rather than any nonlocal influence from $A$ to $L$ and $R$.

\blk

\blk


A final example of how the epistemic view of quantum states has been leveraged in our analysis is how, in the toy field theory, phenomena that are sometimes attributed to backwards-in-time causation can be instead attributed to merely updating one's knowledge about the past.  

For those committed to the idea that a quantum state describes reality, the distinction between the quantum state of a photon being an eigenstate of the which-way observable   (i.e., $\ket{L}$ or $\ket{R}$) and it being a state from a complementary basis (i.e., $\tfrac{1}{\sqrt{2}}(\ket{R} - \ket{L})$ or $\tfrac{1}{\sqrt{2}}(\ket{R} + \ket{L})$) is a distinction concerning the {\em physical state} of the photon.   The conventional discussions of Wheeler's delayed-choice experiment and the quantum eraser experiment demonstrate that some researchers are tempted to endorse this idea {\em even if} the difference in the quantum state assigned at some time is induced by learning the outcome of a measurement implemented {\em at a later time.}  In such a case, one must either grant that there are backwards-in-time influences, or embrace a kind of anti-realism (as Wheeler did).

In an epistemic interpretation of quantum states, by contrast, the fact that learning the outcome of a measurement at one time can lead to an updating of the quantum state for a system {\em at an earlier time} has a very natural interpretation: it is simply an updating of one's {\em knowledge} about the past. Such updates are commonplace in science.  For instance, we update our belief about the existence of dinosaurs in the distant past based on observation of the fossil record today.

From the perspective of the methodology we are advocating, it is fine if a researcher wishes to assume that pure quantum states describe physical states rather than states of incomplete knowledge (i.e., if, in the language of Ref.~\cite{Harrigan2010}, they wish to assume a $\psi$-ontic rather than a $\psi$-epistemic ontological model), but this assumption {\em must be made explicit in the no-go theorem}, so that it is evident that denying it is a way out of the no-go result.
 The point, again, is that whatever one thinks of this assumption, 
 it is ``not forced on us by experimental facts, but by deliberate theoretical choice.''\footnote{Many recent works, including the PBR theorem~\cite{PBR}, have sought to identify principles from which the assumption of $\psi$-ontology can be derived (see Ref.~\cite{LeiferOntology} for a review). The naturalness of these principles is disputed by some of the authors of this article (see, e.g.,  Ref.~\cite{SpekkensPirsaTalk2}).  But regardless of what one thinks about these principles, the important point is that {\em all} the standard no-go theorems---including those of Bell and Kochen-Specker, but also that of PBR---assume the framework of ontological models, and as we noted above, {\em this} is the assumption whose rejection we believe holds the most promise for finding a satisfactory realist interpretation of quantum theory. }

\blk

\subsubsection{Is the toy field theory ad hoc?}
It is worth noting that the argument purporting to show that there is no local causal explanation of the phenomenology of the Mach-Zehnder interferometer never had the same force as the argument that there is no local causal explanation of Bell inequality violations.  This is because in the Mach-Zehnder interferometer, one can always imagine some hypothetical system that propagates from the location of the detector in the $R$ arm to the final detectors (via the second beamsplitter). That is, one can always a imagine a hypothetical system that mediates
 a causal influence from the former to the latter without requiring such influences to be faster than the speed of light.   The argument that is typically presented {\em against} such a local causal explanation proceeds along the following lines: ``Yes, there might be {\em something}---a ghost particle for example---that propagates information from the region where the detector is (or is not) placed in the $R$ arm to the second beamsplitter and that dictates the subsequent motion of the photon, but to posit such a ghostly influence is {\em ad hoc} in the sense that it has nothing to recommend it except the evidence that it is specifically designed to explain.''

Note that this response explicitly {\em concedes} that the inference from the TRAP phenomenology to the failure of a local causal explanation is not logically sound, but is rather only a plausibility argument.  Nonetheless, an ad hoc explanation is indeed an implausible one. 
Therefore, if our proposed account of the TRAP phenomenology consisted of nothing more than to posit a ghostly influence of the sort just described, then it would be appropriate to criticize it as ad hoc and thus implausible.  It is therefore worthwhile for us to explicitly respond to this potential criticism.  


To see why the toy field theory is {\em not} an ad hoc explanation, one must consider its origin.  It arose as a natural application of the ideas of Ref.~\cite{Spekkens2007} to quantum interference phenomena.\footnote{The basics of the toy field theory and its implications for the interpretation of interference phenomena were worked out in 2005.}  As we noted above, what is critical to achieving a local causal explanation is the fact that the counterpart of the vacuum quantum state in the toy field theory is a probability distribution with support on many different physical states.  But this fact is an unavoidable consequence of the epistemic restriction, since the latter dictates that {\em every} pure quantum state  has such a probability distribution as its counterpart. 
 In other words, the possibility of sending information through a mode which in the quantum account is described by the vacuum quantum state 
   is an {\em inevitable} consequence of the epistemic restriction.  Given that toy theories based on an epistemic restriction of this type can also reproduce a long list of phenomena that are conventionally regarded as uniquely quantum (see, e.g., Table II of Ref.~\cite{Spekkens2016}), it follows that the epistemic restriction underlying the toy field theory has much more to recommend it than the fact that the toy field theory can reproduce the TRAP phenomenology of quantum interference.  Recalling the definition of an ad hoc explanation which we presented above---having nothing to recommend it except the evidence that it is specifically designed to explain---it is clear that the toy field theory explanation of interference phenomena is not ad hoc. \footnote{Another fact worth mentioning regarding the charge of {\em ad hocness} for the toy field theory's treatment of the quantum vacuum state is that there are many  phenomena in quantum electrodynamics that suggest that the quantum vacuum ought not to be conceptualized as empty, but rather as containing ``fluctuations''.
Some researchers who sought to formalize this idea engaged in
  a research program~\cite{SED,marshall1963random,marshall_1965,quantumdice} that took as its starting point that the quantum vacuum state should be modelled as a probability distribution over classical field configurations.  
 Consequently, the set of phenomena that motivate this research program also serve to motivate the manner in which the quantum vacuum state is modelled in the toy field theory.\blk}
\blk


\subsubsection{Shifting the goal post}
No doubt those researchers who are sympathetic to the view that interference captures the essence of quantum theory will be tempted to respond to the arguments of this article as follows: 
\begin{quote}
Sure, you have reproduced {\em some} of the phenomenology of quantum interference, but you haven't reproduced {\em all} of it.  What about all of the experiments involving beamsplitters that are not 50-50, or involving phase shifts other than $\phi=0$ and $\phi=\pi$?  You can't make sense of {\em those} in the toy field theory.\footnote{In a similar vein, Yakir Aharonov's response to a presentation on this topic \cite{SpekkensPirsaTalk} 
 was to concede that the toy field theory provides an adequate explanation of interference phenomena for {\em bosons}, but to insist that such an explanation will not be possible for {\em fermions}.} 
\end{quote}
Our response is as follows.  
If someone wishes to claim that aspects of interference beyond the TRAP phenomenology demonstrate the impossibility of maintaining a classical worldview, then not only must they specify precisely which aspects they have in mind and how they propose to formalize the notion of classicality,
they must also {\em back up their claim} with a rigorous no-go theorem, following the methodology we endorsed above. 
Until they do, the view that the phenomena in question resist explanation in terms of a classical worldview is mere speculation, and might only indicate a ``lack of imagination'', to recall Bell's phrase.\footnote{ In Appendices~\ref{DestructiveMeasurements} and \ref{MirrorRemoved}, we consider two examples of additional interference phenomenology,  based on minor modifications to the Mach-Zehnder interferometer, that were speculated to resist classical explanation in Ref.~\cite{hance2022comment}.  We demonstrate that the phenomenology of both examples can be easily accommodated within the toy field theory, and therefore that they do not, in fact, resist classical explanation.}

Our own preferred notion of classical explainability is explainability in terms of a Leibnizian (hence noncontextual) ontological model and, as noted in Section~\ref{beyondTRAP}, recent work has identified some aspects of the phenomenology of interference that {\em do} resist such an explanation~\cite{catani2022aspects,wagner2022coherence}.  In particular, Ref.~\cite{catani2022aspects} established that the incorporation of beamsplitters that are not 50-50 into the standard interference experiment is, in fact, sufficient for establishing such a no-go result. 



\subsection{Send-off} In an article based on a 1981 conference presentation~\cite{feynman2018simulating}, Feynman
 returned to the question of what was the essential difficulty of quantum theory.
The article concerned whether quantum dynamics could be simulated using a classical computer with local connections, and Feynman concluded that it could not using an argument that is in essence the same as the one appearing in Bell's theorem.  (In Feynman's approach, the proof proceeds by showing that the quantum correlations can only be reproduced using local connections if the probabilities can go negative.)  He notes furthermore that the relevant experiment has been done and that it agrees with the quantum predictions.\footnote{Feynman did not cite Bell's work explicitly, suggesting that he might have rederived the result independently.  However, as noted by Whitaker~\cite{whitaker2016richard}, the fact that he was aware of the experimental tests of Bell inequalities implies that he must at least have been aware of the existence of Bell's work.
}.  In this article, therefore, Feynman takes the essential difficulty to be the existence of a Bell-type no-go result.  
How does one square this with his earlier claim, in the Lectures on Physics, that ``interference holds the essential difficulty''?
 (Especially given that, as we have shown here, the TRAP phenomenology of interference is {\em not} associated to a violation of a Bell inequality.)  Did Feynman {\em change his mind} about what was the essential difficulty?  
 
 No.  It seems, rather, that he did not think that there was any significant difference
 between Bell inequality violations and the phenomenology of interference---both appeared to him to be classically inexplicable.  
Indeed, a quote by Feynman regarding Bell's theorem from a 1983 article reprinted in Ref.~\cite{hey2018feynman} makes it clear that he in fact {\em denied} that the difficulty uncovered by Bell was of a different nature than the difficulty associated with understanding other quantum phenomena (including, presumably, interference).  Speaking of the ways in which quantum phenomena cannot be accounted for in a classical worldview, he states:
\begin{quote}
Bell's Theorem is a contribution to [that set of ideas], which is to point out mathematically that it has to happen. People knew it had to happen before; all he did was to demonstrate it. It is not a theorem that anybody thinks is of any particular importance. We who use quantum mechanics have been using it all the time. It is not an important theorem. It is simply a statement of something we know is true---a mathematical proof of it.
\end{quote}

The obvious criticism to make of Feynman here is that he is wrong in his
assessment of the degree of significance of Bell's theorem.  The criticism we wish to emphasize, however, is not of this assessment, but of the   {\em reasons} he articulates in support of it.
The cavalier attitude towards foundational questions that is expressed in this quote is precisely the sort of  methodological looseness
 that we are advocating against.  Interpretational claims which practitioners of quantum theory  believe to be true in the absence of a mathematical proof are not  ``known to be true''.  Indeed many such claims
are plainly mistaken. The claim that interference phenomenology captures the essence of quantum theory is a prime example.  A mathematically rigorous no-go theorem is the bar that must be passed by any argument purporting to support some interpretational claim. This is {\em why} Bell's theorem is, in fact, a singularly important result in twentieth-century physics.  
The {\em absence} of a parallel no-go theorem establishing 
 the classical nonexplainability of the TRAP phenomenology of quantum interference is what left open the possibility of the sort of classical account presented in this article.  Feynman's failure to recognize this sort of possibility---which led to his mistaken claim regarding interference phenomena being the essence of quantum theory---illustrates 
why it is critical to back up interpretational claims with mathematically rigorous no-go theorems.  

\section*{Acknowledgements}

RWS would like to acknowledge the important contribution of Elliot Martin, who worked as an undergraduate student on an early version of the ideas described in this article as part of a research project in the summer and fall of 2005, and to thank Jon Barrett for helpful discussions about the project at that time. 
The authors also acknowledge discussions with Sabine Hossenfelder, Tim Palmer, and Jonte Hance which served to motivate the inclusion of certain discussions in the appendices.
This research was supported by Perimeter Institute for Theoretical Physics. Research at Perimeter Institute is supported by the Government of Canada through the Department of Innovation, Science and Economic Development Canada and by the Province of Ontario through the Ministry of Research, Innovation and Science. 
ML was additionally supported, in part, by grant number FQXi-RFP-IPW-1905 from the Foundational Questions Institute and Fetzer Franklin Fund, a donor advised fund of Silicon Valley Community Foundation. ML and LC were supported, in part, by the Fetzer Franklin Fund of the John E. Fetzer Memorial Trust. LC was also supported by the Army Research Office (ARO) (Grant No. W911NF-18-1-0178) and acknowledges funding from the Einstein Research Unit `Perspectives of a Quantum Digital Transformation'. DS also acknowledges support by the Foundation for Polish Science (IRAP project, ICTQT, contract no.2018/MAB/5, co-financed by EU within Smart Growth Operational Programme).

\bibliographystyle{unsrturl}

\bibliography{ToyFieldTheory_Biblio}

\appendix

\section{A more formal account}\label{FormalAccount}

\subsection{The toy field theory}\label{toyfieldtheoryformalaccount}


We begin by describing our diagrammatic representation, which follows the one introduced in Ref.~\cite{Spekkens2007}. The set of physical states of a mode is represented diagrammatically as a set of 4 boxes, and a particular physical state as a marked box.  Thus, for instance, the physical state wherein $N=0$ and $\Phi=1$ is depicted as:
\begin{align}
\centering
\includegraphics[width=0.19\textwidth]{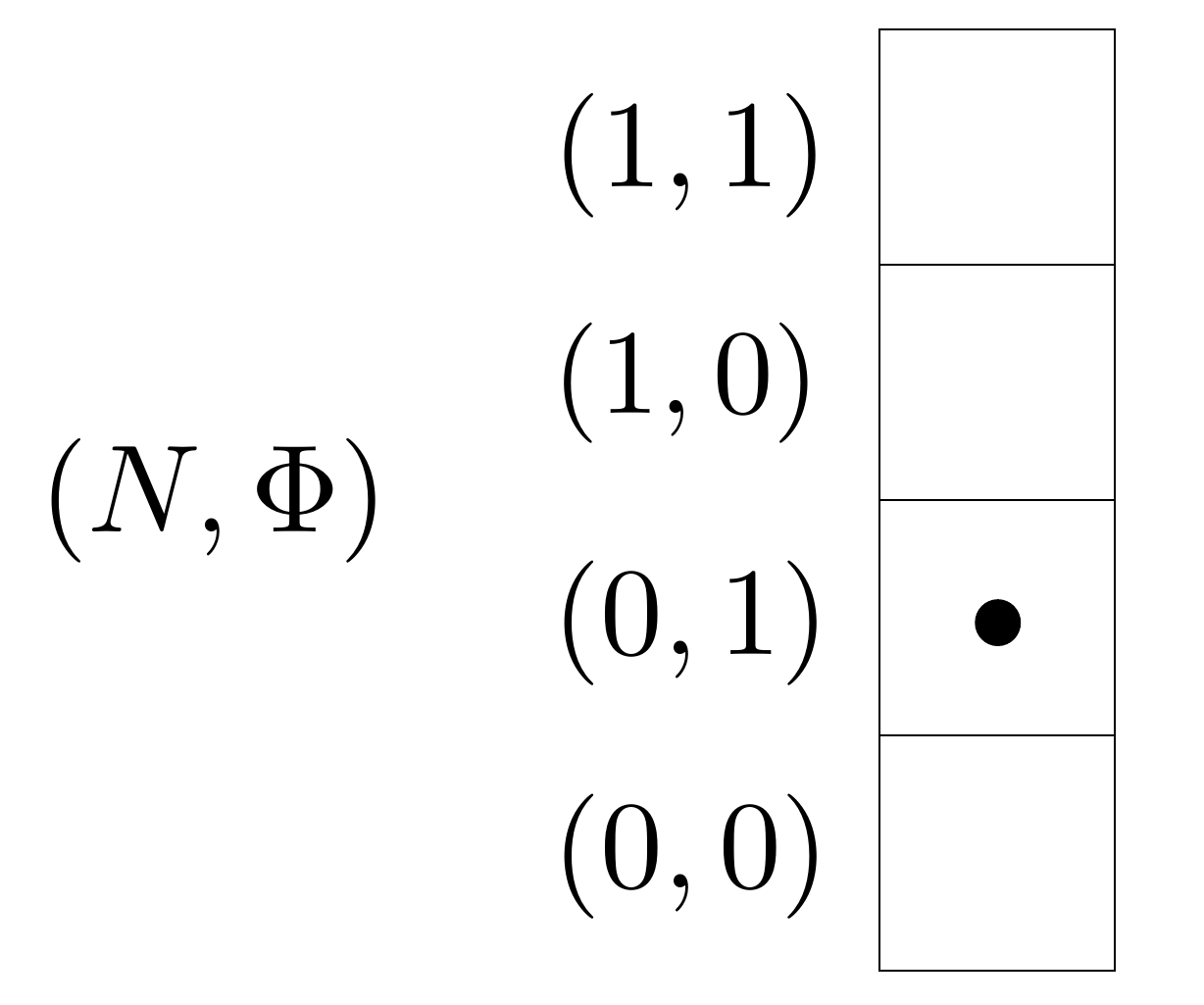}.
\end{align}
The set of physical states of a {\em pair} of modes is represented diagrammatically by a $4\times4$ grid of boxes.  For instance, the physical state defined by $N_L=1, \Phi_L=0, N_R=0, \Phi_R=1$ is represented as
\blk
\begin{align}
\centering
\includegraphics[width=0.32\textwidth]{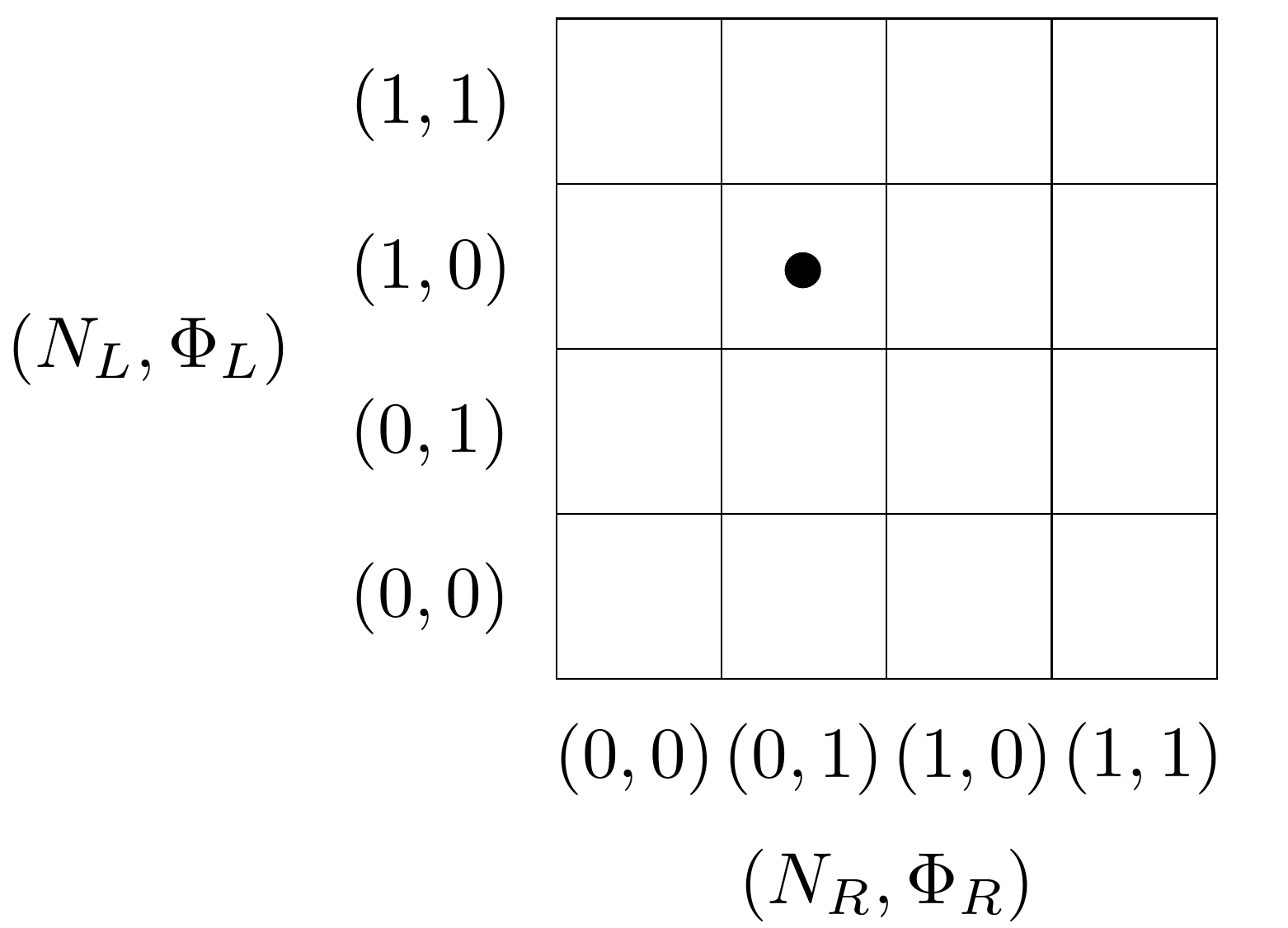}.
\end{align}

%

An agent's knowledge of a mode can be completely specified by a probability distribution over $N$ and $\Phi$, denoted $p(N,\Phi)$.  We will write $p(X)=[x]$ to indicate that the distribution over the variable $X$ is a {\em point} distribution with all weight on the value $x$.  In particular, if $X$ is a binary variable, then a distribution that assigns probability $w$ to $X=0$ can be written as $p(X)=w[0]+(1-w)[1]$.  

%

In the case where a mode is known to be unoccupied, so that $N=0$, while the phase is unknown, the probability distribution is
$$p(N,\Phi) = [0] \left( \tfrac{1}{2}[0]+\tfrac{1}{2}[1] \right),$$
for which the marginal distribution on $N$ is a point distribution on the value $0$, $p(N)= [0]$, and the marginal distribution on $\Phi$ is uniform, $p(\Phi)=\tfrac{1}{2}[0]+\tfrac{1}{2}[1]$.  (It follows that the marginal distribution on the parity $N\oplus \Phi$ is also uniform, $p(N \oplus \Phi)=\tfrac{1}{2}[0]+\tfrac{1}{2}[1]$, corresponding to the fact that the parity is also unknown.)

This distribution is depicted diagrammatically as
\begin{align}
\centering
\includegraphics[width=0.19\textwidth]{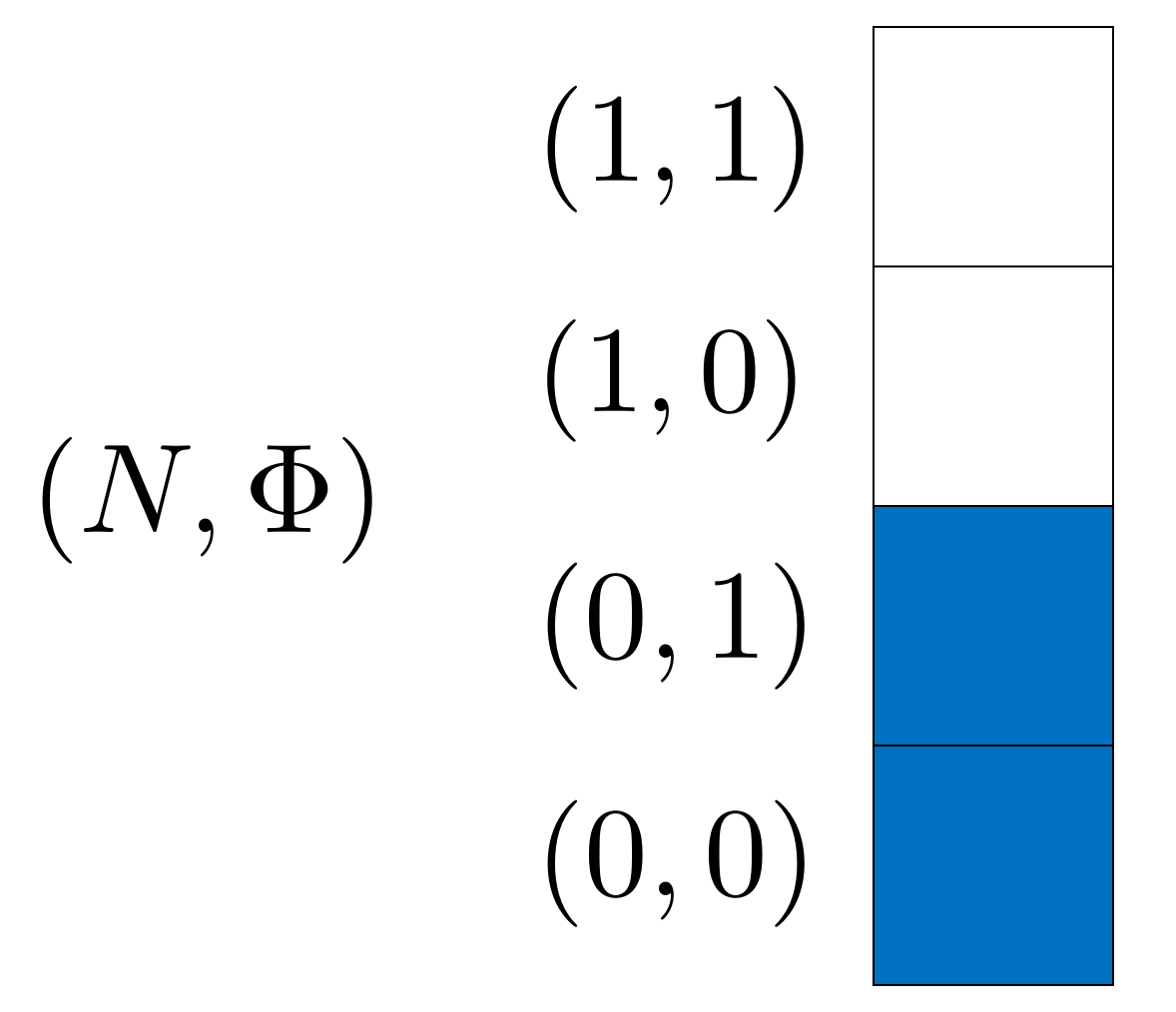}.
\end{align}
For the epistemic restriction that defines the toy field theory,  the probability distribution describing a valid state of knowledge is always  a flat distribution,  so that one can get away with merely specifying which physical states are  known {\em not} to describe the actual physical state (depicted as unfilled) and which are deemed to be possible options for the actual physical state (depicted as blue).\blk




Formally, therefore, the analogue in the toy field theory of the quantum state $|1\rangle_L |0\rangle_R$
 at the input of the Mach-Zehnder interferometer is:
\begin{align}\label{EpistemicInput}
 &p(N_L,\Phi_L,N_R,\Phi_R)\nonumber\\
 &=[1]\left(\tfrac{1}{2}[0]+\tfrac{1}{2}[1]\right)[0]\left(\tfrac{1}{2}[0]+\tfrac{1}{2}[1]\right),
 \end{align}
which is depicted diagrammatically as: 
\begin{align}\label{EpistemicInputDiagram}
\centering
\includegraphics[width=0.32\textwidth]{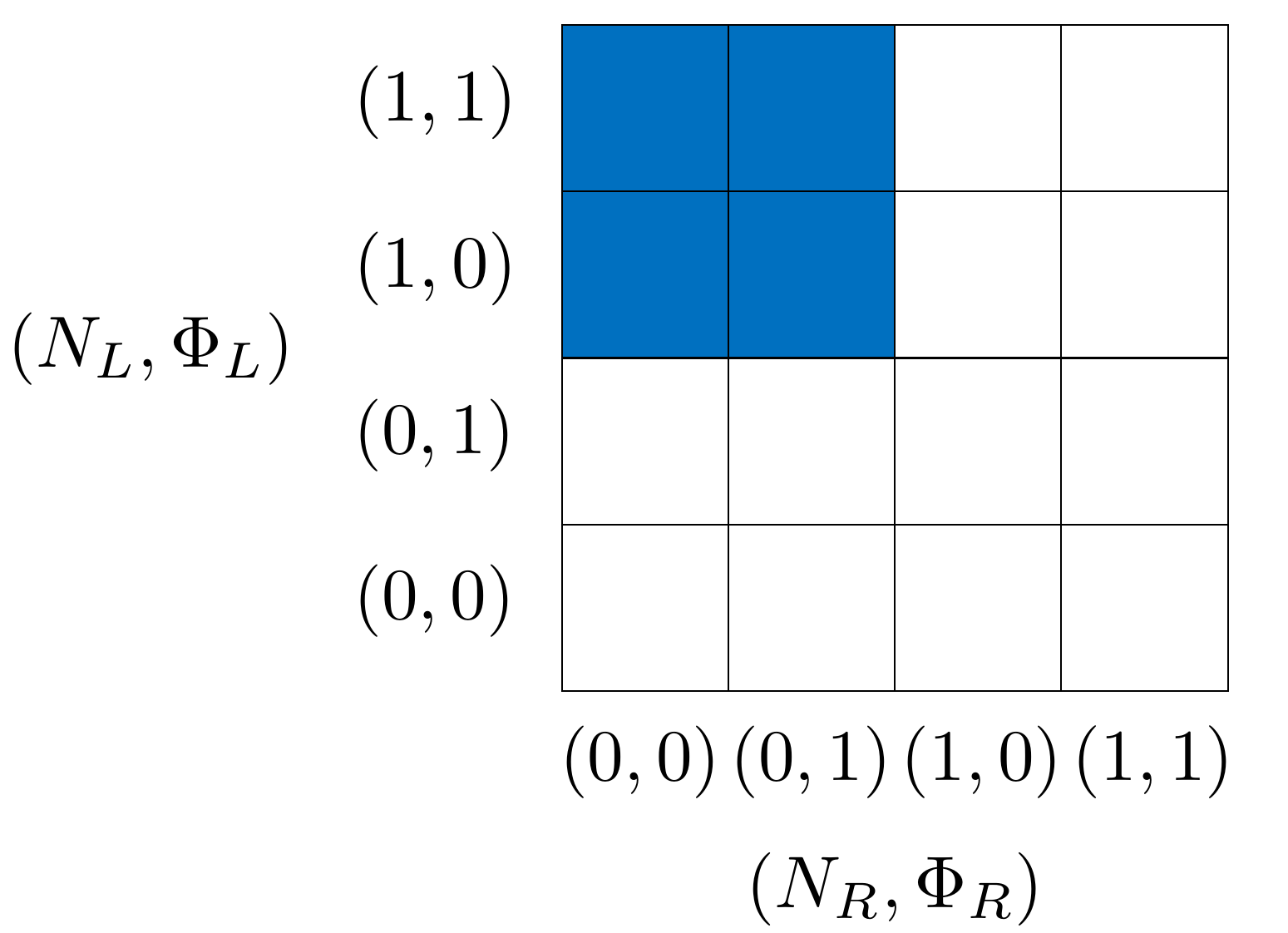}.
\end{align}



The analogue in the toy field theory of the unitary associated to a 50-50 beamsplitter is the function described in Eq.~\eqref{BSfunction}, which we repeat here:
\begin{align}\label{BSfunction2}
&N^{\rm out}_{L} = \Phi_L^{\rm in} \oplus \Phi_R^{\rm in} \nonumber 
 \\
&N^{\rm out}_{R} =  N_L^{\rm in} \oplus N_R^{\rm in} \oplus \Phi_L^{\rm in} \oplus \Phi_R^{\rm in}\nonumber
  \\
&\Phi^{\rm out}_{L} =  N_L^{\rm in} \oplus \Phi_R^{\rm in}\nonumber
 \\
&\Phi^{\rm out}_{R} =  \Phi_R^{\rm in}. 
\end{align}
This corresponds to a permutation of the 16 possible physical states, depicted diagrammatically as follows:
\begin{align}\label{BSdiagram}
\centering
\includegraphics[width=0.32\textwidth]{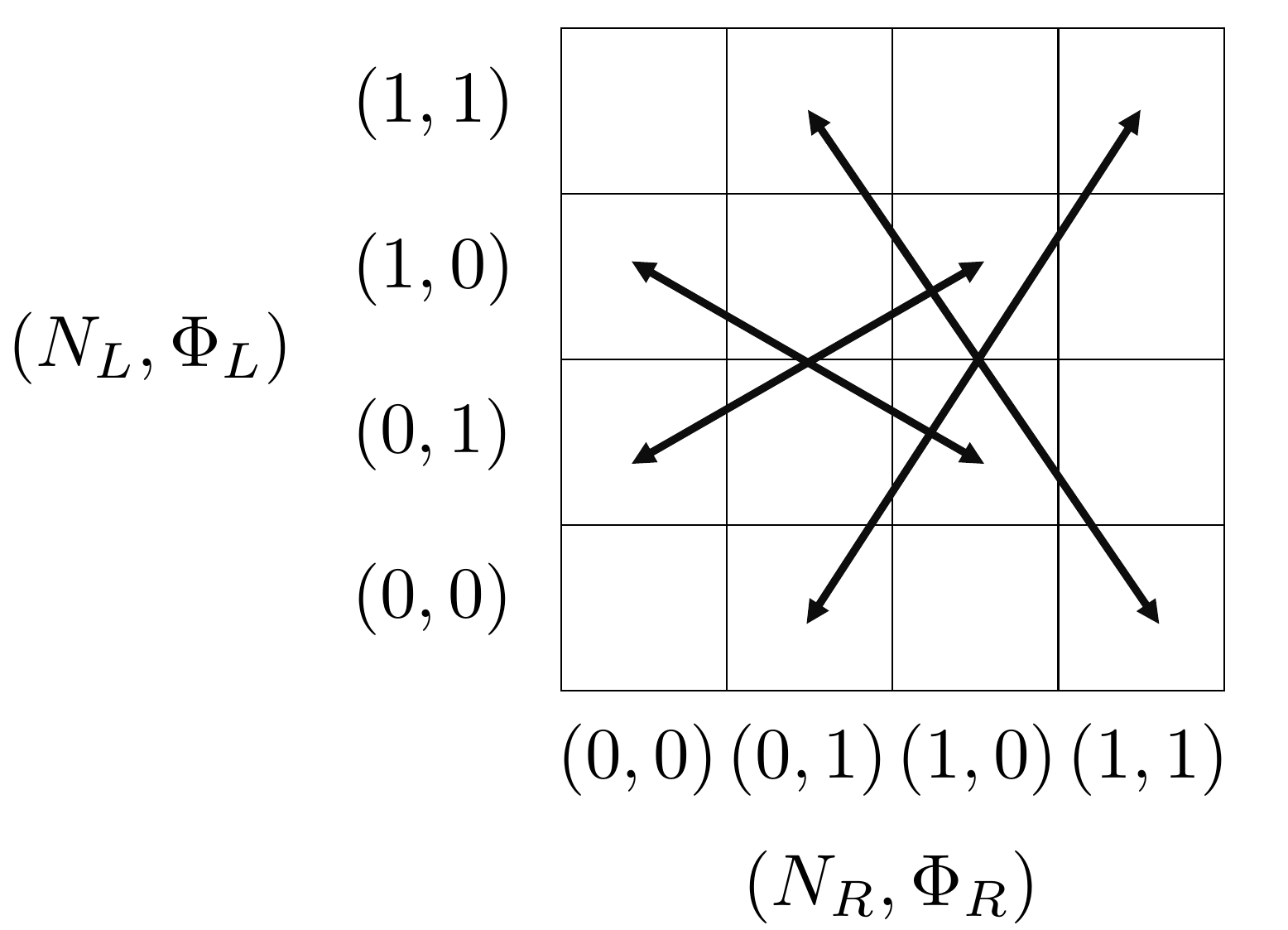}.
\end{align}
We refer to this function as the {\em beamsplitter dynamics}.

If one changes coordinates from $N_L,N_R, \Phi_L,\Phi_R$ to  $N_L,\Delta N,  \Phi_L,\Delta\Phi $, where 
\begin{align}
\Delta N \equiv N_L\oplus N_R\nonumber\\
\Delta\Phi \equiv \Phi_L\oplus \Phi_R,
\end{align}
then the beamsplitter dynamics can be expressed as:
\begin{align}\label{BSfunction3}
&N^{\rm out}_{L} = \Delta\Phi^{\rm in} \nonumber 
 \\
&\Delta N^{\rm out} =  \Delta N^{\rm in}\nonumber
  \\
&\Delta\Phi^{\rm out} =  N_L^{\rm in} \nonumber
 \\
&\Phi^{\rm out}_{R} =  \Phi_R^{\rm in},
\end{align}
i.e., it acts as identity on $\Delta N$ and $\Phi_R$, and swaps the values of $N_{L}$ and $\Delta\Phi$.  In this form, the function is seen to be simply the Swap Rule, articulated in subsection~\ref{ToyTrap}.  It is clear, therefore, that the Swap Rule is simply another way of describing the function defined by Eq.~\eqref{BSfunction}.


The impact of the first beamsplitter is determined by applying the function of Eq.~\eqref{BSfunction} to the distribution described in Eq.~\eqref{EpistemicInput}.  One thereby finds that the probability distribution after the first beamsplitter
 is 
\begin{align}\label{epistemicstate2}
 &p(N_L,N_R,\Phi_L,\Phi_R)\nonumber\\
&=\left( \tfrac{1}{2}[0][1]+\tfrac{1}{2}[1][0] \right) \left( \tfrac{1}{2}[0][1]+\tfrac{1}{2}[1][0] \right),
\end{align} 
which is depicted diagrammatically as:
 \begin{align}\label{diagramepistemicstate2}
 \centering
\includegraphics[width=0.32\textwidth]{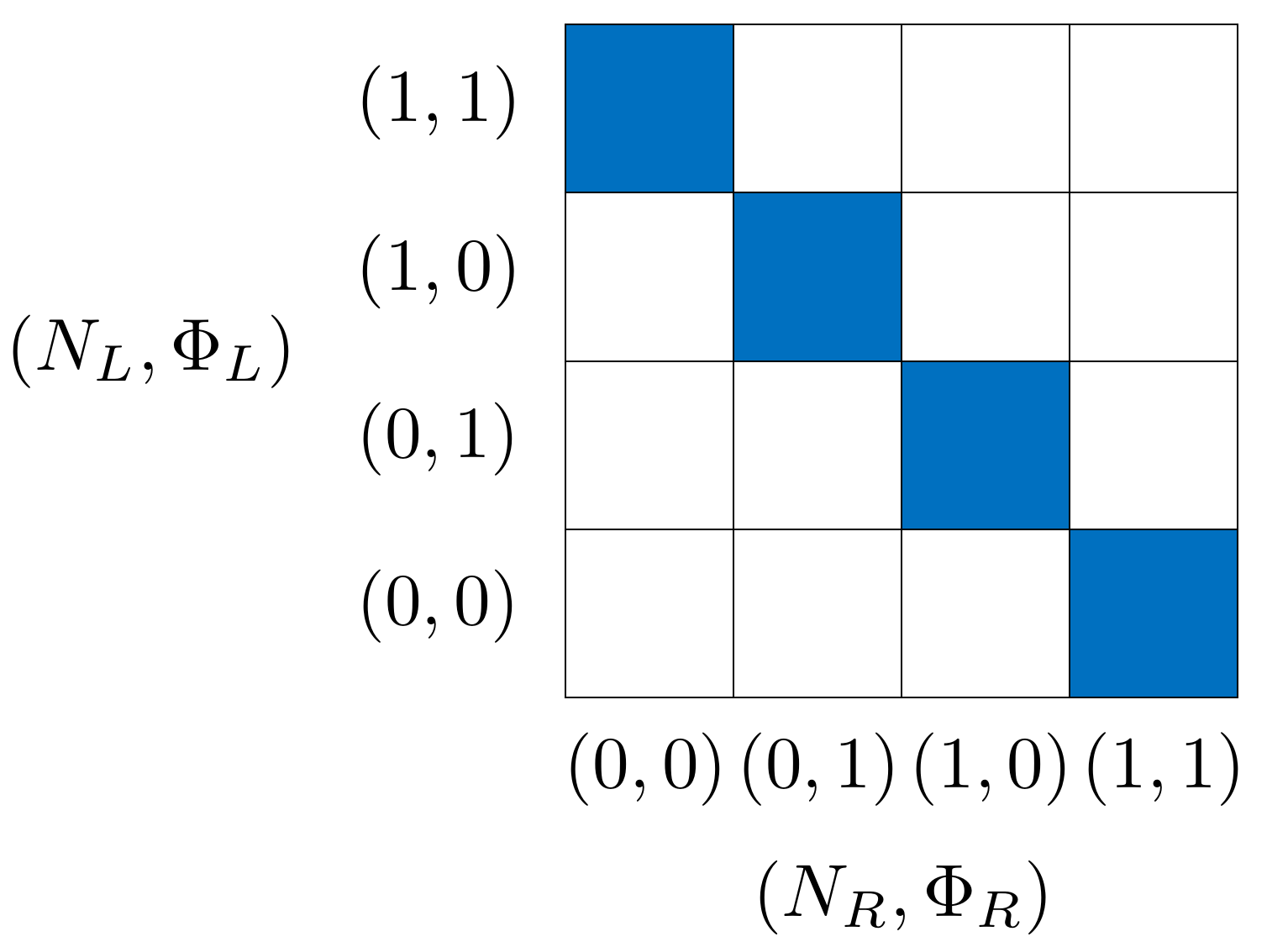}.
 \end{align}
This can also be inferred diagrammatically by working out the action of the permutation depicted in Eq.~\eqref{BSdiagram} on the distribution represented in Eq.~\eqref{EpistemicInputDiagram}.

\blk

{\bf The case of the Mach-Zehnder interferometer with phase shifter.} We now consider the toy field theory account of a Mach-Zehnder interferometer with a phase shifter in arm $R$, of the type depicted in Fig.~\ref{MachZehnder}. \blk

 A $\pi$ phase shift to a mode is represented by the function that flips the value of the discrete phase of that mode,   while preserving its occupation number:
\begin{align}\label{phasefipdynamics}
&N^{\rm out} =  N^{\rm in}\nonumber \\
&\Phi^{\rm out} =  \Phi^{\rm in} \oplus 1,
\end{align}
which is represented diagrammatically as:
\begin{align}
\centering
\includegraphics[width=0.2\textwidth]{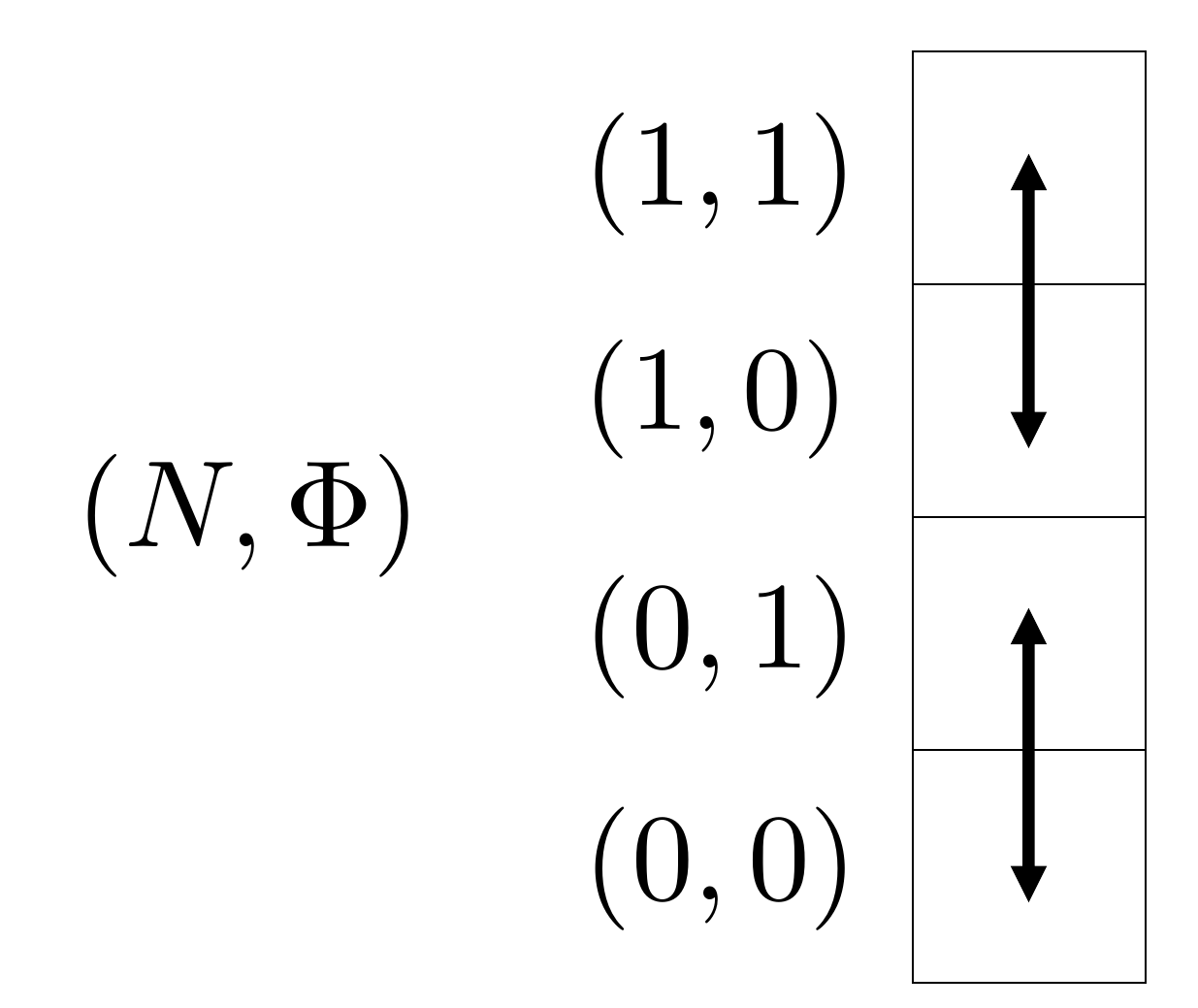}.
\end{align}

Note that the presence of a phase shifter on arm $R$ is presumed to leave the physical state of mode $L$ unchanged, so that it is the identity function that acts on the latter.
  
It follows that, as a permutation on the physical states of the {\em pair} of modes, a phase shifter on arm $R$ that implements a $\pi$ phase shift is represented as:
\begin{align}\label{PhaseShift2modes}
\includegraphics[width=0.32\textwidth]{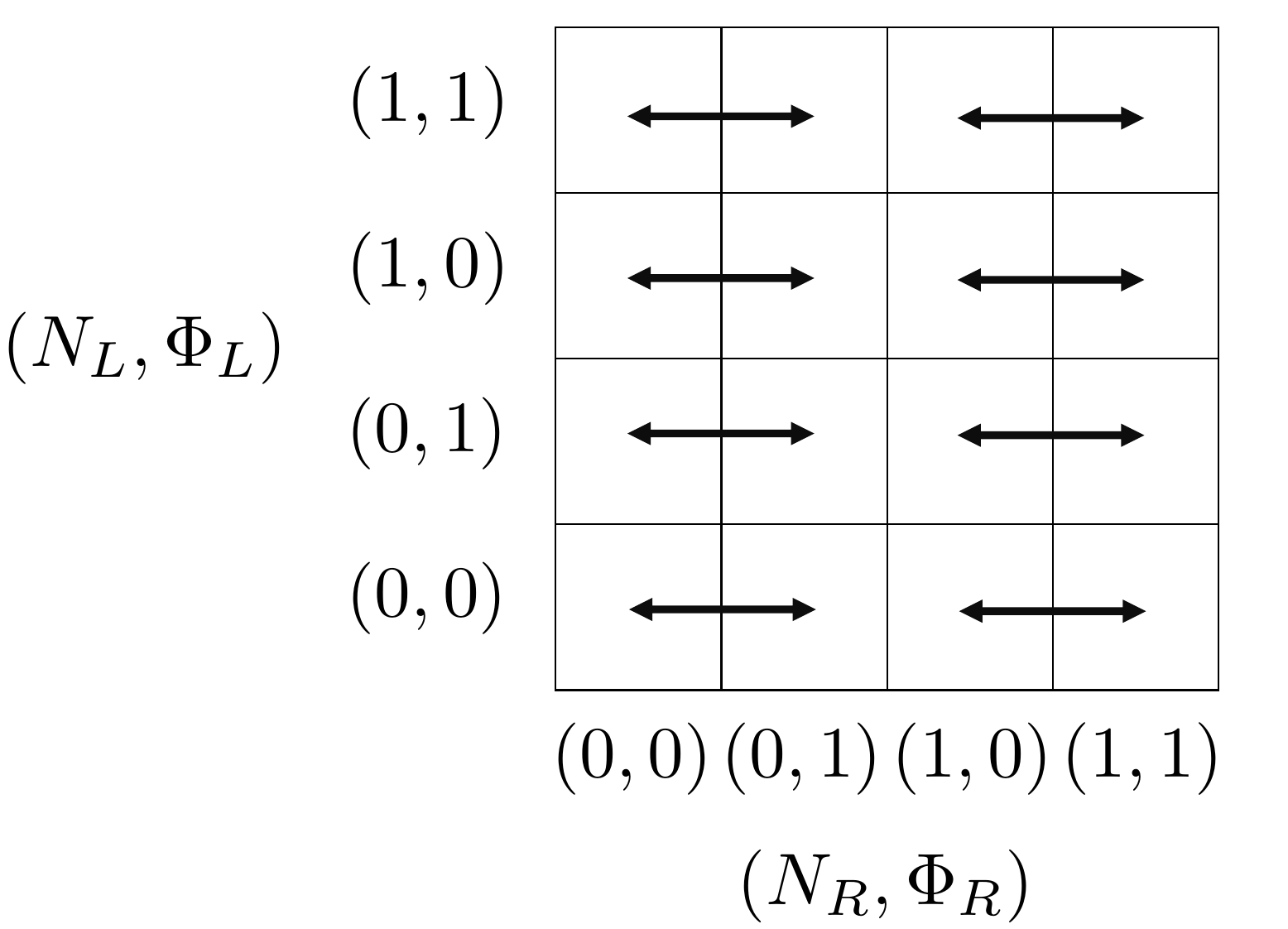}.
\end{align}
\blk

%

The phase flip dynamics of Eq.~\eqref{phasefipdynamics} maps the distribution of Eq.~\eqref{epistemicstate2} to 
\begin{align}\label{epistemicstate3}
&p(N_L,N_R,\Phi_L,\Phi_R)\nonumber\\
&=\left( \tfrac{1}{2}[0][1]+\tfrac{1}{2}[1][0] \right) \left( \tfrac{1}{2}[0][0]+\tfrac{1}{2}[1][1] \right),
\end{align}  
which is represented diagramatically as
 \begin{align}\label{diagramepistemicstate3}
 \centering
\includegraphics[width=0.32\textwidth]{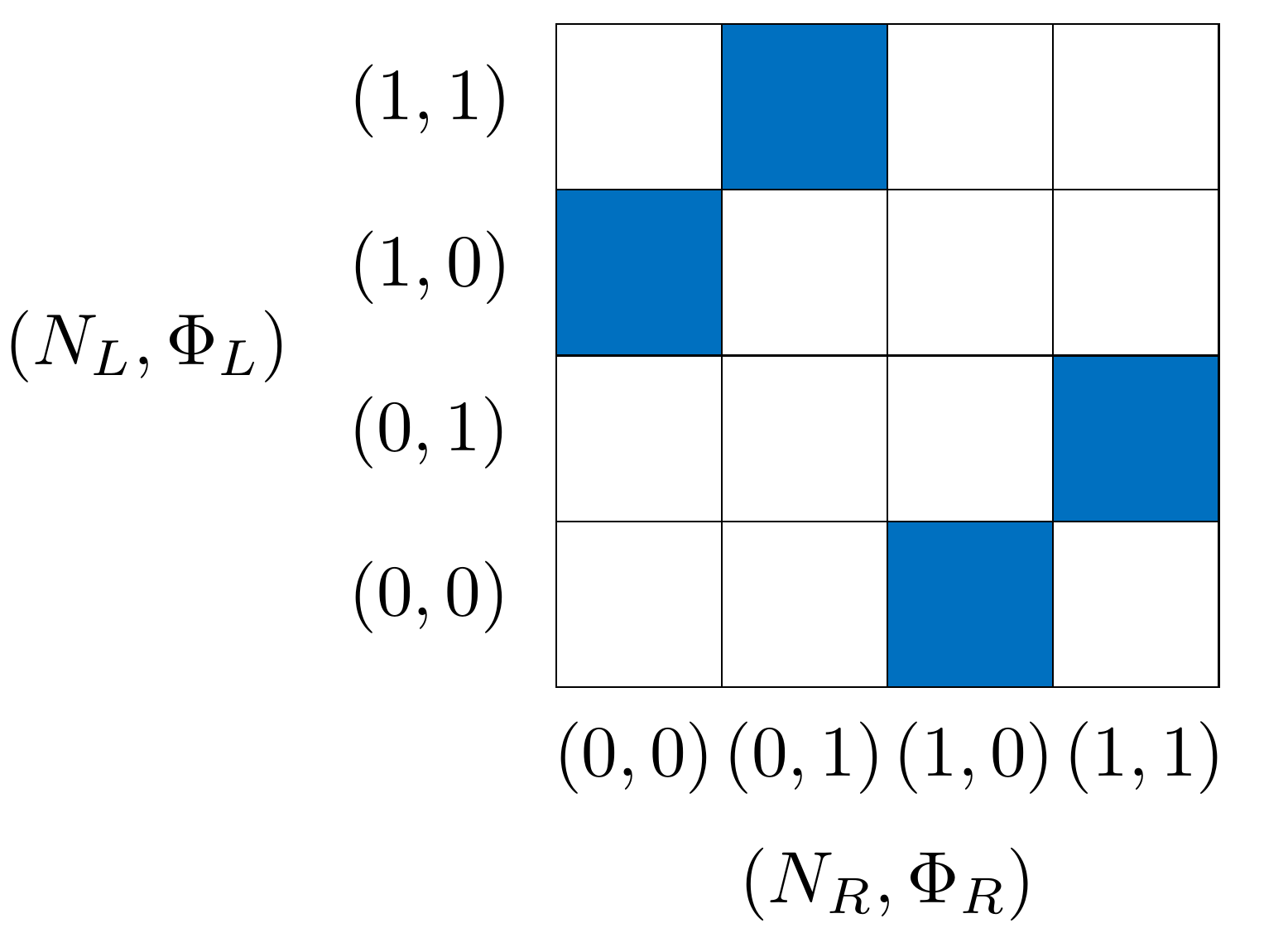}.
 \end{align}
This can also be inferred diagrammatically by applying the permutation of Eq.~\eqref{PhaseShift2modes} to Eq.~\eqref{diagramepistemicstate2}.

The second beamsplitter then maps the two possible distributions,
 Eq.~\eqref{epistemicstate2} and Eq.~\eqref{epistemicstate3}, via the beamsplitter dynamics of Eq.~\eqref{BSfunction}, back to the distribution of Eq.~\eqref{EpistemicInput} (depicted diagrammatically in Eq.~\eqref{EpistemicInputDiagram}) or to the distribution 
\begin{align}\label{EpistemicStateRoccupied}
 &p(N_L,\Phi_L,N_R,\Phi_R)\nonumber\\
 &=[0]\left(\tfrac{1}{2}[0]+\tfrac{1}{2}[1]\right)[1]\left(\tfrac{1}{2}[0]+\tfrac{1}{2}[1]\right),
 \end{align}
 which is depicted diagrammatically as:
\begin{align}\label{diagramepistemicstateRoccupied}
\centering
\includegraphics[width=0.32\textwidth]{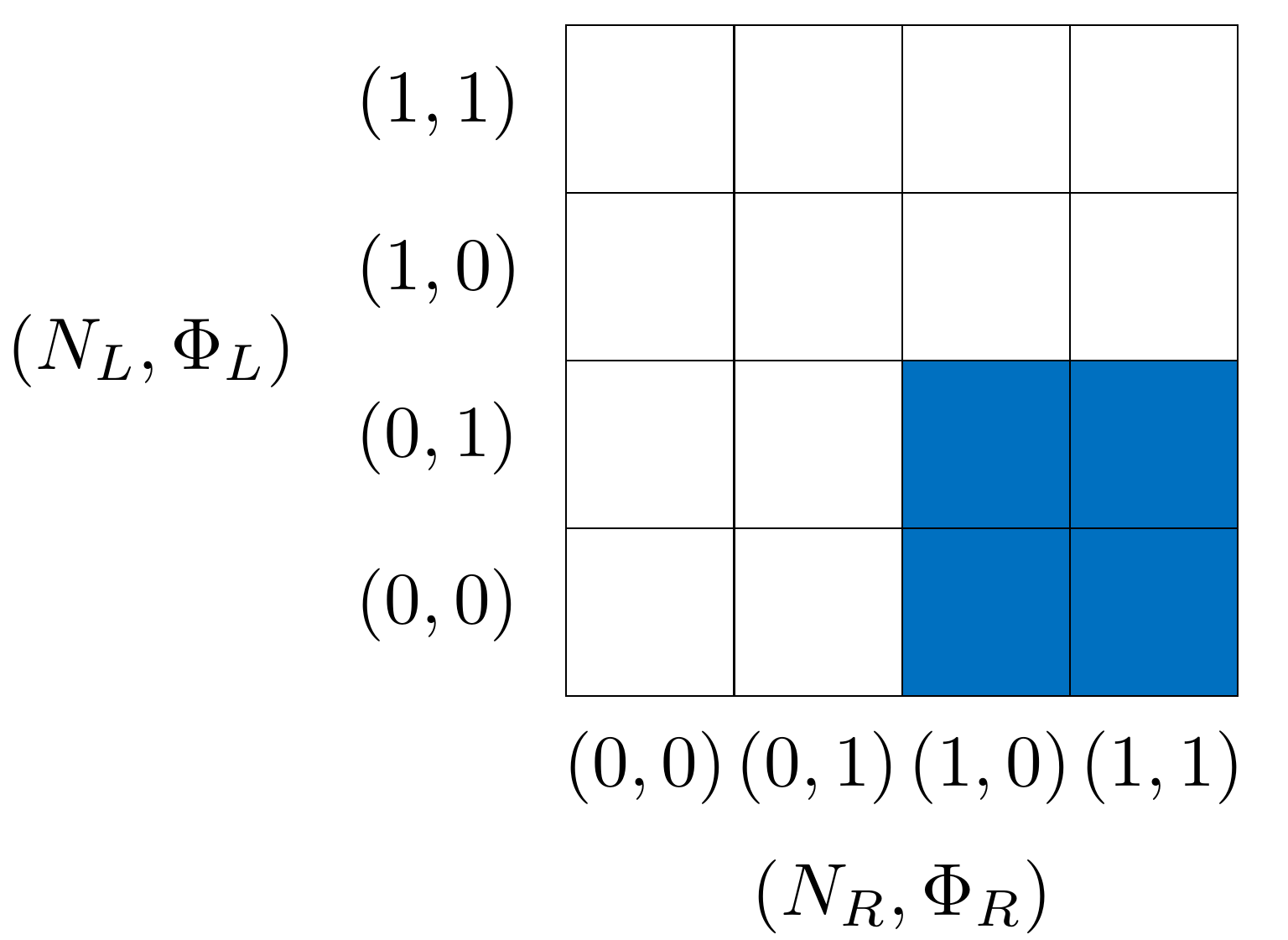}.
\end{align}
These two cases correspond respectively to the $L$ mode or the $R$ mode being occupied at the output of the second beamsplitter.  

\begin{figure*}[htb!]
\centering
{\includegraphics[width=.95\textwidth]{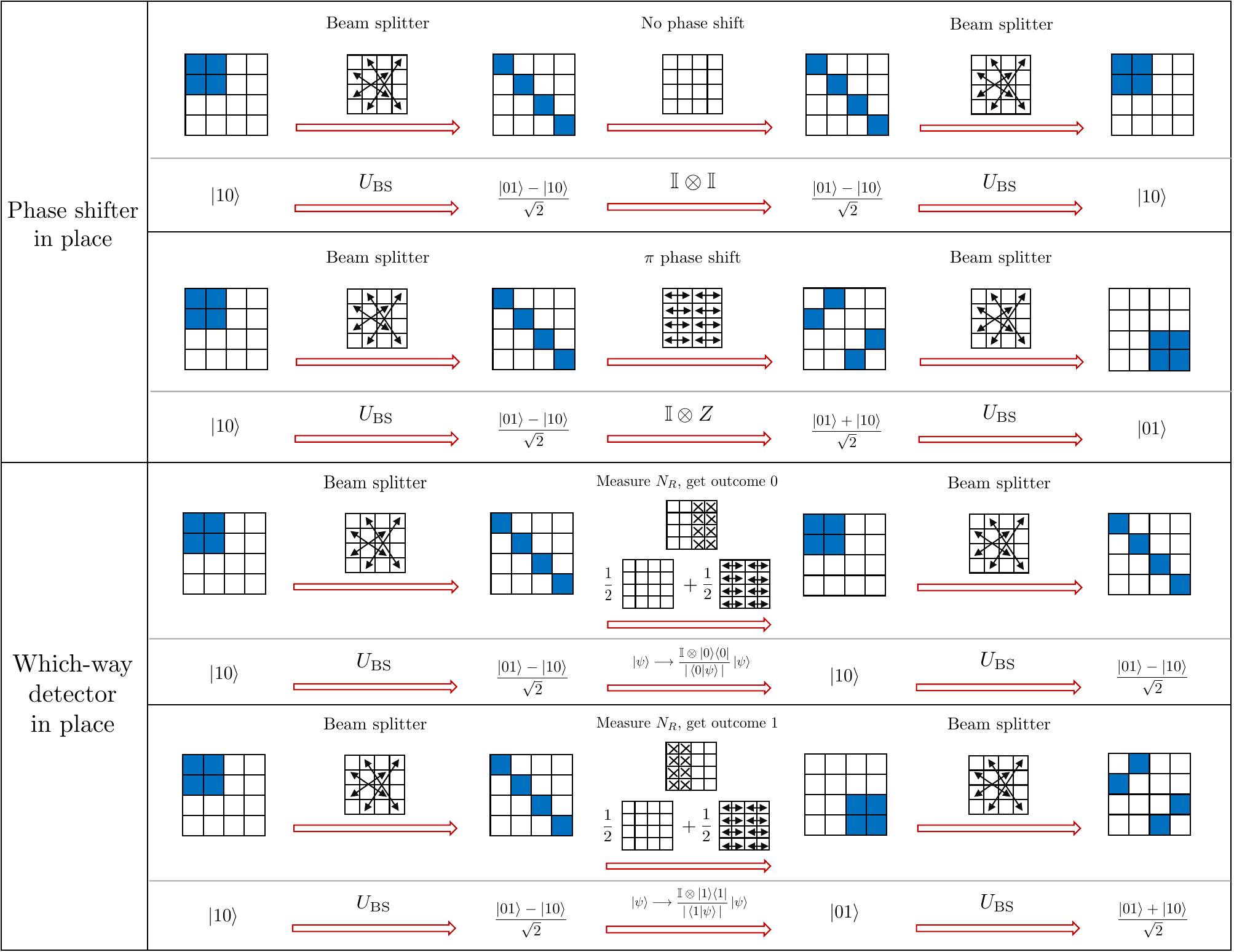}}
\caption{
 The toy field theory account of the Mach-Zehnder interferometer experiment with a phase shifter in arm $R$, implementing no phase shift or a $\pi$ phase shift, and with a detector in arm $R$.  
The evolution of the state of knowledge over the physical states of the two modes is depicted in each case. 
We also provide the quantum evolution for direct comparison.  Here, $U_{\rm BS}$ denotes the unitary associated to the beamsplitter action, described in Eq.~\eqref{BSunitary}, and $Z$ denotes the unitary associated to a $\pi$ phase shift, described in Eq.~\eqref{phaseshift}. \blk
}
\label{BigFigure}
\end{figure*}

A diagrammatic account of the full evolutions of the probability distribution for no phase shift and for a $\pi$ phase shift are summarized in the first two rows of Fig.~\ref{BigFigure}. 


\begin{multicols}{2}

\end{multicols}

The detectors in the output ports of the Mach-Zehnder interferometer implement a measurement of the  occupation numbers of the $L$ and $R$ modes.  In the toy field theory, such measurements simply reveal the possessed values of $N_L$ and $N_R$, so we recover the prediction that when there is no phase shift, it is the detector in the $L$ output port that always fires, while in the case of a $\pi$ phase shift, it is the detector in the $R$ output port that always fires.

{\bf The case of the Mach-Zehnder interferometer with which-way measurement.} We now turn to the case where there is a detector in the $R$ arm of the Mach-Zehnder interferometer, as in Fig.~\ref{MachZehnder}b). 



We begin by providing a formal account of the measurement update rule which was described in subsection~\ref{ToyTrap}.

If the initial probability distribution is $p(N,\Phi)$  and one  obtains the $N=0$ outcome in the measurement of occupation number (note that this can only occur if $p(N,\Phi)$ assigns a nonzero probability to $N=0$), then the update rule can be understood as consisting of two steps.  Step 1 is the updating of knowledge implied by the fact that learning the outcome allows one to resolve some of one's uncertainty regarding the physical state: 
\begin{align}
p(N,\Phi) \mapsto p'(N,\Phi)= \frac{1}{\mathcal{N}} \delta_{N,0} p(N,\Phi)
\end{align}
where $\mathcal{N}$ is a normalization constant that ensures that $p'(N,\Phi)$ is a normalized probability distribution.  We depict this diagrammatically as 
\begin{align}\label{MmtUpdateRuleOntic}
\centering
\includegraphics[width=0.06\textwidth]{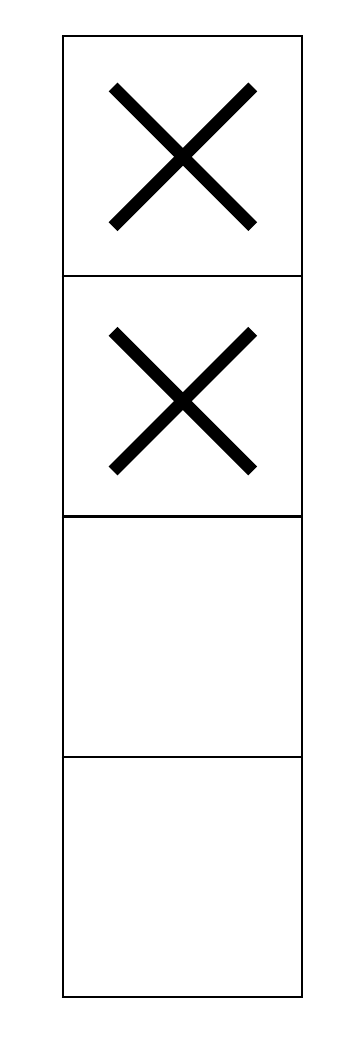},
\end{align}
where the cross symbol on a physical state represents the fact that it has been ruled out by the measurement and so its probability is set to zero.
Step 2 is the updating of knowledge that is implied by the fact that the phase is randomized, 
\begin{align}\label{MmtUpdateRuleOnticAlgebraic}
p'(N,\Phi) \mapsto p''(N,\Phi) 
\equiv \tfrac{1}{2}p'(N,\Phi)  + \tfrac{1}{2}p'(N,\Phi\oplus 1),
\end{align}
which we depict diagrammatically as
\begin{align}\label{MmtUpdateRuleOntic}
\centering
\includegraphics[width=0.19\textwidth]{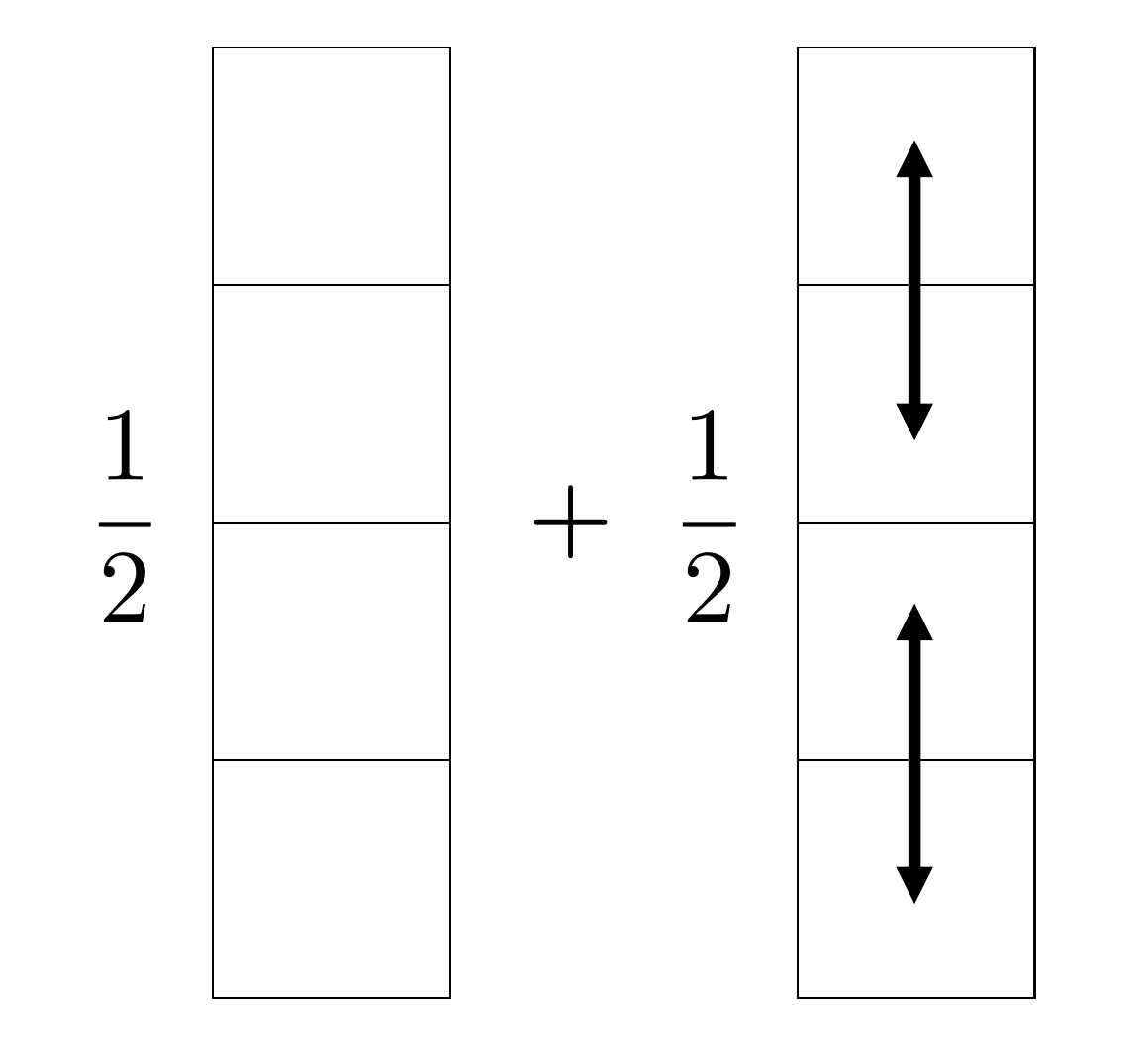}.
\end{align}

The toy field theory analogue of the quantum state update rule described in Eq.~\eqref{stateupdateFock} is obtained by composing these two steps:
\begin{align}\label{MmtUpdateRule}
p(N,\Phi) \mapsto [0]\left(\tfrac{1}{2}[0] + \tfrac{1}{2}[1] \right),
\end{align}
which stipulates that regardless of the initial distribution $p(N,\Phi)$, the final distribution is the one wherein $N$ is known to be 0 and $\Phi$ is unknown.




Note that the measurement update rule implies that even if one has perfect knowledge of the phase at the input of the measurement, one can infer nothing about the phase at the output of the measurement.  Also, even if one has perfect knowledge of the phase at the {\em output} of the measurement (by knowing the outcome of a subsequent measurement of phase for instance), then one can infer nothing about the phase at the input of the measurement.   


\blk

We now consider how this impacts a probability distribution over the physical states for the {\em pair} of modes.

Again, consider a measurement of occupation number on mode $R$ that yields the outcome 0.  The probability distribution $p(N_L,\Phi_L,N_R,\Phi_R)$, is then updated in two steps as before.  The knowledge-updating step associated to resolving uncertainty  is:
\begin{align}\label{Bayesianupdating}
p(N_L,\Phi_L,N_R,\Phi_R) \mapsto &p'(N_L,\Phi_L,N_R,\Phi_R)\nonumber\\
&\equiv \frac{1}{\mathcal{N}} \delta_{N_R,0} p(N_L,\Phi_L,N_R,\Phi_R),
\end{align}
where $\mathcal{N}$ is a normalization constant. It is depicted diagrammatically as
\begin{align}\label{mmtupdaterulediagram1}
\centering
\includegraphics[width=0.19\textwidth]{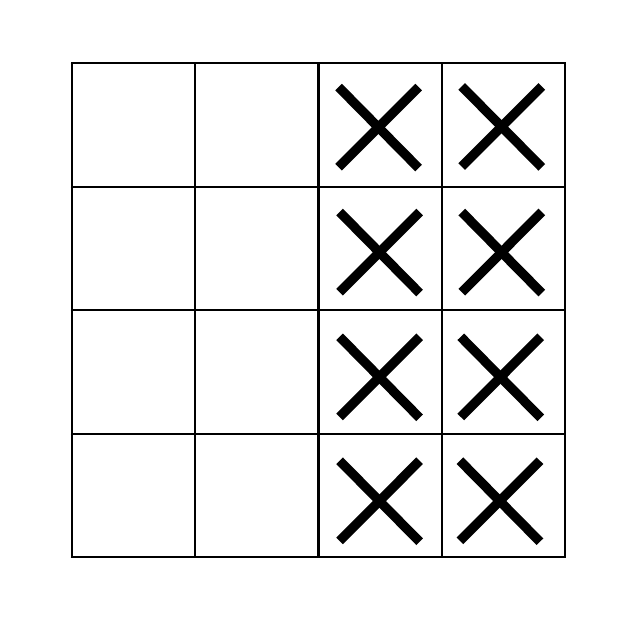}.
\end{align}
The knowledge-updating step associated to phase randomization is
\begin{align}\label{randomization}
p'(N_L,\Phi_L,N_R,\Phi_R) \mapsto &\tfrac{1}{2} p'(N_L,\Phi_L,N_R,\Phi_R)\nonumber\\
&+ \tfrac{1}{2} p'(N_L,\Phi_L,N_R,\Phi_R\oplus 1),
\end{align}
which is depicted diagrammatically as
\begin{align}\label{mmtupdaterulediagram2}
\centering
\includegraphics[width=0.4\textwidth]{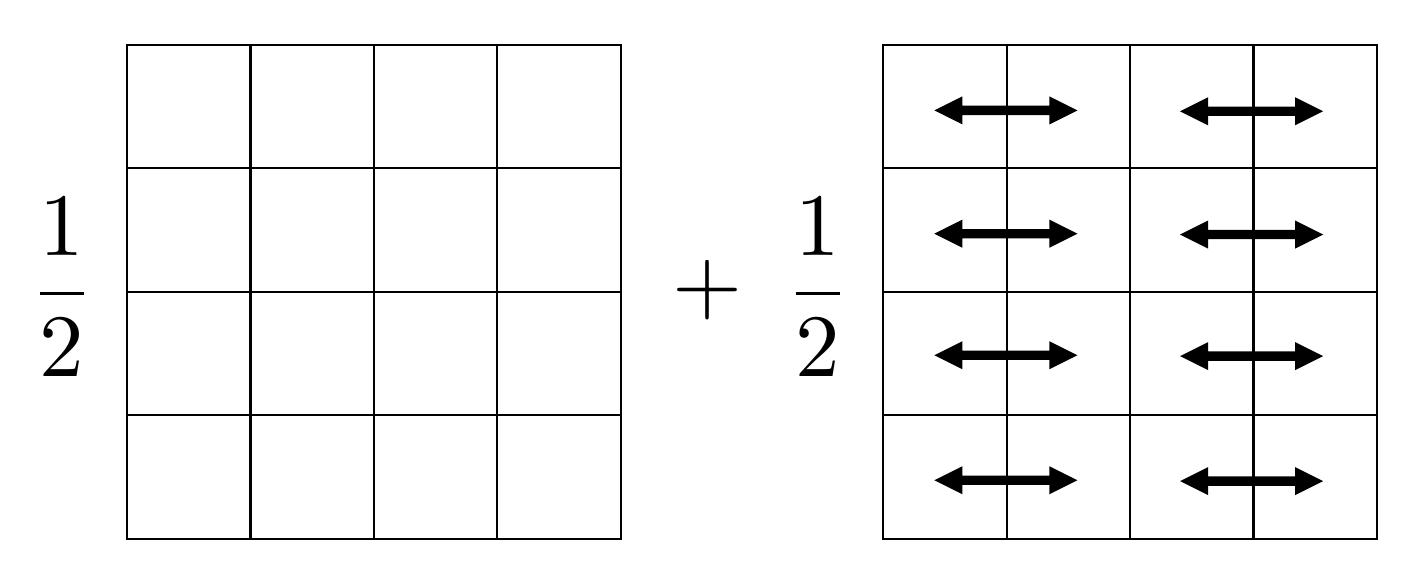}.
\end{align}

The toy field theory analogue of the quantum state update rule, Eq.~\eqref{stateupdateFock0}, is obtained by composing these two steps, 
\begin{align} \label{mmtupdateruleformula}
p(N_L,\Phi_L,N_R,\Phi_R) \mapsto p'(N_L,\Phi_L) [0]\left(\tfrac{1}{2}[0] + \tfrac{1}{2}[1] \right),
\end{align}
where $p'(N_L,\Phi_L)$ is the marginal over $N_L$ and $\Phi_L$ of $p'(N_L,\Phi_L,N_R,\Phi_R)$.

%


We can now complete the analysis of the Mach-Zehnder inteferometer with a detector in the $R$ arm.  Consider the case where the detector {\em does not} fire.  We first apply the knowledge-updating step associated to resolving uncertainty by virtue of  
  learning that $N_R=0$, that is Eq.~\eqref{Bayesianupdating}, to the distribution in Eq.~\eqref{epistemicstate2}, obtaining
\begin{align}\label{epistemicstateBayesianupdate}
 &p(N_L,N_R,\Phi_L,\Phi_R)= [1][0] \left( \tfrac{1}{2}[0][1]+\tfrac{1}{2}[1][0] \right),
\end{align} 
which is depicted diagramatically as
\begin{align}
\label{epistemicstateBayesianupdatediagram}
\centering
\includegraphics[width=0.31\textwidth]{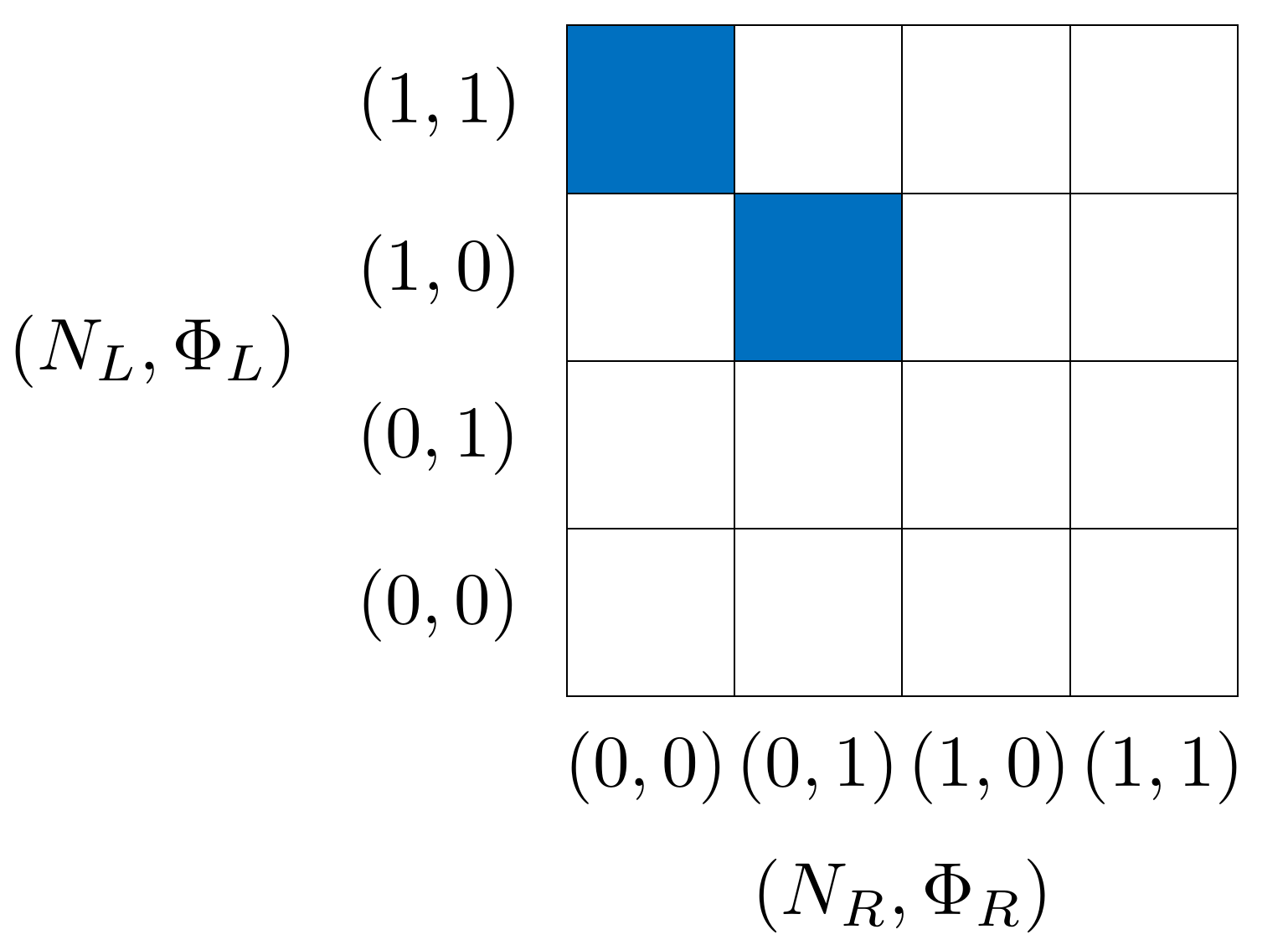}.
\end{align}
We then apply the knowledge-updating step associated to randomization of the local phase $\Phi_R$, that is, Eq.~\eqref{randomization}, to this distribution to obtain (notice that we've reordered the variables)
 \begin{align}\label{epistemicstaterandomized}
 &p(N_L,\Phi_L,N_R,\Phi_R)= [1] \left( \tfrac{1}{2}[0]+\tfrac{1}{2}[1] \right)[0] \left( \tfrac{1}{2}[0]+\tfrac{1}{2}[1] \right),
\end{align} 
which is the same distribution as in Eq.~\eqref{EpistemicInput}, depicted diagrammatically in Eq.~\eqref{EpistemicInputDiagram}.

A diagrammatic account of the full evolution of the probability distribution when there is a which-way detector on arm $R$ which does not fire is given in the third row of Fig.~\ref{BigFigure}. 
 
The case where the detector {\em does} fire, so that one implements the resolving of uncertainty
  appropriate for learning that $N_R=1$ (rather than $N_R=0$) is treated similarly.

\subsection{The lesson of the toy field theory for the difference between first-quantized and second-quantized descriptions}\label{lessontoyfieldtheoryformalaccount}

In Sec.~\ref{significanceofthemove}, we described an alternative to the toy field theory.  Specifically, we described a toy theory that stands to the first-quantized description of the Mach-Zehnder interferometer as the toy field theory stands to its second-quantized description.

%

This theory takes the system of interest to be a photon, and presumed that it has two properties: a discrete which-way property describing whether the photon is in the left or right arm, $W\in \{L,R\}$, and a discrete momentum that is canonically conjugate to this which-way property,  $\Theta \in \{ 0,1\}$.  

The physical state space can be depicted as a $2\times 2$ grid of boxes, and a particular physical state as a marked box.  Thus, for instance, the physical state wherein $W=L$ and $\Theta=0$ is depicted as:
\begin{align}
\centering
{\includegraphics[width=.2\textwidth]{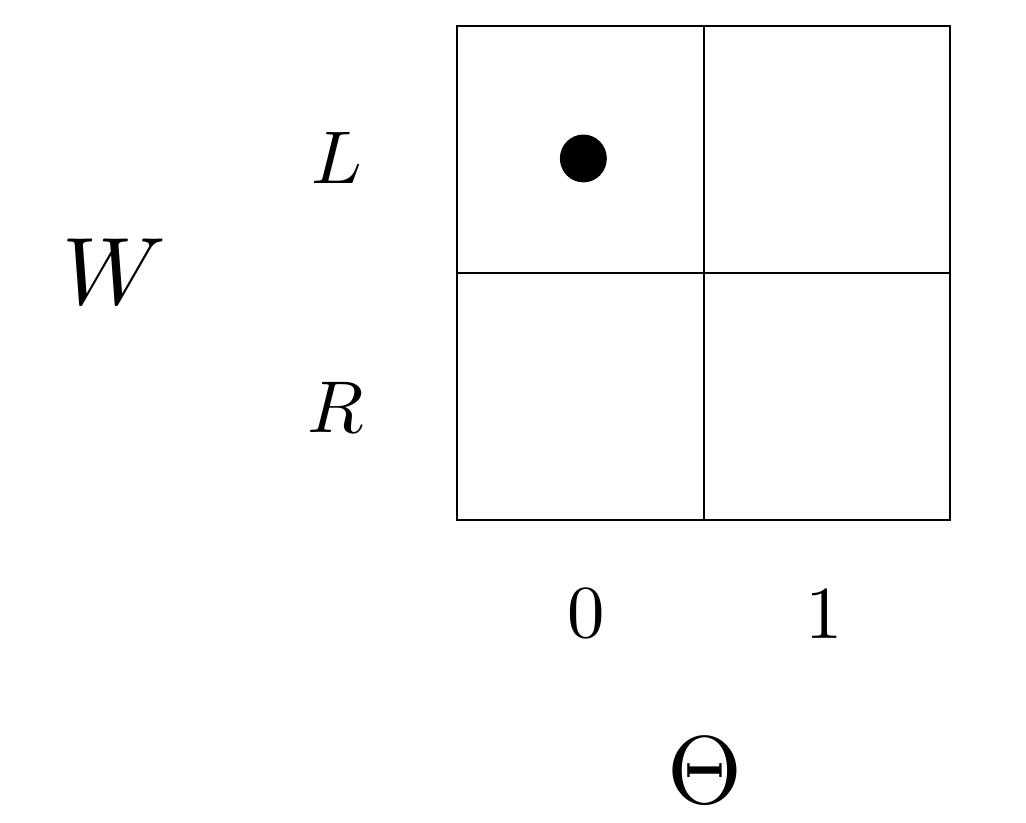}}.
\end{align}
The probability distributions $p(W,\Theta)$ that satisfy the epistemic restriction are then denoted diagrammatically in the obvious way, by filling in two of the four boxes.

In Sec.~\ref{significanceofthemove}, we noted that the translation from the toy field theory to the first-quantized toy theory is given by the map of Eq.~\eqref{translationtoy}. This can be represented diagrammatically as follows:
\begin{align}\label{TranslationToyFig}
\centering
{\includegraphics[width=.45\textwidth]{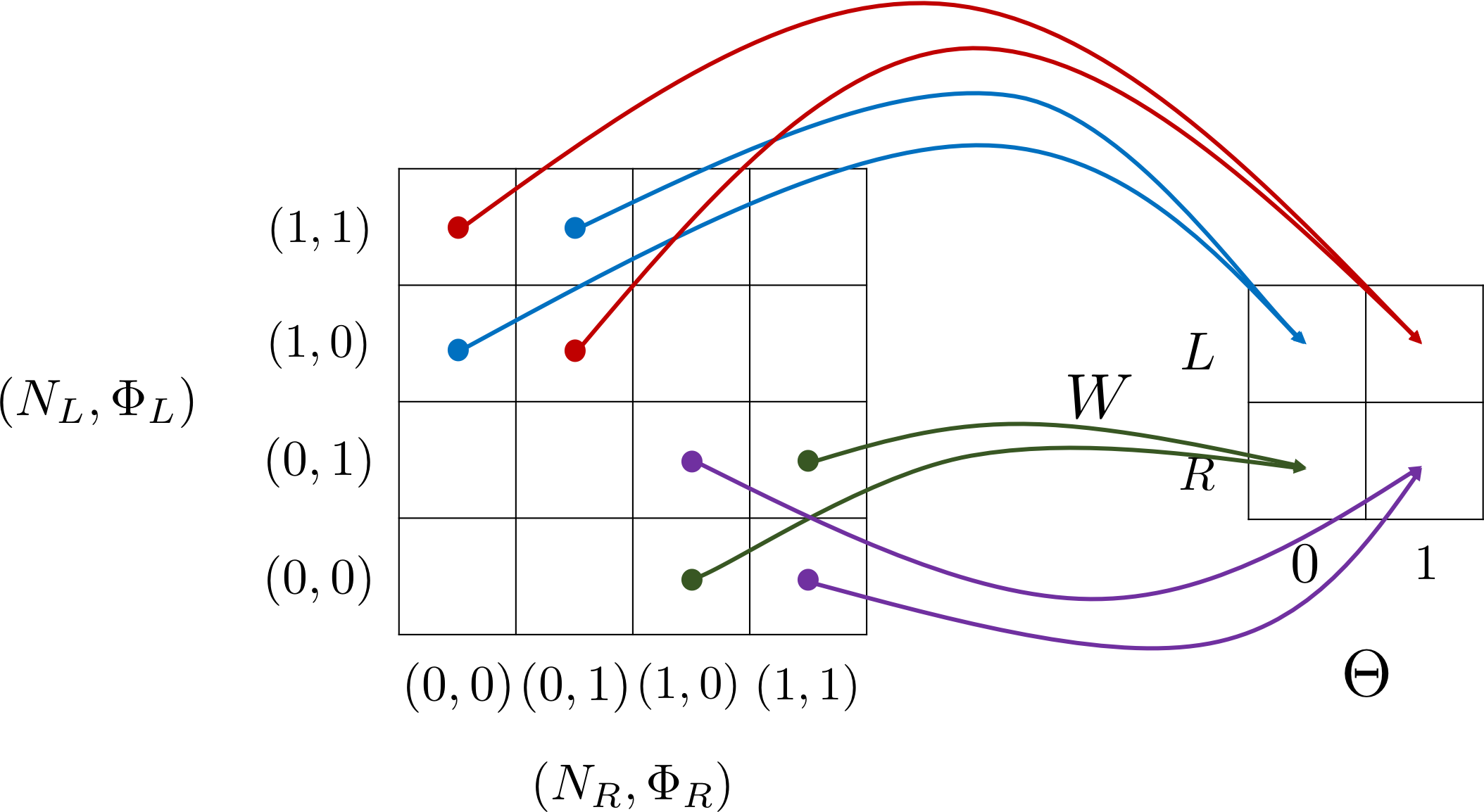}}.
\end{align}

\begin{figure*}[htb!]
\centering
{\includegraphics[width=0.95\textwidth]{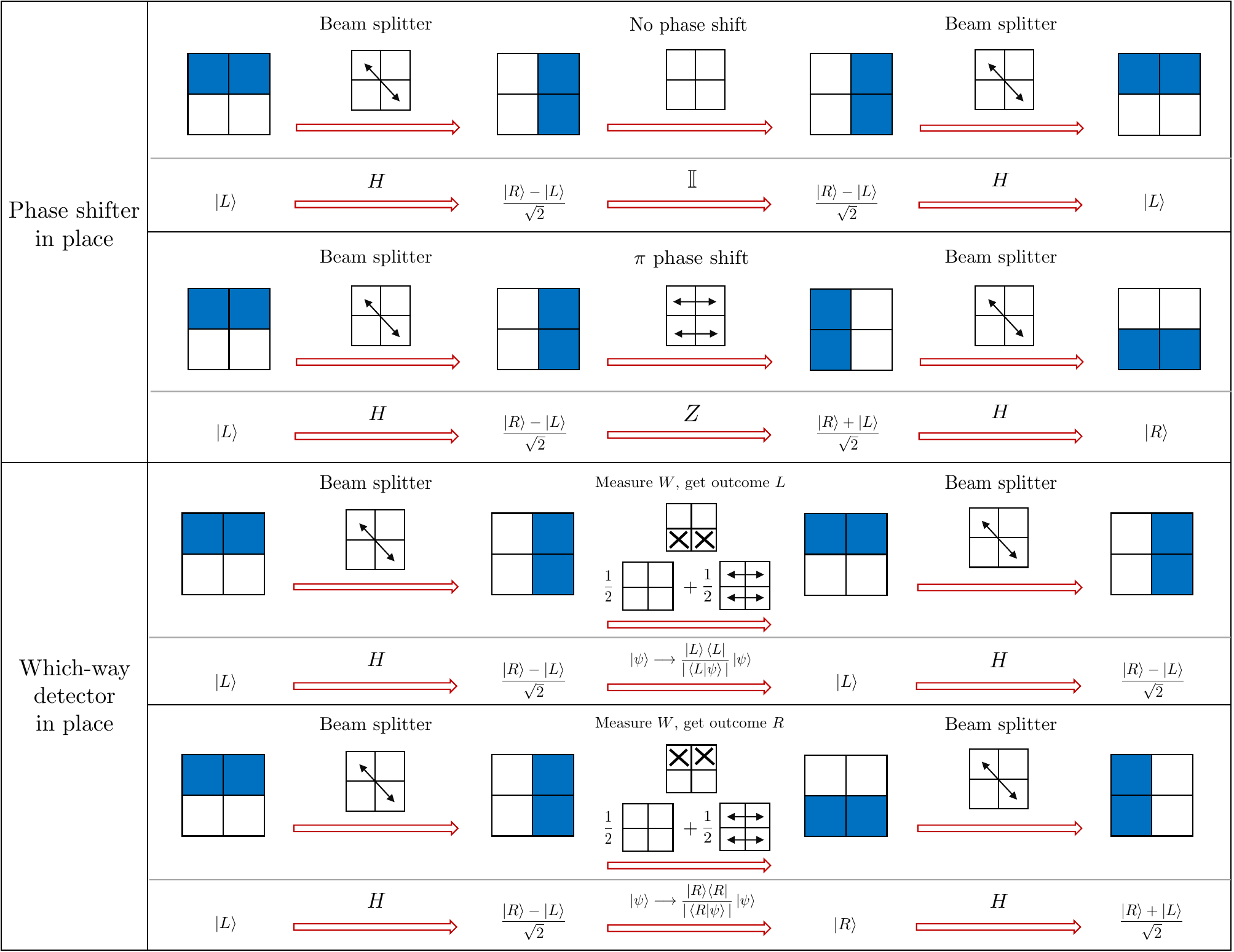}}
\caption{ 
A first-quantized description of quantum interference phenomena and how these are reproduced using the coarse-grained variables of the first-quantized toy theory. At each time step, the state of knowledge over the coarse-grained which-way variable $W$ and relative phase variable $\Theta$ is depicted just above the corresponding quantum state. For the standard interferometer (with and without the phase shift), the photon always ends up in the same output port ($L$ and $R$, respectively). In the case where a which-way measurement is carried out, the photon is equally likely to appear in either output port.
 We also provide the quantum evolution for comparison.  Here, $H$ denotes the unitary associated to the beamsplitter action, described in Eq.~\eqref{BSunitary}, and $Z$ denotes the unitary associated to a $\pi$ phase shift, described in Eq.~\eqref{phaseshift}. \blk
}
\label{BigFigureRelational}
\end{figure*}

This allows one to translate the second-quantized toy theory description of evolution through the Mach-Zehnder interferometer, presented diagrammatically in  Fig.~\ref{BigFigure}, into the first-quantized toy theory description of this evolution, presented diagrammatically in Fig.~\ref{BigFigureRelational}.
\begin{multicols}{2}
\end{multicols}

\subsection{Formal toy-field-theoretic account of the quantum eraser experiment}\label{formalquantumeraser}

In Sec.~\ref{Section_7}, we described the toy field theory account of the quantum eraser experiment.  Here, we present this account a bit more formally


The joint probability distribution of the pair of modes at the output ports of the first beamsplitter 
  is the one described in Eq.~\eqref{epistemicstate2}.  The auxiliary system $A$ is presumed to be prepared such that $Q_A$ is known, while $P_A$ is unknown, described by the distribution $p(Q_A,P_A) = [a_0] (\tfrac{1}{2}[a_+] +\tfrac{1}{2}[a_-])$.
     It follows that the joint probability distribution for the pair of modes together with auxiliary system $A$ is
\begin{align}\label{toytimestep2all}
&p(N_L,N_R, Q_A,\Phi_L,\Phi_R,P_A)\nonumber\\
&=\left( \tfrac{1}{2}[0][1]+\tfrac{1}{2}[1][0] \right) [a_0]\nonumber\\
&\times \left( \tfrac{1}{2}[0][1]+\tfrac{1}{2}[1][0] \right) (\tfrac{1}{2}[a_+] +\tfrac{1}{2}[a_-]).
\end{align} 
The CNOT dynamics, described in Eq.~\eqref{cnot}, maps this to 
\begin{align}\label{toytimestep3all}
&p(N_L,N_R,Q_A,\Phi_L,\Phi_R,P_A)\nonumber\\
&=\left( \tfrac{1}{2}[0][1][a_1]+\tfrac{1}{2}[1][0][a_0] \right) \nonumber\\
&\times \tfrac{1}{2} 
\left\{
\left( \tfrac{1}{2}[0][1]+\tfrac{1}{2}[1][0] \right) [a_+]
+ \left( \tfrac{1}{2}[0][0]+\tfrac{1}{2}[1][1] \right) [a_-]
\right\}.
\end{align} 
By considering the marginal $p(N_R,Q_A)$ of Eq.~\eqref{toytimestep3all}, it is clear 
  that a measurement of $Q_A$ allows one to infer the value of $N_R$,
  and so constitutes an indirect measurement of $N_R$.  
Similarly, by considering the marginal $p(\Phi_L,\Phi_R,P_A)$ of Eq.~\eqref{toytimestep3all}, it is clear that 
  a measurement of $P_A$ allows one to infer the value of  $\Phi_L \oplus \Phi_R$ 
   at the end of the measurement.  
  
  The argument for the reproduction of the phenomenology of the quantum eraser then proceeds as follows. 
 If one sorts the data by whether a measurement of $Q_A$ yields $a_0$ or $a_1$, then one is sorting by whether $N_R$ has value 0 or 1 and consequently in each subset one sees no interference.  On the other hand, if one sorts the data by whether a measurement of $P_A$ yields $a_+$ or $a_-$, then one is sorting by whether $\Phi_L \oplus \Phi_R$ has value 0 or 1, and consequently in each subset one sees interference.

It is worth noting that one can also understand these predictions by means of a graphical analysis.  In order for the physical states of each of the three systems (mode $L$, mode $R$, and auxiliary system $A$) to be represented on different axes, the graphical representation of the joint distribution must be three-dimensional.  This is achievable 
 but it is less intuitive to work with than its two-dimensional counterpart and does not provide much additional insight relative to the analysis just given, so we do not provide such an analysis here. 
 \blk

\section{Contrasting the toy field theory with other realist accounts of quantum interference}
\label{OnticQInterference}

For the first two interpretational claims outlined in the introduction, there are other ways of seeing that these are not forced on us by the phenomenology of the Mach-Zehnder interferometer and its variants.
 
This is especially true with regards to the claim of the observer-dependence of reality.  Almost all researchers motivated to work on realist interpretations of quantum theory are keen to eliminate any appeal to ``psycho-kinetic effects''.  (The only prominent exceptions are certain proposals for dynamical collapse models where the consciousness of the observer is implicated in the mechanism that leads to collapse~\cite{wigner1995remarks,London1939observation}.)

Some realist interpretations also do away with the notion of wave-particle complementarity.  Bohmian mechanics~\cite{Bohm1952suggested}, for instance, provides an account of interference phenomenology where there is a particle {\em and} a wave~\cite{Philippidis1979}. Note, however, that only if the system of interest is a single particle is it the case that the wave is defined over familiar 3-dimensional {\em physical} space.  For two particles, for instance, the wave is defined on
  the 6-dimensional space of spatial configurations of the pair (specifying the 3 spatial coordinates of each particle).   
 Other $\psi$-ontic models~\cite{Harrigan2010}, such as the one proposed recently by Blasiak~\cite{Blasiak2015}, can also be understood as providing an account of the single-particle phenomenology in terms of a particle and a wave in physical space.  Like Bohmian mechanics, such an account is not possible for two or more particles. \blk
 
Moreover, even in the case of a single particle, we noted earlier (in footnote~\ref{Bohmianfootnote2}) that having a particle {\em and} a wave is quite different from the way in which wave-particle complementarity is rejected in the toy field theory, where there are modes that have particle-like and wave-like properties which are canonically conjugate to one another.

It is with the third interpretational claim, stating that the TRAP phenomenology implies the failure of explainability in terms of local causes, that the account provided by the toy field theory differs most from the other accounts that deny the first two interpretational claims. 
 The latter accounts generally {\em do} require a rejection of the possibility of local causal explanations for one or more of the interference phenomena discussed in this article.  

Consider in particular the quantum eraser experiment.  \
The fact that the toy field theory can account for the quantum eraser without nonlocality demonstrates that nonlocality is not {\em required} to explain this phenomenology.  However, every $\psi$-ontic model, in particular, Bohmian mechanics, dynamical collapse theories~\cite{ghirardi1986unified,pearle1989combining,RevModPhys.85.471}, Blasiak's model, etcetera, {\em do} require nonlocality to account for it.  This follows from the fact that the quantum eraser is simply a version of the Einstein-Podolsky-Rosen (EPR) experiment~\cite{EPR1935} and every $\psi$-ontic model is incapable of providing a local account of the correlations obtained in the EPR experiment~\cite{Harrigan2010}.  
By contrast, $\psi$-epistemic models (such as the toy field theory and other classical statistical theories with an epistemic restriction~\cite{Spekkens2007,Spekkens2016}) {\em are} capable of doing so.  This is possible because the correlations in the EPR experiment (and those in the quantum eraser experiment) do not violate any Bell inequalities.  In other words, the reason that there are experimental scenarios for which a local account is possible in the  toy field theory but not in any $\psi$-ontic model is that the latter sort of model posits strictly more nonlocality than is required 
by Bell's theorem. \blk

It is worth noting a subtlety here.  From the Bohmian perspective, the case of the basic Mach-Zehnder interferometer is special among interference experiments insofar as the Bohmian account in this case {\em does not involve any nonlocality}.  In the Bohmian account, the wavefunction of the photon branches, with one branch of the wavefunction 
  propagating through the $L$ arm and the other through the $R$ arm.  Meanwhile, the Bohmian particle takes one particular arm, falling in the support of only one of these two branches of the wavefunction.  The branch of the wavefunction (or `wave') containing the Bohmian particle is termed `occupied', while the other  is termed `empty'.  Finally, a detector can have a local influence on the dynamics of the empty wave, which in turn has an influence on the particle at the second beamsplitter.  In this sense, the empty wave plays a role analogous to the phase of the unoccupied mode in the toy field theory.  
The analogy between unoccupied modes and empty waves breaks down, however, in the quantum eraser experiment.  This is because the latter experiment involves not just the photon, but the auxiliary 2-level system.  The fact that these interact via a CNOT gate implies that the pilot wave in the Bohmian account is the wavefunction of this pair of systems, which unlike the wavefunction for the photon {\em cannot} be understood as a wave over 3-dimensional physical space.  Via the dynamics of this wavefunction, the choice of what measurement to make on the auxiliary system comes to determine the dynamics of the Bohmian particle associated to the photon in a nonlocal fashion.  

The model of Blasiak \cite{Blasiak2015} is analogous to Bohmian mechanics insofar as it avoids nonlocality in the case of the basic Mach-Zehnder interferometer, but cannot provide a local account of the quantum eraser experiment.

There is another nuance worth noting concerning the possibility of securing local causal explanations of interference phenomena in Bohmian mechanics.  Whether a given interference phenomena can be explained locally or not actually depends on whether one is pursuing the Bohmian interpretational methodology   to quantum theory in its first-quantized description (which yields standard Bohmian mechanics) or in its second-quantized description (which yields a Bohmian version of quantum field theory~\cite{Bohm1952suggested,struyve2007minimalist}). In the latter, rather than having a particle that has a spatial location which is guided by a wave defined over physical space, one has modes that have local field amplitudes which are guided by a wave defined over the {\em joint configuration space} of those field amplitudes.  The guiding wave again fails to be a wave in 3-dimensional physical space 
 and one can verify that this sort of Bohmian account requires nonlocality in its account of the phenomenology of even the basic Mach-Zehnder experiment.  
  Unlike Bohmian mechanics, the toy field theory manages to provide a local causal account of the phenomenology of the Mach-Zehnder interferometer  whether one uses the first-quantized or second-quantized formalisms.  The ontological account of the experiment provided in the first-quantized toy theory 
  is merely a {\em coarse-graining} of the description of the experiment provided in the second-quantized toy theory (the toy field theory).
    In the Bohmian approach, on the other hand, the nature of the ontology and the explanation is {\em completely different} depending on whether one develops a pilot-wave interpretation at the first-quantized or second-quantized levels.  
  
Aharonov {\em et al.}~\cite{Aharonov2017} also provide an account of interference that does not endorse wave-particle complementarity but does endorse nonlocality.  
In their account, a particle has a definite location (therefore lies on one arm of the interferometer) but it also has a property of {\em modular momentum} and the dynamics of the latter can depend on the situation on the other arm of the interferometer, and is therefore nonlocal.

It is also useful to consider the extent to which the Everett interpretation of quantum theory \cite{Everett1957}(also known as the `many-worlds interpretation') satisfies or violates the different interpretational claims. 
Whether wave-particle complementarity holds or not seems to us to depend on which version of the Everett interpretation one endorses. 

The Everett interpretation assumes that the wavefunction is an exhaustive description of the ontology.  In this sense, it is like Bohmian mechanics, but without the Bohmian particle positions. 
The wavefunction is thought to define branches, like in Bohmian mechanics, but one does not distinguish some branches as actual and others as `empty' based on the positions of the Bohmian particles.  Rather, one considers all branches to be simultaneously actual, but to be describing parallel worlds.  
The way to define the branches (which in the jargon of Everettians corresponds to solving the ``preferred basis problem'') is to select a preferred basis as one relative to which useful patterns arise \cite{Wallace}, and this is typically picked out by decoherence. 
  Relative to this view, one could argue that  in the Mach-Zehnder interferometer with the detector in place, the preferred basis is such that the which-way property is well-defined, whereas in the Mach-Zehnder interferometer without the detector, the preferred basis is such that the relative phase property is well-defined.   In other words, if the properties obtaining in a given branch are just those observables for which the wavefunction of that branch is an eigenstate, then this version of the Everettian interpretation would seem to endorse wave-particle complementarity.

However, the description of the Everett interpretation provided by Deutsch in Ref.~\cite{deutsch1998fabric} suggests an ontology of particles alone, but wherein particles can interact with their counterparts in parallel worlds. (In this sense, it seems analogous to the ``many interacting worlds'' view of Ref.~\cite{WisemanInteractingWorlds}.)  This version would seem to {\em not} endorse wave-particle complementarity.

 
 In either case, it is generally argued that the Everett interpretation restores the possibility of local explanations.  Indeed, it achieves locality in a similar way to how, in Penrose's account of the Elitzur-Vaidman bomb tester, one can imagine restoring locality by allowing the counterfactual world to influence the actual world.  The difference is that in the Everettian approach, all the worlds are actual.
 
 The achievement of the toy field theory that is certainly {\em not} reproduced by any version of the Everett interpretation  is to account for the TRAP phenomenology of interference {\em while maintaining the classical worldview}.  In other words, what we have shown here is that even if one grants that the Everett interpretation salvages locality for the TRAP phenomenology, the price it pays
  to do so---a radical deviation from a classical worldview---  is not necessitated by the phenomena. 

Finally, we note that among all of the realist accounts of the TRAP phenomenology of interference,
{\em only} the one provided by the toy field theory manages to preserve the classical paradigm of kinematics and dynamics (sets for kinematics and functions for dynamics) while also preserving {\em the principle of Leibnizianity} (described in Sec.~\ref{beyondTRAP}). This is because the alternative realist accounts are all $\psi$-ontic, and it is known that every $\psi$-ontic model violates the  principle of generalized noncontextuality and thus also the principle of Leibnizianity~\cite{Spekkens2005}.  In Sec.~\ref{beyondTRAP}, we noted that our preferred notion of classicality includes assuming {\em both} the classical notions of kinematics and dynamics as well as the principle of Leibnizianity, and so only the account provided by the toy field theory manages to salvage our preferred notion of classicality.  

\section{Further elaborations}

\subsection{Probabilities as credences}
\label{EnsembleEpistemicTalk}

%
%
%
%
%
%
%
%
%

In the epistemically restricted classical statistical theories of Refs.~\cite{Spekkens2007,Bartlett2012,Spekkens2016} and in the one presented in this article (i.e., the toy field theory), probabilities are conceptualized as quantifying the degrees of belief of an agent, that is, they are conceptualized as {\em credences}.  It is sometimes suggested that this means that these theories are not really tested by the observation of relative frequencies in the data.  We here respond to this claim. 


First, we note that it is unavoidable that there is some degree of uncertainty about the parameters in a physical model.  Therefore, the view that a theory cannot be confirmed if it includes descriptions of an agent's degrees of beliefs implies a rejection of almost all scientific hypothesis-testing.
In other words, such a view about theory confirmation is clearly too extreme.

We therefore consider
 a less extreme view, namely,
 that assigning equal credence to one measurement outcome or another
{\em  does not imply} that one should believe that in an arbitrarily large number of repetitions of the experiment, the relative frequency is likely to be one half.  

The critic might argue as follows.  To view probability assignments in these theories as credences implies that they are viewed as degrees of belief about {\em a single system in a sequence of repetitions of an experiment}, and such a belief does 
  not imply anything about the relative frequencies one expects to see in the full sequence of repetitions. 
The following is a more formal version of the claim.
Let  ${\bf p}$  be the probability distribution assigned in a single run, say the $k$th run. For concreteness, imagine a sequence of coin flips, so that ${\bf p} = (p_0,p_1)$ where $p_0$ is the probability of heads and $p_1$ is the probability of tails.  The critic's view 
can be formalized as the suggestion that the distribution ${\bf p}$ merely describes the marginal on the $k$th system in the sequence (i.e., it arises from the full distribution over all elements of the sequence by marginalization of all but the $k$th system).  Since the marginal on the $k$th system in a sequence does not constrain the marginals for any other system, it is indeed the case that, under such an interpretation of ${\bf p}$, nothing is implied about which relative frequencies in the sequence are likely to be observed.

Our response is as follows. 
The view just articulated is {\em not} the interpretation of the distribution ${\bf p}$ that is intended in our work. Rather, we stipulate that the systems in the sequence of experiments are identical and independently distributed (i.i.d.), so that to stipulate a distribution ${\bf p}$ is to stipulate that the i.i.d. distribution is of the form ${\bf p}\otimes {\bf p} \cdots \otimes {\bf p}$.  In this case, the form of ${\bf p}$ {\em does} determine which relative frequencies in a sequence are most likely.  The law of large numbers implies that in the limit of an arbitrarily large ensemble, the most likely relative frequencies are given by ${\bf p}$.
For instance, if an agent assigns equal credence to heads and tails coming up for some coin and believes that a string of coin flips are i.i.d., then a run of all heads {\em is} considered much less likely than one where the relative frequency of heads is 1/2. 


The observed relative frequencies can make an agent reject their hypothesized probability distribution in this circumstance.  The question one must ask is simply whether the probability distribution hypothesized in the i.i.d. model makes the observed relative frequencies likely or not. 
 One can assess this, for instance, by whether the hypothesis has a good $p$-value for the observed finite-run relative frequencies.

  
  Another claim one sometimes hears is that 
  one must interpret probabilities as objective chances rather than credences if one wishes to test these against the observed data.
For instance, a critic might say that 100 tosses of a coin {\em do} serve to test the hypothesis of a certain {\em weighting} of the coin, where the weighting is considered as a property of the coin that determines the objective chance for each outcome, but they do not serve to test a hypothesis about an agent's credences regarding the coin.

Our response to this claim is the standard one that is made in favour of interpretations of probability as credences, which appeals to the de Finetti representation theorem~\cite{deFinetti1990theory,caves2002unknown}.  
de Finetti showed that if an agent takes an arbitrarily large sequence of coin flips to be {\em exchangeable}---i.e., the agent's credences are invariant under any permutation of elements of the sequence---then the agent's credences about the sequence must have the form $\int_{{\bf p} \in \Delta}  \; {\bf p}\otimes {\bf p} \cdots \otimes {\bf p}\;  {\rm d}\mu({\bf p})$ where $\Delta$ is the simplex of possibilities for the distribution ${\bf p}$ and $\mu$ is a probability measure over $\Delta$. 

The key thing to note about the form of the probability distribution on the full sequence implied by the assumption of exchangeability is that it is exactly the same as what one would write down if one assumed that ${\bf p}$ described {\em objective chances}, characterizing properties of the coin, and that $\mu$ described the agent's ignorance of these properties.  The fact that the {\em same} probability vector ${\bf p}$ is assigned to each coin flip in a sequence does not, therefore, imply that there is some property of the coin (possibly unknown) that ${\bf p}$ describes.

Note that in the i.i.d. scenario, the agent's beliefs about the {\em full sequence of runs}  is ${\bf p}\otimes {\bf p} \cdots \otimes {\bf p}$, which is simply a distribution of de Finetti form where the measure $\mu$ is a Dirac-delta measure centred on ${\bf p}$.   It corresponds to an exchangeable sequence wherein there are no correlations among the different runs.  Again, it is of precisely the same form as the distribution one would assign if ${\bf p}$ described some {\em property of the coin} for which there was no uncertainty.

\begin{multicols}{2}
\begin{figure*}[htb!]
\centering
{\includegraphics[width=0.85\textwidth]{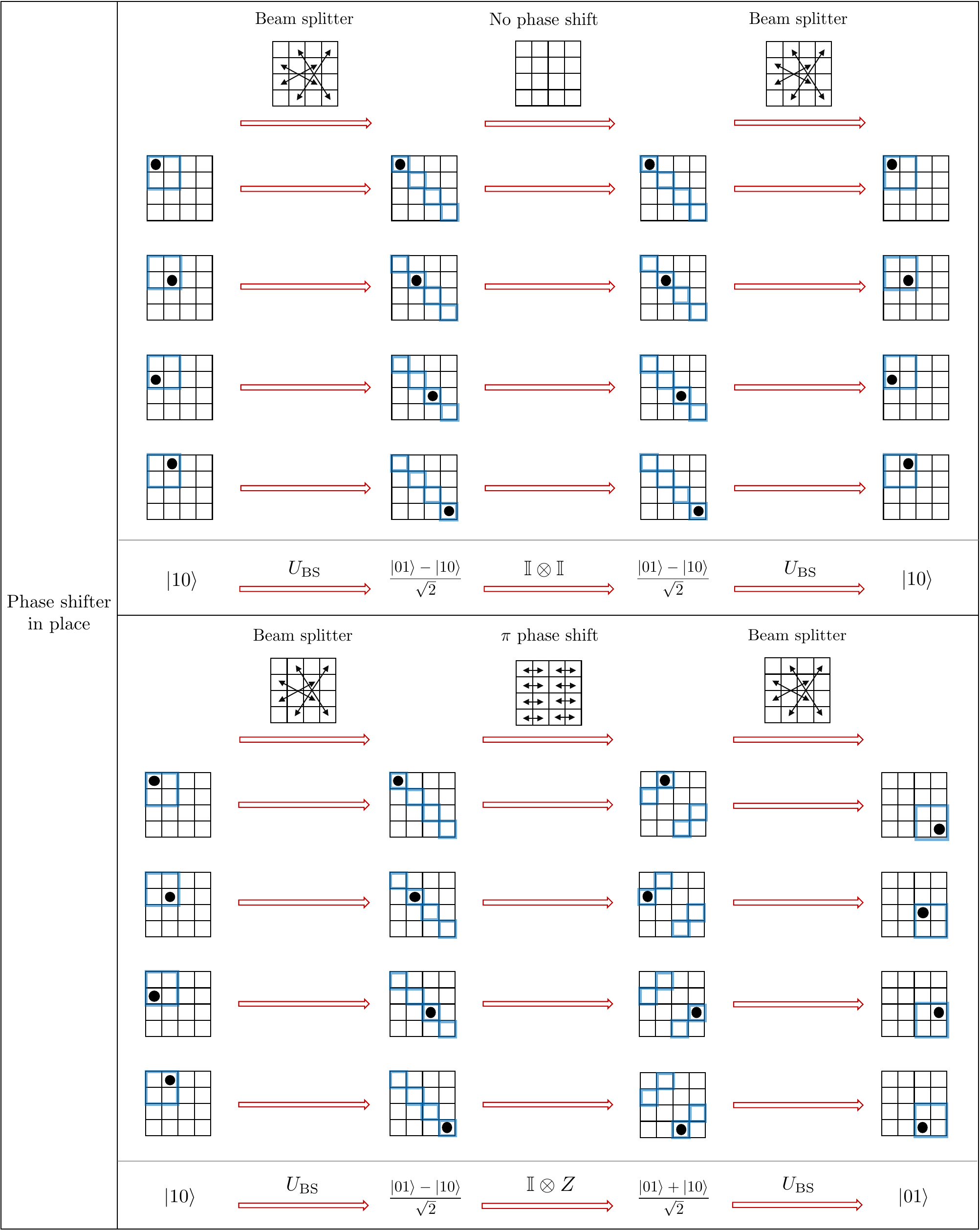}}
\caption{
A depiction of how the physical states describing the different subensembles of the full 
 ensemble of experimental runs  transform in the Mach-Zehnder interferometer with a phase-shifter in place.  The first row of this figure is the counterpart of the top row of the ``phase shifter in place'' row of Fig.~\ref{BigFigure}, while the second row is the counterpart of the bottom row of the latter. 
}
\label{BigFigure2}
\end{figure*}
\end{multicols}

For an i.i.d. distrubtion  ${\bf p}\otimes {\bf p} \cdots \otimes {\bf p}$, the law of large numbers implies that in the limit of an infinitely large ensemble, the relative frequencies converge to ${\bf p}$ with high likelihood. Consequently, one can associate the distribution ${\bf p}$ to the  likely relative frequencies in this ensemble rather than to credences for single systems in an i.i.d. sequence.

Although the concrete example we have used above is a sequence of coin flips, the same can be said of probability distributions over the physical states of a system in a sequence of repetitions of the same experiment.   In the next section, we show how to provide a description of 
 the toy field theory account of the TRAP phenomenology of the Mach-Zehnder interferometer in terms of an ensemble of repetitions of the experiment.  

\subsection{Reframing our account in terms of an ensemble of repetitions of the experiment}\label{ensemblereframing}

The discussion in the previous section
 implies that although we have up until now conceptualized 
 the probability distributions in the toy field theory as states of incomplete knowledge about the physical state of a system in a single run of an i.i.d. experiment, one  can {\em also} conceptualize these distributions as describing the likely relative frequencies of different physical states {\em in an infinite ensemble of runs of the experiment}.   Reframing the predictions of the toy field theory in terms of relative frequencies can be useful insofar as it provides a slightly different perspective on the features of the theory.  We will focus, in particular, on how our analysis of locality appears in this reframing.

 \blk


To begin with, consider the ensemble corresponding to the probability distribution describing the pair of modes prior to the first beamsplitter, depicted in the first column of Fig.~\ref{BigFigure} (this is the analogue of the initial quantum state $\ket{10}$).
This ensemble includes  four subensembles of equal size wherein $N_L=1$ and $N_R=0$ in each of these while the local phases take every possible pair of values: $(\Phi_L,\Phi_R)=(0,0)$ in one subensemble, $(\Phi_L,\Phi_R)=(1,0)$ in another, $(\Phi_L,\Phi_R)=(0,1)$ in another, and $(\Phi_L,\Phi_R)=(1,1)$ in the fourth.  In other words, the four subsensembles are associated to the four physical states
$(N_L,\Phi_L,N_R,\Phi_R)=(1,0,0,0), (1,1,0,0), (1,0,0,1)$, and $(1,1,0,1)$, which are depicted diagramatically in the first column of the first four rows in the top half of Fig.~\ref{BigFigure2}.  We have included a light blue outline around the set of four physical states in the depictions of each physical state
 to remind the reader of
  the full ensemble of which each subensemble is a part. 

\begin{figure*}[htb!]
\centering
{\includegraphics[width=0.85\textwidth]{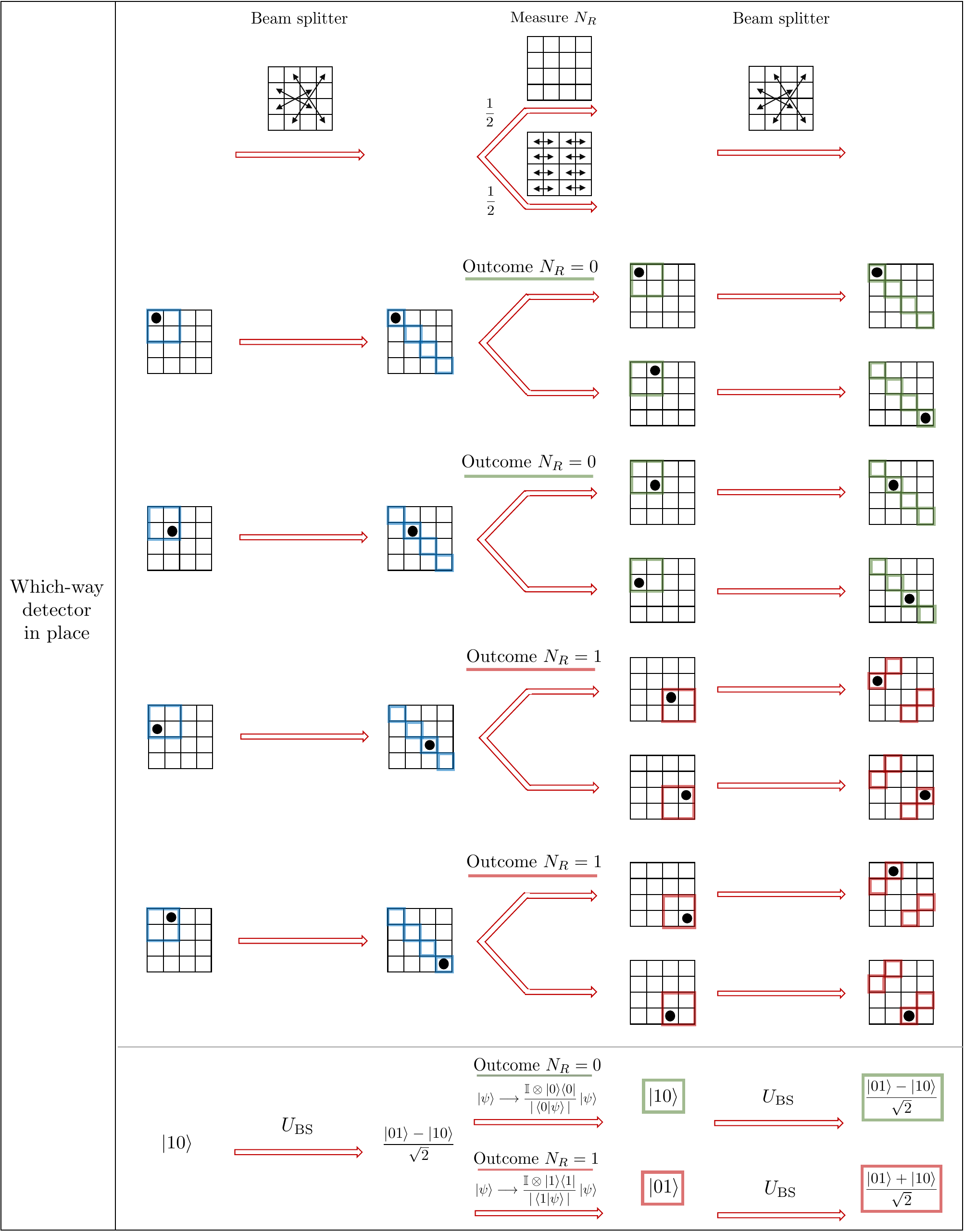}}
\caption{
A depiction of how the physical states describing the different subensembles of the full  ensemble of experimental runs  transform in the Mach-Zehnder interferometer with a measurement of occupation number occuring in the $R$ arm.  The parts of this figure describing the two possible outcomes, $N_R=0$ and $N_R=1$, are the counterparts of the two rows in the ``Which-way detector in place'' part of Fig.~\ref{BigFigure}.  
}
\label{BigFigure3}
\end{figure*}

Note that this decomposition of the ensemble into four subensembles occurs also in the bottom half of Fig.~\ref{BigFigure2} and in Fig.~\ref{BigFigure3}, since all three versions of the Mach-Zehnder experiment start with the same ensemble. 
 
For each of the four subsensembles, the first beamsplitter transforms the associated physical state of the pair of modes according to the beamsplitter transformation (which is depicted in \eqref{BSdiagram} and repeated in Fig.~\ref{BigFigure2}). 
 For each physical state of the pair of modes at the input to the first beamsplitter (depicted in the four rows of Fig.~\ref{BigFigure2}), its image under the beamsplitter transformation is depicted 
in the second column of the same row. 
It follows that the ensemble associated to the pair of modes at the {\em output} of the first beamsplitter comprises four subensembles associated to the four physical states depicted in that column.
We have again indicated with a blue outline this set of four physical states, thereby making evident the association between this ensemble and the probability distribution depicted in the second column of Fig.~\ref{BigFigure}. 

How the physical states associated to each of the four subensembles within the full ensemble are transformed under the two possible phase shifts---no phase shift or a $\pi$ phase shift---is depicted in the top and bottom halves of Fig.~\ref{BigFigure2} respectively.  
In both cases, the set of four physical states associated to the ensemble after the phase-shifter is depicted with a blue outline.
In the case of no phase  shift, it corresponds to the probability distribution depicted in the third column of the first row of Fig.~\ref{BigFigure}, while in the case of a $\pi$ phase shift, it corresponds to the one depicted in the third column of the {\em second} row of Fig.~\ref{BigFigure}.


Finally, within each of these two ensembles, how the physical states associated to each of the four subensembles are transformed under the second beamsplitter is also depicted in Fig.~\ref{BigFigure2}.  Again, in both cases, the set of four physical states associated to the final ensemble is depicted with a blue outline and corresponds to the probability distributions in the final column of the first two rows of Fig.~\ref{BigFigure}.



Now we turn to the case where a measurement of occupation number occurs in the $R$ arm of the Mach-Zehnder interferometer. 

\begin{multicols}{2}
\end{multicols}

Suppose one seeks to make predictions about what will occur at later times in the experiment given that one has observed a particular outcome.  Conditioning on an outcome corresponds to {\em updating} the ensemble that is relevant for making predictions.  This updating can be understood as consisting of two steps.  First, one {\em selects} from the pre-existing ensemble the subensemble that is consistent with the outcome that was observed.    Second, the fact that a measurement leads to a random disturbance---that is, one of several different transformations to the physical state---implies that elements of the ensemble selected in the first step get split into distinct elements (bifurcated in the case of interest here), thereby leading to an increase in the number of distinct subensembles. 

Consider, in particular, the case wherein one conditions on obtaining the $N_R=0$ outcome in the measurement of occupation number of the $R$ arm. 

Recall that, after the first beam splitter, the ensemble of runs consists of four subensembles, associated to physical states $(N_L,\Phi_L,N_R,\Phi_R)=(1,1,0,0), (1,0,0,1),(0,1,1,0),(0,0,1,1)$ (depicted in the second column of Fig.~\ref{BigFigure3}).  The first step of the ensemble update associated to conditioning on $N_R=0$ is the {\em selection} step, consisting of redefining the ensemble to include only the two subensembles that are consistent with $N_R=0$, namely, the ones associated to the physical states $(N_L,\Phi_L,N_R,\Phi_R)= (1,1,0,0), (1,0,0,1)$, which are the top two cases in the second column of Fig.~\ref{BigFigure3} (i.e., the two rows that are labelled ``outcome $N_R=0$'').
  This is the ensemble counterpart of the knowledge-updating step associated to resolving uncertainty, depicted in Eq.~\ref{mmtupdaterulediagram1}, such that, prior to taking into account the disturbance to the physical state, the ensemble is updated to the one associated to the probability distribution depicted in \eqref{epistemicstateBayesianupdatediagram}.

With this description of the selection step of the ensemble update procedure in hand, we can now explain why it 
does not involve any nonlocal influences. 

When, based on the outcome of the measurement on the $R$ mode, one selects the subensemble of physical states of the pair of modes that is consistent with this outcome, it can happen that not only does the selection procedure exclude certain physical states {\em of the $R$ mode} from the subensemble, it excludes certain physical states {\em of the $L$ mode} as well.  This is because the original ensemble prior to the selection (the one describing the pair of modes at the outputs of the first beamsplitter) embodies correlations between the properties of the $L$ mode and those of the $R$ mode. Indeed, in the example we are considering, where one selects the subensemble of physical states of the pair of modes consistent with $N_R=0$, the fact that the original ensemble has $N_L$ and $N_R$ taking opposite values ($N_L \oplus N_R =1$) implies that by excluding from the subensemble  all physical states of the pair of modes having $N_R=1$, one thereby also excludes from the subensemble all physical states of the pair of modes having $N_L=0$.  The key point to note is that this
reduction in the set of properties of the $L$ mode that are present in the selected subensemble (i.e., the elimination of any physical states with $N_L=0$) is {\em induced by the selection on the properties of the $R$ mode} and consequently 
{\em does not imply a nonlocal influence} from the $R$ arm to the $L$ arm.  

\blk

We now turn to the second step of the ensemble update procedure, the bifurcation step, and explain why it also does not involve any nonlocal influences. 



As noted above, the subensemble after the selection process 
consists of two physical states for the pair of modes.   While both of these physical states have the same occupation numbers for the modes, namely, $N_L=1$ and $N_R= 0$, they have different phases for the modes.  Specifically, one has $\Phi_L=0$ and $\Phi_R=1$ and the other has $\Phi_L=1$ and $\Phi_R=0$.  These two physical states are depicted in the second column of Fig.~\ref{BigFigure3} in the first and second rows respectively 
 (i.e., the two rows that are labelled ``outcome $N_R=0$'').  The fact that the phase randomization associated to the measurement of occupation number is presumed to be local to the $R$ mode implies that $\Phi_R$ is randomized while $\Phi_L$ remains undisturbed.   It follows that the element of the pre-randomization ensemble where the physical state has $\Phi_L=1$ and $\Phi_R=0$ (the first row in Fig.~\ref{BigFigure3}) is mapped to a {\em pair} of elements in the post-randomization ensemble, where both still have $\Phi_L=1$, but where now one has $\Phi_R=1$ and the other has $\Phi_R=0$. This bifurcation is depicted in the third column of the first row of Fig.~\ref{BigFigure3}.  Similarly, the other element of the pre-randomization ensemble, where the physical state has $\Phi_L=0$ and $\Phi_R=1$ (the second row in Fig.~\ref{BigFigure3}), is also mapped to a pair of elements in the post-randomization ensemble.  In this case, both still have $\Phi_L=0$, while one has $\Phi_R=1$ and the other has $\Phi_R=0$.   This bifurcation is depicted in the third column of the second row of Fig.~\ref{BigFigure3}.  The key point to note here is that in both cases, the measurement on the $R$ mode has no effect on the physical properties of the $L$ mode.  Graphically, this is seen by the fact that {\em along the vertical axis} of the discrete physical state space, which is the axis that describes properties of the $L$ mode, the coordinate of the physical state is unchanged. It is only along the horizontal axis of the discrete physical state space, which is the axis that describes properties of the $R$ mode, that the coordinate of the physical state changes.  

The rest of the story regarding the evolution through the Mach-Zehnder interferometer is relatively straightforward.  We provide the details for completeness. \blk

The ensemble we are left with after these two steps of the updating procedure is back to being comprised of four subensembles.  These are associated to the four physical states depicted in the third column of Fig.~\ref{BigFigure3}, which we have indicated with a green outline.  Clearly, this ensemble corresponds to the probability distribution depicted in the third column of the third row of Fig.~\ref{BigFigure}. 

Fig.~\ref{BigFigure3} also depicts how the physical states associated to each of the four subensembles are transformed under the second beamsplitter.  
The ensemble describing the output of the second beamsplitter is depicted in the final column of Fig.~\ref{BigFigure3}, with the set of four physical states outlined in green.  It corresponds to the probability distribution 
depicted in the final column of the third row of Fig.~\ref{BigFigure}.

In the case wherein  one obtains the outcome $N_R=1$ rather than $N_R=0$, a similar analysis is possible: the ensembles describing the pair of modes before and after the second beamsplitter are depicted with a red outline in Fig.~\ref{BigFigure3}, and correspond to the probability distributions described in the third and fourth columns of the fourth row of Fig.~\ref{BigFigure}.

So we see that for both possible outcomes, we recover the TRAP phenomenology of interference in the Mach-Zehnder interferometer without recourse to nonlocal influences. \blk

\subsection{The case where the measurement is destructive}
\label{DestructiveMeasurements}

In the article thus far, we have considered only one kind of detector in quantum theory, the type
 that implements a \textit{nondestructive} measurement of the occupation number of a mode.  For these, the occupation number at its output mode is presumed to be whatever the occupation number at the input mode was found to be.  But according to quantum theory, the TRAP phenomenology of interference can be reproduced even if the detector implements a \textit{destructive} measurement of the occupation number, that is, one that absorbs the excitation if it happens to be present, such that 
the occupation number of the input mode may be found to be $1$, but the output mode is nonetheless left in a quantum state with occupation number $0$.  

Explicit in Feynman's account (and to an even greater degree in Elitzur and Vaidman's account) is that the destruction of interference occurs even in the case where the measurement {\em does not detect the excitation on its arm}.  But in this case, {\em there is no difference} in the quantum state update rule between destructive and nondestructive measurements, as the output of the measurement device is left in the quantum vacuum state in both cases.  For this reason, the distinction between destructive and nondestructive measurements is not significant for discussions of the TRAP phenomenology.
 It follows that if one can reproduce this phenomenology for {\em either} type of measurement in a classical local model, one has undermined the claim that the phenomenology necessitates a departure from the classical worldview. Hence, an explicit consideration of destructive measurements is not required to establish our thesis.



Nonetheless, the question arises of whether the toy field theory can provide an explicit local classical model for the case of a destructive measurement as well.  We here show that it can.


One way to implement a destructive measurement of occupation number on a given arm of the interferometer is to place a photodetector there.
(For a perfect measurement of occupation number, one requires an ideal photodetector; it is the latter that we have in mind for the purposes of this discussion.)

Another way to implement such a measurement on a given arm is to put a brick or other absorbing medium there.  Unlike a photodetector, the macroscopic properties of a brick do not change in a noticeable way if it absorbs a single excitation of the field mode.  Nonetheless, because the microscopic properties {\em do} register the absorption, the brick 
  still implements a destructive measurement of occupation number; it is just one for which the experimenter does not learn the outcome.  In this sense, having an arm blocked by a brick is equivalent to having a photodetector in place and ignoring its outcome.  
Moreover, the TRAP phenomenology concerns those cases where a detection is registered at one of the output ports of the second beamsplitter, so that the excitation is known {\em not} to be present in the $R$ arm, and for these cases it is irrelevant whether it is a photodetector or a brick that implements the destructive measurement.

We turn now to demonstrating how the toy field theory reproduces the TRAP phenomenology of interference when the measurement of occupation number on the $R$ arm is destructive.
 
 
First, we note explicitly why it is intuitively clear from the outset that the toy field-theoretic model of a destructive measurement ought to be different from its model of a nondestructive measurement in the second step of the updating map, the step described as {\em disturbance} in the main text (see Eqs.~\eqref{MmtUpdateRuleOnticAlgebraic} and \eqref{MmtUpdateRuleOntic} and the surrounding discussion) and described as {\em bifurcation} in the ensemble-reframing of  Sec.~\ref{ensemblereframing}.


If $f$ denotes the function relating the phase of the output mode of a measurement to the phase of its input mode, then one can describe the disturbance/bifurcation step of the update map for a nondestructive measurement, as follows: $f$ is sampled uniformly at random from the set consisting of the identity function and the flip function and is applied to the binary phase variable. 


What would be the problem if a {\em destructive} measurement of occupation number were also modelled in this fashion?  The answer is that it is implausible that such measurements implement the identity or flip functions on the phase.
For instance, if the measurement is implemented by placing a brick in an arm of the interferometer, then it is implausible that the phase at the far side of the brick (its ``output port'') should be determined by the phase at the near side of the brick (its ``input port''). 

It follows  that the toy-field-theoretic model of a destructive measurement should differ  from that of a nondestructive measurement in such a way that an implausible dependence of this sort is not required.  We now show that this can indeed be achieved, while still reproducing the TRAP phenomenology of interference.
 
It is sufficient to take the disturbance/bifurcation step of the update rule for a destructive measurement in the toy field theory to be defined as follows.  There are again two possibilities for the function  $f$ that  relates the phase of the output mode of a measurement to the phase of its input mode.  But they are not the identity and flip functions.  Rather, they correspond to the other pair of functions from a binary variable to a binary variable (there are only four such functions), namely, the function that is constant with value 0 (the ``reset-to-0'' function) and the function that is constant with value 1 (the ``reset-to-1'' function).  As before, which of these functions is applied is sampled uniformly at random. 

Consider now what this implies for $\Phi_R$, the phase variable on the $R$ arm, at the output of the destructive measurement (for instance, the far side of the brick).
Because it is equally likely to be the image of the reset-to-0 map or the reset-to-1 map, it is simply sampled uniformly at random from the values 0 and 1.  What is key here is that  its value is sampled {\em independently} of its value at the measurement's input (for instance, the near side of the brick).  There is consequently no dependence of the output phase on the input phase.  This harmonizes with the intuition that such a dependence would be implausible for certain implementations of the destructive measurement, such as placing a brick in the arm.\footnote{It is useful to also consider the occupation number on the $R$ arm, $N_R$, and the update rule for it.  Recall that in the case of a  measurement that is {\em nondestructive}, $N_R$ at the output is equal to $N_R$ at the input.  By contrast, in the case of a measurement that is {\em destructive}, $N_R$ at the output is always zero, regardless of the value of $N_R$ at the input.  So there is also no dependence of the output occupation number on the input occupation number.}



In summary, then, the disturbance/bifurcation step of the update map for a destructive measurement is as follows: the physical state of the output mode of the measurement is sampled uniformly at random from the pair of physical states $(N,\Phi)=(0,0)$ and $(N,\Phi)=(0,1)$  {\em completely independently} of the physical state of the input mode.

Finally, we can explain why, in spite of this difference in the disturbance/bifurcation step of the update rule for the destructive case, one can still reproduce the TRAP phenomenology.

The key is that the two different bifurcations result in the {\em same} rule for updating the probability distribution over physical states, which is the only thing that is relevant for the operational phenomenology.

To see this, consider the stochastic map on $\Phi_R$ associated with (i) an equal probabilistic mixture of the identity function and the phase-flip function, and (ii) an equal probabilistic mixture of the reset-to-0 function and the reset-to-1 function.  In both cases, it is the stochastic map describing complete randomization of $\Phi_R$, that is, the map that prepares the uniform probability distribution over the two values of $\Phi_R$ regardless of the probability distribution fed into the stochastic map. (For more details about this type of equivalence, see Sec. V.C of Ref.~\cite{Schmid2021unscrambling}.)


What does this imply? In the case where the occupation number on the $R$ arm is found to be $N_R=0$, it implies that the probability distribution over physical states at the measurement's output is an equal mixture of the cases where $\Phi_R=0$ and $\Phi_R=1$ 
 independently of the value of $\Phi_R$ at the measurement's input.  And this is the case {\em regardless} of whether the measurement is of the destructive or nondestructive variety.  In short, despite the differences in the bifurcation step of the update maps for destructive and nondestructive measurements, both imply phase randomization. 

As we explained in Sec.~\ref{MZIwWhichWay}, this phase randomization is what guarantees that there is a randomization of where the detection is registered at the output ports of the second beamsplitter and hence it is what guarantees that the toy field theory reproduces the relevant phenomenology---that a measurement of occupation number on an arm of the interferometer causes a loss of interference.  So, despite the differences in the disturbance/bifurcation step of the update maps for destructive and nondestructive measurements, they both reproduce the relevant TRAP phenomenology of interference.  

Finally, the difference in the disturbance/bifurcation step does not change any of the interpretational commitments of the toy field theory. In particular, no nonlocal influences are required to reproduce the phenomenology.



We end with a few more comments intended to head off possible confusions.

In the toy field theory, if a given field mode is known to be unoccupied, then by the epistemic restriction, it is necessarily known to be in a uniform mixture of the two possible phases, 
and it is also known that its physical state is uncorrelated with the physical state of any other system.  This is true regardless of the reason for the mode being unoccupied, and it follows that predictions that one makes about the outcomes of subsequent measurements involving that mode are insensitive to these reasons.  In particular, these predictions are insensitive to whether a mode is known to be unoccupied by virtue of the source in the experiment being turned off or by virtue of the mode in question being at the output of a device that has absorbed the field excitation that was produced at the source.

It follows in particular, that the output mode of a destructive measurement device is described by such a probability distribution {\em at all times}.  In other words, it is the {\em default situation} for any mode that is unoccupied to have an unknown phase that is not correlated with other systems.  As such, this situation does not need any special explanation within the toy field theory.   The far side of a brick, for instance, is always described by such a probability distribution, regardless of whether it has just absorbed an excitation on its near side or not. 




The probability distribution just described---the uniform mixture of $(N,\Phi)=(0,0)$ and $(N,\Phi)=(0,1)$ with no correlations with any other system---is the counterpart within the toy field theory of the {\em vacuum state} in quantum theory.
Furthermore, the features of the toy field theory we have just outlined are in fact {\em required} if this probability distribution is to emulate the analogous features of the quantum vacuum state.   Recall that in quantum theory, the reason that a given mode is in the quantum vacuum state is irrelevant to the predictions about the outcomes of subsequent measurements involving that mode, and the vacuum state is the default state of any mode, appropriate for describing the far side of a brick regardless of what is happening on its near side.

\subsection{The case where a mirror is removed}
\label{MirrorRemoved}

\begin{figure}[htbp] 
   \centering
   \includegraphics[width=0.98\columnwidth]{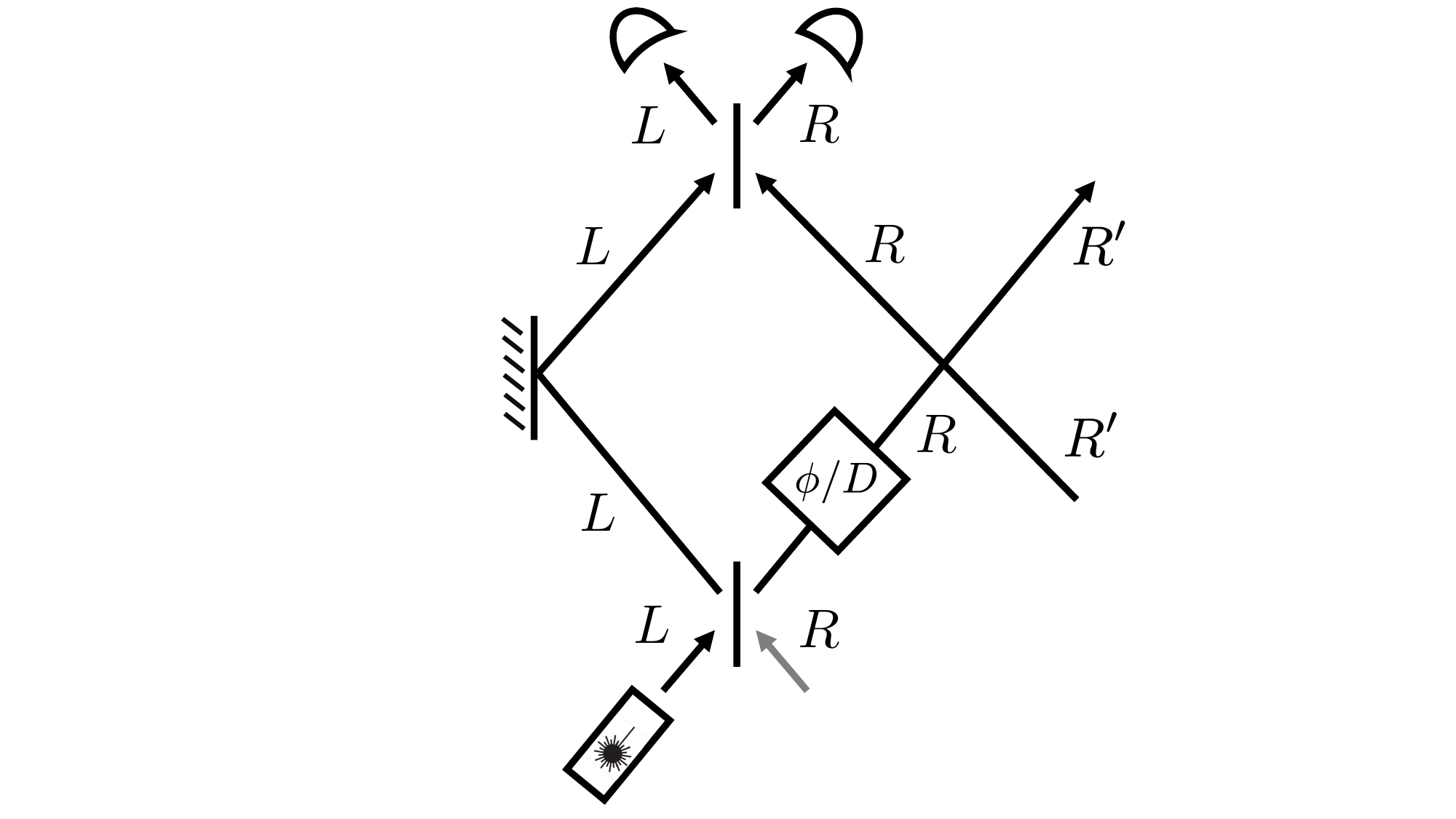} 
   \caption{The Mach-Zehnder interferometers of Fig.~\ref{MachZehnder}, but with the mirror on the $R$ mode removed.}
   \label{fig:nomirror}
\end{figure}

Consider the case of the Mach Zehnder interferometer in the set-up of Fig.~\ref{MachZehnder}(a) or Fig.~\ref{MachZehnder}(b),
 but where the mirror on the right path has been removed.  This is depicted in Fig.~\ref{fig:nomirror}.
 In this case, the $R$ mode at the input of the second beamsplitter is no longer determined by the physical state of the $R$ mode at the output of the phase shifter or which-way detector, but rather it tracks the physical state of another mode, the one aligned with the right input of the second beamsplitter, but on the back side of the mirror, denoted $R'$.
In other words, the transformation implemented in the case of no mirror on the physical states of the $R$ mode  and the $R'$ mode is simply a swap operation, i.e., the physical state of the $R$ output tracks that of the $R'$ input and the physical state of the $R'$ output tracks that of the $R$ input.  This is clearly analogous to the sort of transformation one would use to describe the absence of the mirror in quantum theory.

Given that no source was shining onto the backside of the mirror in the setup of the interferometer, the $R'$ mode is known to be unoccupied.  Hence, by the epistemic restriction, the probability distribution over the physical states of the $R'$ mode is the one that is point distributed on $N_{R'}=0$ and uniformly distributed over values of the binary phase $\Phi_{R'}$.  (If one traces the line made by the $R'$ mode back to its origin, it will meet some inert object, such as a wall of the laboratory, and hence this case is really equivalent to the one wherein the input to the second beamsplitter tracks the mode on the far side of a brick.)  

It follows that the distribution over physical states of the mode $R$ at the input of the second beamsplitter is the one that is point distributed on $N_{R}=0$ and uniformly distributed over values of the binary phase $\Phi_{R}$.  
 
It is straightforward, therefore, to determine what our theory predicts in the case where the mirror is removed.  Recall that after the first beamsplitter, the $L$ mode and the $R$ mode are equally likely to be the one that is occupied.
 In those cases where it is the $R$ mode that is occupied, 
 the absence of the mirror on the right path implies that the excitation escapes from the interferometer and thus neither detector at the output ports of the second beamsplitter are found to fire.  In those cases where it is the $L$ mode that is occupied, the excitation remains in the interferometer and consequently one of the detectors at the output ports of the second beamsplitter fires.  Which detector fires is determined (via the beamsplitter update rule) by the relative phase at the input ports.  But given that $\Phi_{R}$ is uniformly distributed at this input, the relative phase $\Phi_{L}\oplus \Phi_{R}$ is also uniformly distributed and consequently the excitation is equally likely to be detected at either of the two output ports.  Thus, in the ensemble of cases where the excitation is detected, there is no interference pattern.
 
Thus, the toy field theory reproduces the predictions of quantum theory for the case of a Mach-Zehnder interferometer with one of the mirrors removed. 

\blk

\subsection{An ontology of field modes rather than an ontology of particles}
\label{ModesNotParticles}

The fundamental systems in the toy field theory are field modes, not particles.  We here elaborate a bit on the differences between modes and particles in order to prevent possible confusions in interpreting the toy field theory. As we already said in the main text, a particle is characterized by having position and momentum as properties.  These are aptly described as {\em motional} degrees of freedom.  A mode, on the other hand, is characterized by having occupation number and phase as properties.  These are best characterized as {\em excitational} degrees of freedom.  Spatially localized modes are \textit{parametrized} by position.  As such, under a change in viewpoint from particles to modes, the notion of position changes  from being a dynamical variable whose value may evolve over time as the particle evolves to being a parameter that indexes which mode one is talking about, and it is the occupation numbers and the phases of modes that are the dynamical variables.  The notions of modes and their excitational degrees of freedom are familiar and uncontroversial in the study of field theories.



\subsection{A description in terms of spatially localized modes}
\label{SpatiallyLocalizedModes}


The fact that the ontology of the toy field theory is an ontology of modes rather than particles is critical to providing a {\em local} causal explanation of the TRAP phenomenology.  
To further elucidate this point, it is useful to consider the analogy between a classical field theory and a  cellular automaton.


 There are no particles in a cellular automaton;  the fundamental systems are the cells and the dynamics consists of changes of the internal states of these cells (which are often taken to be binary but need not be).   
 This is akin to how, in a classical field theory, the systems 
 are spatially localized modes and the dynamics consists of changes of the internal states of these modes (which are typically taken to be scalar or vector fields, but need not be).  


In a cellular automaton, some set of cells may exhibit a distinctive pattern that persists over time and propagates to adjacent sets of cells at every time step.  The motion of a glider in Conway's Game of Life is the paradigmatic example. Although this sort of motion may have the appearance of a particle propagating through a discrete space, patterns of excitations (such as Conway's gliders) are not, strictly speaking, fundamental systems having position as a property.  
Motion in a classical field theory is analogous, with modes playing the role of cells.



A cellular automaton is said to invoke only local causal influences if the update rule for the excitational state of a given cell (at a given time-step) depends only on its state and that of its neighbours at the previous time-step.  The same is true of a classical field theory regarding the rule for updating the physical states of a set of modes---the rule is local if the state of a mode at one time depends only on its state at the earlier time and that of spatially contiguous modes. 

In the main text, we have emphasized that the toy field theory is local in the sense that the $L$ and $R$ modes only exert causal influences on one another when they become spatially contiguous, such as at a beamsplitter.\footnote{Another example is when a mode and another system (such as another mode) undergo the CNOT dynamics, as in Eq.~\eqref{cnot}, which also requires them to be spatially contiguous.}
However, there is one detail about our account of the experiment that might still raise concerns regarding our claim of having a local model.  
Specifically, we have conceptualized ``the $L$ mode after the first beamsplitter'' as a single monolithic thing. We have not distinguished the points on the left path that are at different distances from the first beamsplitter.   To claim that a field theory is truly local, it is reasonable to require that one be able to provide an account in terms of modes that are {\em spatially localized}, in addition to requiring that all interactions among modes are local interactions.  In other words, in addition to {\em dynamical locality}, one must also have {\em kinematical locality}~\cite{spekkens2015paradigm}.
As such, it is useful to demonstrate explicitly that our toy field theory can be expressed in terms of spatially localized modes. We now turn to that demonstration. 
 
 Given that our model involves a discrete space of physical states for each mode, it makes sense to also discretize space and time in our model.  Thus we model the paths through the interferometer as a finite sequence of modes and we model time as a finite sequence of time-steps.

Recall that it is common practice in physics to model a spatially localized device in an experiment on some system by {\em a local modification of the Lagrangian that dictates the system's evolution} rather than trying to model the device as a large number of fundamental systems (its atomic constituents) in some particular configuration which then couple to the system of interest by the universal Lagrangian.
We now note that for models based on cellular automata, there is a natural analogue of this practice, 
namely, to model the device by a local modification of the cellular automaton's update rule. 



For simplicity, we will describe the experimental configuration as a circuit.  The cellular automaton model can then be described as follows: 
each wire in the circuit is modelled as a 1-dimensional array of cells of the cellular automaton, while each gate in the circuit is encoded as a modified update rule for the cells that are the inputs and outputs of that gate.

\begin{figure}[htbp] 
   \centering
   \includegraphics[width=0.98\columnwidth
   ]{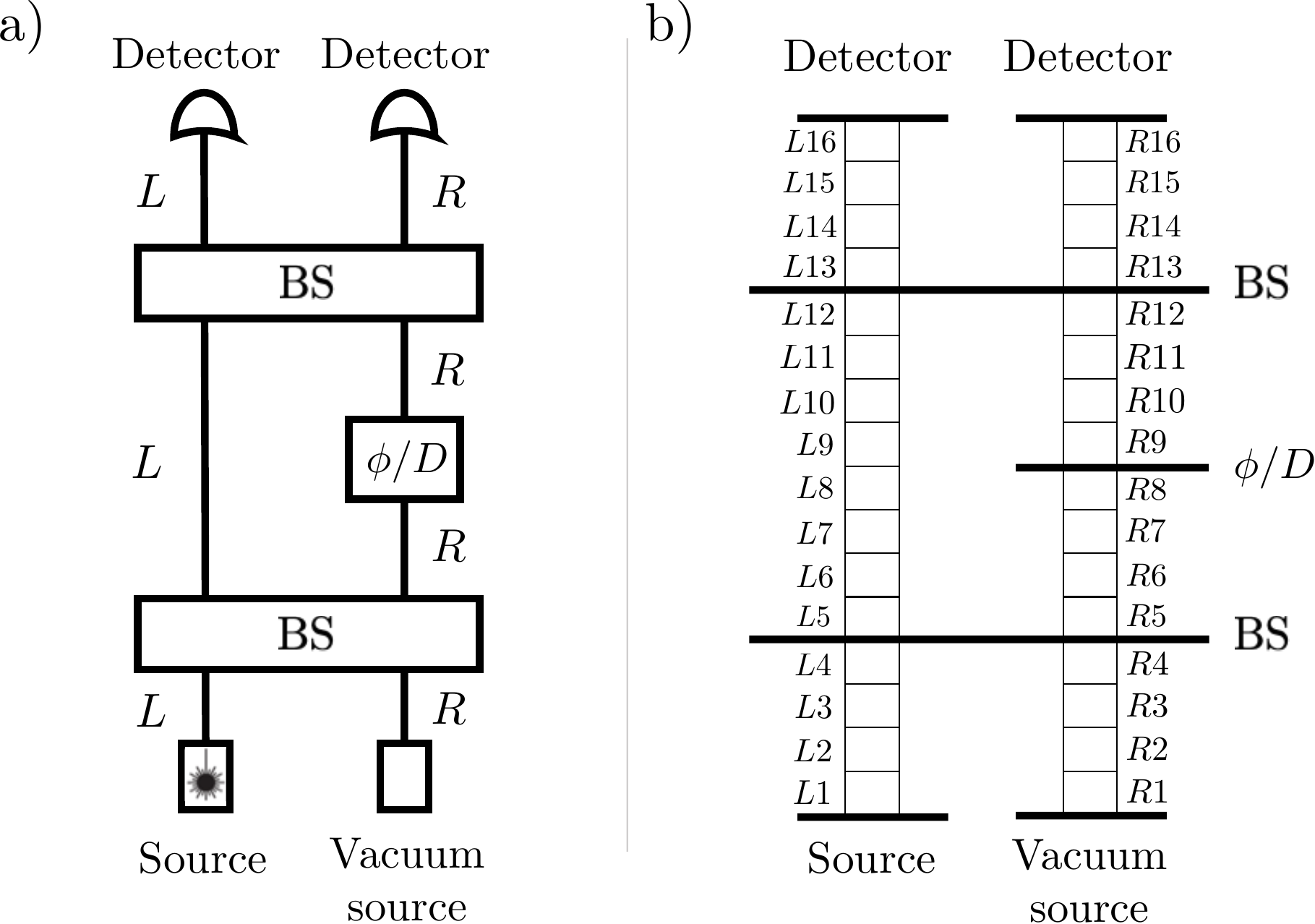} 
   \caption{(a) The circuit representation of the experimental setup of the Mach-Zehnder interferometer of Fig.~\ref{MachZehnder}. BS denotes a beamsplitter. (b) Modelling the circuit as a cellular automaton.}
   \label{MZIAsCircuit}
\end{figure}

We now redescribe the Mach-Zehnder interferometer of Fig.~\ref{MachZehnder} (with either the phase shifter or which-way detector in place) in this manner.  To begin with, we recast the interferometer as a circuit, which we depict in Fig.~\ref{MZIAsCircuit}(a).
 Next, we decompose each wire in this circuit into 
 a 1-dimensional array of cells, as depicted in Fig.~\ref{MZIAsCircuit}(b).  We have numbered the cells on the left wire by $L1, L2, \dots, L16$, and those on the right wire by $R1, R2, \dots, R16$.

 For the case of our toy field theory, a given cell $j$ has a physical state described by a binary occupation number $N_j$ and a binary phase $\Phi_j$.   We denote the physical state of cell $j$ by $\Lambda_j \equiv (N_j,\Phi_j)$. 

 The first update rule that needs to be stipulated for a circuit cellular automaton is the one describing the time evolution of cells that are not the input or output of any gate.  
 We call this the {\em free propagation} update rule.

\begin{figure}[htbp] 
   \centering
   \includegraphics[width=0.7\columnwidth]{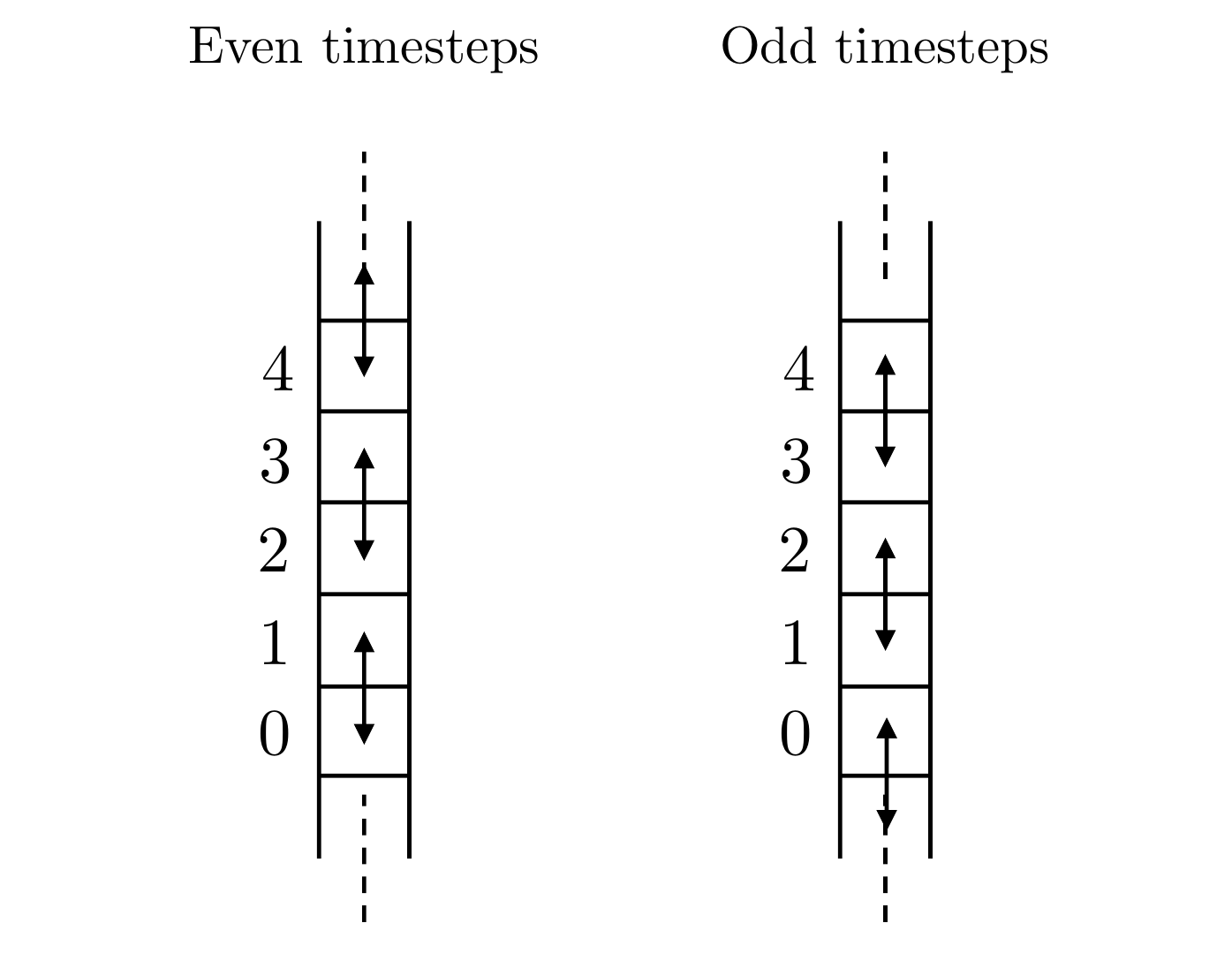} 
   \caption{The update rule for free propagation.}
   \label{FreeProp}
\end{figure}

For the cellular automaton version of our toy field theory, we will use a particular free propagation update rule which is an instance of what is called the Margolus partitioning scheme, specialized to the case of a 
 1-dimensional cellular  automaton~\cite{margolus1984physics}.  It is a rule that varies in its action from one time-step to the next.  Specifically, it is of the following form, which is depicted schematically in Fig.~\ref{FreeProp},
\begin{align}\label{freepropUpdateRule}
&\textrm{for $t$ even}:\nonumber\\
&\;\;\textrm{for $j$ even}: \Lambda_{j}(t+1) = \Lambda_{j+1}(t),\nonumber\\
&\;\;\textrm{for $j$ odd}: \Lambda_{j}(t+1) = \Lambda_{j-1}(t).\nonumber\\
&\textrm{for $t$ odd}:\nonumber\\
&\;\;\textrm{for $j$ even}: \Lambda_{j}(t+1) = \Lambda_{j-1}(t),\nonumber\\
&\;\;\textrm{for $j$ odd}: \Lambda_{j}(t+1) = \Lambda_{j+1}(t).
\end{align}

  The idea is that the internal state of a cell is swapped with one neighbour at one timestep and the other neighbour at the next.  In this way, an excitation can move in either direction along a wire depending on the correlation between the parity of its cell index and the parity of the time index.  Such an update rule for free propagation has the benefit of leading to dynamics that is time-reversal symmetric.

\begin{figure}[htbp] 
   \centering
   \includegraphics[width=\columnwidth]{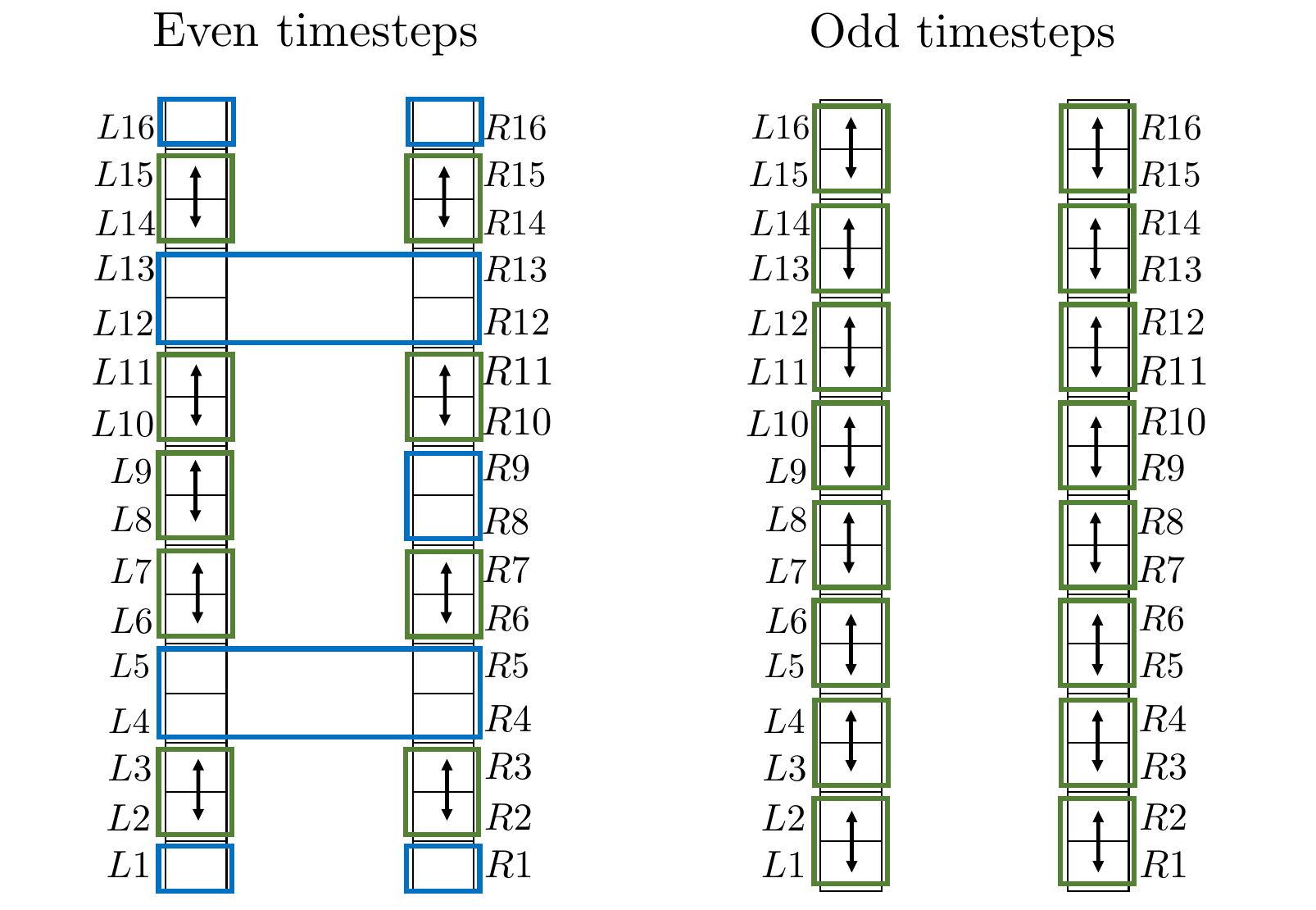} 
   \caption{The overall update rules for all cells.}
   \label{CAUpdateRules}
\end{figure}

The overall update rules for all cells in the circuit model of the Mach-Zehnder interferometer are as depicted in Fig.~\ref{CAUpdateRules}.

Here, cell-groupings that are coloured green indicate places where the update rule coincides with the free propagation update rule described above.  Consider, for instance, the cell $L8$. At even timesteps, it is grouped with the cell $L9$ and the pair are subject to the update rule that simply swaps their physical states:
\begin{align}
&N_{L9}(t+1) = N_{L8}(t),\nonumber\\
& \Phi_{L9}(t+1) = \Phi_{L8}(t), \nonumber\\
&\textrm{and}\nonumber\\
&N_{L8}(t+1) = N_{L9}(t),\nonumber\\
&\Phi_{L8}(t+1) = \Phi_{L9}(t).\label{freepropUpdateRule1}
\end{align}
At odd timesteps, by contrast, the cell $L8$ is grouped together with the cell $L7$ and it is now this pair whose physical states are swapped:
\begin{align}
&N_{L8}(t+1) = N_{L7}(t),\nonumber\\
& \Phi_{L8}(t+1) = \Phi_{L7}(t), \nonumber\\
&\textrm{and}\nonumber\\
&N_{L7}(t+1) = N_{L8}(t),\nonumber\\ \label{freepropUpdateRule2}
&\Phi_{L7}(t+1) = \Phi_{L8}(t).
\end{align}

Cell-groupings that are coloured blue in Fig.~\ref{CAUpdateRules} indicate places where the update rule is a modification of the free propagation rule to the functions that describe the action of the associated gate.  
We have here adopted a convention wherein the update rule for all gates differs from free propagation only at even timesteps.  The cells at the boundaries of the circuit of Fig.~\ref{CAUpdateRules} interact at even timesteps with the internal degree of freedom of a source or detector, and so there is just a single cell from the respective wires in the blue-coloured cell grouping in these cases. 

We can infer the modified update rules for gates from the functions that describe the action of each gate and the fact that all of these functions are time-reversal invariant. 

For a gate on a single wire, for instance, time-reversal invariance means that the function relating the state of cell $j+1$ at $t+1$ to the state of cell $j$ at time $t$ is the same as the function that relates the state of cell $j$ at $t+1$ to the state of cell $j+1$ at time $t$.  The analogous property holds for the two-input two-output gates.  





Consider the gate associated to the phase shifter and the pair of cells $R8$ and $R9$ that constitute its input and its output. From Eq.~\eqref{phasefipdynamics}, one can infer that the modified update rule for this pair of cells (at even timesteps) is: 
\begin{align}
&N_{R9}(t+1) = N_{R8}(t),\nonumber\\
&\Phi_{R9}(t+1) = \Phi_{R8}(t)\oplus 1 \nonumber\\
&\textrm{and}\nonumber\\
&N_{R8}(t+1) = N_{R9}(t),\nonumber\\
&\Phi_{R8}(t+1) = \Phi_{R9}(t)\oplus 1. \nonumber\\
\end{align}

Next, consider the gate associated to the first beam splitter and the quadruple of cells $L4, R4, L5, R5$ constituting its inputs and outputs.  
From Eq.~\eqref{BSfunction}, one can infer that the modified update rule for these cells (at even timesteps) is:
\begin{align}
&N_{L5}(t+1) = \Phi_{L4}(t) \oplus \Phi_{R4}(t),\nonumber\\
&\Phi_{L5}(t+1) = N_{L4}(t) \oplus \Phi_{R4}(t),\nonumber\\
&N_{R5}(t+1) = N_{R4}(t) \oplus N_{L4}(t) \oplus \Phi_{L4}(t) \oplus \Phi_{R4}(t),  \nonumber\\
&\Phi_{R5}(t+1) = \Phi_{R4}(t), \nonumber\\
&\textrm{and}\nonumber\\
&N_{L4}(t+1) = \Phi_{L5}(t) \oplus \Phi_{R5}(t),\nonumber\\
&\Phi_{L4}(t+1) = N_{L5}(t) \oplus \Phi_{R5}(t),\nonumber\\
&N_{R4}(t+1) = N_{R5}(t) \oplus N_{L5}(t) \oplus \Phi_{L5}(t) \oplus \Phi_{R5}(t),  \nonumber\\
&\Phi_{R4}(t+1) = \Phi_{R5}(t). \nonumber\\ 
\end{align}



Now consider the cell $L1$, which is the output of the preparation device associated to the source. This cell also has a modified update rule which (on even timesteps) describes a nontrivial interaction with the internal state of the device.  For our purposes, it is irrelevant how the internal state of such a device itself transforms when there are excitations travelling backwards along the wires, since we are not considering these scenarios.  Consequently, it is sufficient to stipulate how $L1$ updates.  If $t_*$ denotes an even timestep at which the source injects an excitation into the interferometer, then 
\begin{align}
&N_{L1}(t_*) = 1,\nonumber\\
&\Phi_{L1}(t_*) \; \textrm{is sampled uniformly at random.}\nonumber
\end{align}
By contrast, if $t$ denotes an even timestep at which the source {\em does not} inject an excitation into the interferometer, then 
\begin{align}
&N_{L1}(t) = 0,\nonumber\\
&\Phi_{L1}(t) \; \textrm{is sampled uniformly at random.}\nonumber
\end{align}

The cell $R1$, which is stipulated to correspond to a source of vacuum simply never emits an excitation.
Consequently, at all even timesteps $t$, we have
\begin{align}\label{sourceofvacuum}
&N_{R1}(t) = 0,\nonumber\\
&\Phi_{R1}(t) \; \textrm{is sampled uniformly at random.}\nonumber
\end{align}


Now consider the gate associated to the which-way detector (labelled by $D$ in Fig.~\ref{MZIAsCircuit}(a))  and the pair of cells $R8$ and $R9$ that constitute its input and its output.   We consider the case where it implements a {\em nondestructive} measurement of occupation number.  
Suppose that the detector fires at even timestep $t_*$. 

Recall that the knowledge-updating rule for such a measurement consisted of two steps, one of which constituted a resolving of uncertainty regarding the physical state and the other of which constituted a phase randomization.   

We assume that the detector is of a type that can detect an excitation travelling in either direction along the wire. (This is the case, for instance, if a photon is detected by atomic absorption.)
It follows that all we can infer based on the fact that the detector fires at timestep $t_*$, is that either $R8$ or $R9$ was excited at the end of the preceding timestep.  However, given that we assume that excitations are only injected into the interferometer at $L1$, it follows that we can infer that it was in fact $R8$ that was excited.  

Meanwhile, the phase randomization step can be described  by the following modified update rule for $R8$ and $R9$ (at even timesteps): 
\begin{align}
&N_{R9}(t+1) = N_{R8}(t),\nonumber\\
&\Phi_{R9}(t+1) \; \textrm{is sampled uniformly at random}\nonumber\\
&\textrm{and}\nonumber\\
&N_{R8}(t+1) = N_{R9}(t),\nonumber\\
&\Phi_{R8}(t+1)  \; \textrm{is sampled uniformly at random.} \nonumber\\
\end{align}

The detectors at the output ports of the second beamsplitter act similarly to the detector on the $R$ arm.  
Consider the cell $L16$, for instance.  Whether the detector fires at even timestep $t$ or not is simply a function of whether $N_{L16}=1$ or not at that timestep.   Regardless of whether it fires or not, this detector acts as a source of vacuum for $L16$.  Hence, the modified update rule for $L16$ at all even timesteps is exactly that of Eq.~\eqref{sourceofvacuum}, namely,
\begin{align}
&N_{L16}(t) = 0,\nonumber\\
&\Phi_{L16}(t) \; \textrm{is sampled uniformly at random.}\nonumber
\end{align}

This concludes our demonstration that the toy field theory can be described in terms of spatially localized modes interacting locally. 
 

\end{document}